\documentclass[aps,pre,onecolumn,11pt,floatfix,superscriptaddress,notitlepage]{revtex4-1}
\usepackage{graphicx, array}
\usepackage{dcolumn}
\usepackage{bm}
\usepackage{latexsym}
\usepackage{epstopdf}
\usepackage{amsmath}
\usepackage{lipsum}
\usepackage{amssymb}
\usepackage{mathtools}
\usepackage{resizegather}
\usepackage[flushleft]{threeparttable}
\usepackage{booktabs}
\usepackage{tabularx}
\newcolumntype{P}[1]{>{\centering\arraybackslash}p{#1}}
\newcolumntype{M}[1]{>{\centering\arraybackslash}m{#1}}

\usepackage[dvipsnames]{xcolor}
\usepackage{color}
\newcommand{\red}[1]{{\bf\textcolor{red}{#1}}}

\begin{document}

\title{Stochastic thermodynamics and modes of operation of\\ a ribosome: A network theoretic perspective} 
\author{Annwesha Dutta}
\affiliation{Department of Physics, Indian Institute of Technology,
Kanpur 208016, India.}
\author{Gunter M Sch\"{u}tz}
\affiliation{Institute of Complex Systems II, Forschungszentrum J\"{u}lich - 52425 J\"{u}lich, Germany.}
\author{Debashish Chowdhury}
\email[]{Corresponding author; e-mail: debch@iitk.ac.in}
\affiliation{Department of Physics, Indian Institute of Technology,
Kanpur 208016, India}
\date{\today}

\begin{abstract}
The ribosome is one of the largest and most complex macromolecular machines in living cells. It polymerizes a protein in a step-by-step manner as directed by the corresponding nucleotide sequence on the template messenger RNA (mRNA) and this process is referred to as `translation' of the genetic message encoded in the sequence of mRNA transcript. In each successful chemo-mechanical cycle during the (protein) elongation stage, the ribosome elongates the protein by a single subunit, called amino acid, and steps forward on the template mRNA by three nucleotides called a codon. Therefore, a ribosome is also regarded as a molecular motor for which the mRNA serves as the track, its step size is that of a codon and two molecules of GTP and one molecule of ATP hydrolyzed in that cycle serve as its fuel. 
What adds further complexity is the existence of competing pathways leading to distinct cycles, branched pathways in each cycle and futile consumption of fuel that leads neither to elongation of the nascent protein nor forward stepping of the ribosome on its track. We investigate a model formulated in terms of the network of discrete chemo-mechanical states of a ribosome during the elongation stage of translation. The model is analyzed using a combination of stochastic thermodynamic and kinetic analysis based on a graph-theoretic approach. We derive the exact solution of the corresponding master equations. We represent the steady state in terms of the cycles of the underlying network and discuss the energy transduction processes. We identify the various possible modes of operation of a ribosome in terms of its average velocity and mean rate of GTP hydrolysis. We also compute entropy production as functions of the rates of the interstate transitions and the thermodynamic cost for accuracy of the translation process.
\end{abstract}
\maketitle

\newpage
\section{Introduction}

The synthesis of proteins, performed by a macromolecular machine, called ribosome, 
is one of the fundamental processes inside every living cell \cite{spirin_ribosomes_2002,aitken_single_2010}. 
The sequence of monomeric subunits of the protein, called amino acid,  is directed by the sequence of the triplets of monomeric subunits of a messenger RNA (mRNA) template; each triplet is called a codon. In the language of information processing, the template-directed polymerization of a protein by a ribosome is called translation (of genetic message). The ribosome hydrolyzes two molecules of GTP and one molecule of ATP (a strongly exergonic or `downhill' reaction) to elongate the nascent protein by one amino acid. The GTP molecules are hydrolyzed in a complete elongation cycle whereas the ATP molecule is hydrolyzed during the prior aminoacylation reaction that results in an activated aminoacyl-tRNA (aa-tRNA). The amino acid brought in by an aa-tRNA at the beginning of an elongation cycle subsequently forms a peptide bond with the nascent growing protein. Simultaneously with the elongation of this polypeptide by one amino acid, the ribosome moves forward by one codon on the mRNA template. Therefore, a ribosome can also be regarded as a molecular motor for which the mRNA template serves as a track; the motor is fueled by GTP and ATP hydrolysis and its step size on the track is a single codon.

Over the last few decades, enormous progress has been made in characterizing the structure of the ribosome and its dynamics during the process of translation by X-ray crystallography, cryo-electron microscopy and combinations of biochemical and biophysical single molecule techniques such as smFRET,  
\cite{frank_structure_2010,wasserman_multiperspective_2016,
frank_molecular_2011,chen_structural_2015,fischer_ribosome_2010,rodnina_ribosomes_2011}. 
There are many theoretical studies of the translation process based on these experimental revelations 
\cite{siwiak_comprehensive_2010,gilchrist_model_2006,xie_model_2012,rudorf_protein_2015,
savir_ribosome_2013,basu_modeling_2007,garai_stochastic_2009,garai_fluctuations_2009,
chowdhury_stochastic_2013,sharma_distribution_2011,sharma_quality_2010,dutta_generalized_2017}. Fluitt et al \cite{fluitt2006}, Rudorf et al. \cite{rudorf_protein_2015}, Vieira et al \cite{hatzimanikatis2016}, Dana et al \cite{tuller2014} developed detailed stochastic kinetic models capturing the translation process in the presence of cognate, near cognate and non cognate aa-tRNA and also considered the inhomogeneity of the mRNA transcript. However, to our knowledge, there is no study of the stochastic thermodynamics and operational modes of ribosome. Our model is designed to provide a clear understanding of the energy transduction processes during translation by decomposing the complex network of distinct states into its cycles, focusing on the energetics and thermodynamic picture in terms of fluxes, their conjugate affinities and entropy change associated with every cycle. It helps us understand what chemical, mechanical and chemo-mechanical cycles compete in the network. It also helps us compute important motor properties like velocity, hydrolysis rate in terms of external parameters like concentration of the different particles which bind to the ribosome. The flux balance relations give us the expression for stall force and balanced potential.

The directed movement of the ribosome on its mRNA track is, however, 
noisy because of the thermal motion  of the surrounding medium 
and the low concentration of the molecules involved in the chemical reactions. 
The ribosome can be regarded as a thermodynamically open system that is coupled 
to various reservoir potentials. More specifically, the reservoirs include 
not only a thermal reservoir at a constant temperature, but also several 
chemical reservoirs maintained at the respective chemical potentials and a 
``force reservoir'' describing a load force acting on the ribosome. 
The main aim of the present work is to improve our theoretical understanding of 
the translation process from the perspective of stochastic thermodynamics 
\cite{seifert12,hwang18,gnesotto18,andrieux06,qian16,rao16,seifert18,gerritsma10,astumian10,
seifert11,loutchko17,julicher97} by a detailed and quantitative description 
of the various stages of the chemo-mechanical cycle that the ribosome undergoes 
in a single translation step.

We describe the kinetics of the ribosome in terms of a Markov network 
of observable mesoscopic states, using experimentally measured interstate 
transition rates \cite{rodnina_ribosome_2017,riccardo_belardinelli_choreography_2016} .  
For the exact analytical treatment of this multiple-pathway discrete-state Markov model we develop a 
graph theoretic framework, following refs.\cite{hill_free_1989,hill_studies_1966,schnakenberg_network_1976}. 
In this approach, the network of the states is represented by a graph consisting of vertices
and edges. 
Vertices correspond to the observable mesostates and the directed edges represent the possible transitions 
between these states. The stationary solution is obtained by studying subgraphs 
of the graph in a systematic analysis that we outline in this paper. 

This approach is powerful since the ratio of the products of the transition rates along a 
cycle and its time reversal, that are of interest from the perspective of stochastic thermodynamics,   
is independent of the mesoscopic states. Cycle fluxes, entropy production rate per cycle, thermodynamic force per cycle  \cite{qian_cycle_2005,liepelt_impact_2010,liepelt_steady-state_2007}
and other important quantities can then be calculated explicitly as a function
of the rates. This allows for discussion of the operation mode of the ribosome
in terms of average velocity and hydrolysis rate.

\section{Markov model of the mechanochemical cycle of a ribosome}

\subsection{The ribosome as a complex nanomachine}
\label{Sec:rcn}

Each ribosome is built from two loosely associated subunits: (1) The small subunit, which is responsible for all the processes related to the deciphering the genetic code present in the mRNA, and (2) the large subunit which serves as the catalytic center where the formation of peptide bonds takes place. The two subunits are joined together by flexible connectors. A class of adaptor molecules, that bring in the amino acid subunits, move along the intersubunit space. One end of the tRNA molecule, that participates in the decoding of the genetic message, interacts with the mRNA. The other end of the tRNA, that brings in the amino acid subunit, interacts with the large subunit. For each end of the tRNA molecule three binding sites are available on the respective subunits of the ribosome. These binding sites are designated by the letters `A' (acceptor site) , `P' (peptidyl site) and `E' (exit site), respectively, in that sequence along the direction of translocation of the tRNAs in the intersubunit space. Each tRNA not only brings in an amino acid whose incorporation elongates the polypeptide (protein), but also holds it transiently thereafter before irreversibly transfering the polypeptide to the next tRNA.

In order to account for the main features of the elongation cycle, we {introduce a model that is an extended version of the model of translation developed earlier by Dutta and Chowdhury \cite{dutta_generalized_2017}. This model explicitly incorporates four competing pathways corresponding to four different types of aa-tRNA, namely, correctly charged cognate tRNA, mischarged cognate tRNA,  correctly charged near-cognate tRNA and correctly charged non-cognate tRNA. Each of these pathways, shown in Fig.~\ref{fig1:translation_pict}, comprises of 5 distinct states, each of which corresponds to a distinct conformational (or `chemical') state of  the ribosome during translation of a single codon. The 5-state subnetwork along all the four pathways looks identical in spite of the fact that these correspond to four different types aa-tRNA. The difference between the four pathways are captured by the difference in the numerical values of the rates of the same inter-state transitions along different pathways. Therefore, we begin by explaining the different conformational changes that the ribosome undergoes along each of these four pathways.

\begin{figure}
\begin{center}
\includegraphics[width=0.7\columnwidth,angle=-90]{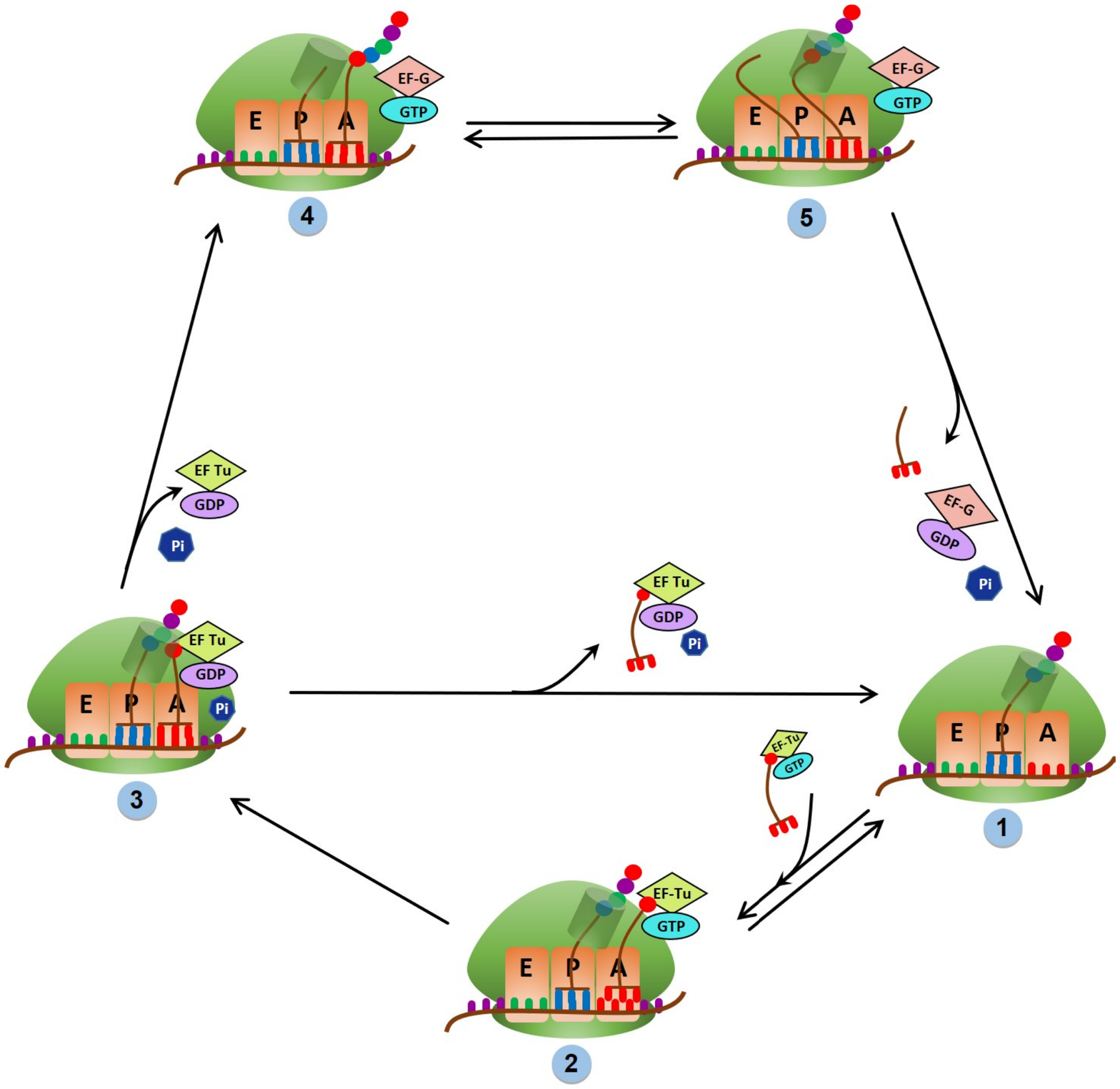}
\end{center}
\caption{Pictorial representation of one complete cycle, along with intra-cycle branched pathways, during the elongation stage of translation at an arbitrary codon on the mRNA transcript.
The ribosome is represented by the three sites where the tRNA binds, viz., the aminoacyl site (A), 
the peptidyl site (P) and the exit site (E). Transition 1 to 2 represents the binding of ternary complex EF-Tu.GTP.aa-tRNA to the A site of the ribosome. 2 to 1 represents the codon-anticodon mismatch. Transition 2 to 3 represents the hydrolysis of GTP for proofreading to ensure whether the aa-tRNA is cognate. 3 to 1 transition represents the rejection aa-tRNA as a result of proof reading. During transition 3 to 4, the peptide bond formation takes place by linking the amino acid to the growing nascent protein. transition between 4 and  represents the brownian ratchet motion and 5 to 1 represents the hydrolysis of GTP and translocation of the ribsosome on the mRNA track. }
\label{fig1:translation_pict}
\end{figure}

In Fig.~\ref{fig1:translation_pict} the state labeled by 1 represents the situation where both the E and A sites of the ribosome are empty while the site P is occupied by the tRNA carrying the nascent protein. In step $1 \rightarrow 2$
the ternary complex EF-Tu.GTP.aa-tRNA with the elongation factor EF-Tu, one GTP, and an aminoacyl-tRNA 
(aa-tRNA) binds to the ribosome. The reverse transition $2 \rightarrow 1$ describes the unbinding of the same ternary complex from the ribosome. While the system in the state $2$, the enzyme GTPase of the EF-Tu is activated, leading to the hydrolysis of a single GTP molecule to GDP and inorganic Phosphate $\mathrm{P_i}$ which is captured by the irreversible transition ($2 \rightarrow 3$). At this stage, the aatRNA may get rejected ($3 \rightarrow 1$). The physical implications of this transition, called kinetic proofreading, will be discussed further later in this section.

However, if the selected aa-tRNA is not rejected along the path $3 \rightarrow 1$, the growing polypeptide is then linked by a peptide bond to the amino acid supplied by the selected aa-tRNA thereby transferring the polypeptide from the tRNA located at the P site to the tRNA at the A site. After the transfer of the polypeptide, the deacylated tRNA remains at the P site. This `peptidyl transferase' activity of the ribosome thus results in the elongation of the polypeptide by one subunit. The composite process comprising the departure of the products of GTP hydrolysis, together with that of the EF-Tu,  and the formation of the peptide bond between the amino acid supplied by the  selected aa-tRNA and the growing polypeptide  is represented by the single transition $3 \rightarrow 4$. 

While the polypeptide gets elongated by one amino acid, a fresh molecule of GTP enters 
bound with an elongation factor EF-G.  Spontaneous Brownian (relative) rotation of the
two subunits of the ribosome coincides with the back and forth transition 
($4 \leftrightharpoons 5$) after the amino acid incorporation 
between the  so-called classical and hybrid configurations of the two tRNA molecules. 
In the classical configuration, both ends of the two tRNA molecules
correspond to the locations of P and A sites. In contrast, in the hybrid configuration, 
the ends of tRNA molecules interacting with the large subunit are found at the locations 
of E and P sites, respectively, while their opposite ends interacting with the small subunit 
continue to be located at the P and A sites, respectively. 

Finally the hydrolysis of the fresh GTP drives the irreversible transition 
$5 \rightarrow 1$ along the pathway for correct amino acid incorporation, thereby completing a cycle. 
This involves the translocation of the ribosome on its track by one codon and, 
simultaneously, that of the two tRNAs inside the ribosome 
by one binding site also on the small subunit, followed by the deacylated (i.e. bare) tRNA exiting from the 
E site. The EF-G.GDP complex dissociates from the ribosome and the initial state 
1 is again attained. The deacylated tRNA is then aminoacylated (`charged') by an enzyme, called aminoacyl tRNA synthetase, by hydrolyzing a molecule of Adenosine Triphosphate (ATP) into Adenosine Monophosphate (AMP) and inorganic pyrophosphate (PPi). It is often believed that the energy of the chemical bond between the amino acid and tRNA is later used by the ribosome for the formation of peptide bond between the amino acid and the nascent polypeptide \cite{spirin_ribosomes_2002}.

For clarity, Fig.~\ref{fig2_spatial_model} indicates how these internal processes along each of the four pathways relate to the translocation of the ribosome along the mRNA template. The forward movement by one codon 
occurs from state 5 and leads to state 1. The total displacement from the start codon
at time $t$ is an integer multiple $n$ of the average length ${\ell} \approx 1$nm of a codon where 
$n$ is the number of elongation cycles completed up to time $t$. Hence the average
ribosome velocity $v$ along the mRNA is proportional to the elongation rate 
$e = v/{\ell}$ which is the average number of completed elongation cycles per time unit
which is identical to the average rate of elongation of the nascent protein. Since in this paper
we treat the ribosome as a molecular motor, we use the term velocity instead of the rate of elongation.

\begin{figure}
\begin{center}
\includegraphics[width=0.7\columnwidth,angle=-90]{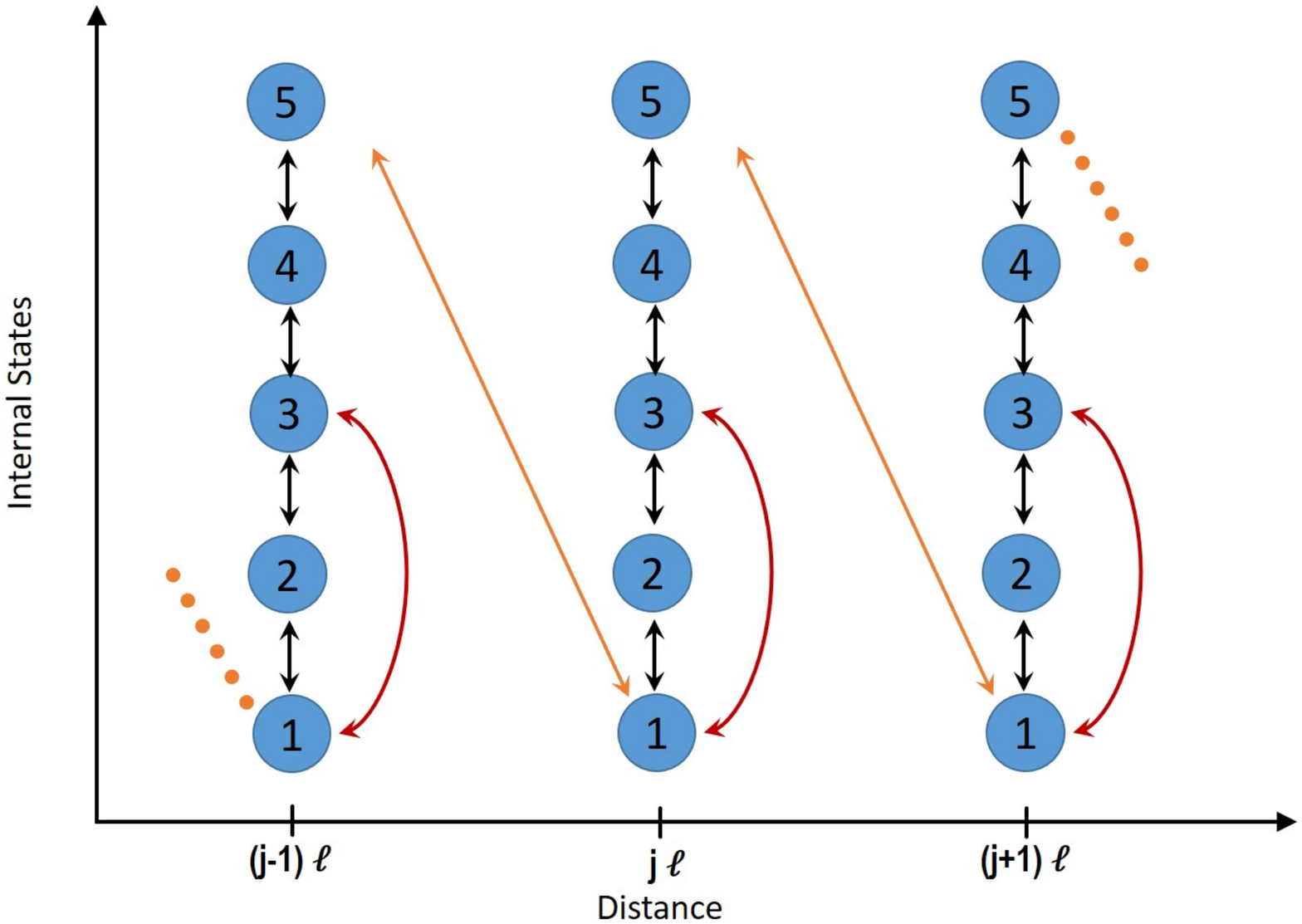}
\end{center}
\caption{A schematic representation of the Markov model of translation. Both the chemical/conformational states of the ribosome at a given codon as well as the inter-state transitions are shown, alongwith the sequence of codons on the mRNA template that also indicate the positions of the ribosome on its track. The circular discs labelled by the indices $1,\dots, 5$ represent the chemical/conformational states of the ribosome at a codon while the tick marks on the horizontal axis labelled by $\dots, (j-1){\ell}, j , (j+1){\ell}, \dots$ denote the positions of the successive codons.}
\label{fig2_spatial_model}
\end{figure}

\begin{figure}
	\begin{center}
		\includegraphics[width=0.7\columnwidth]{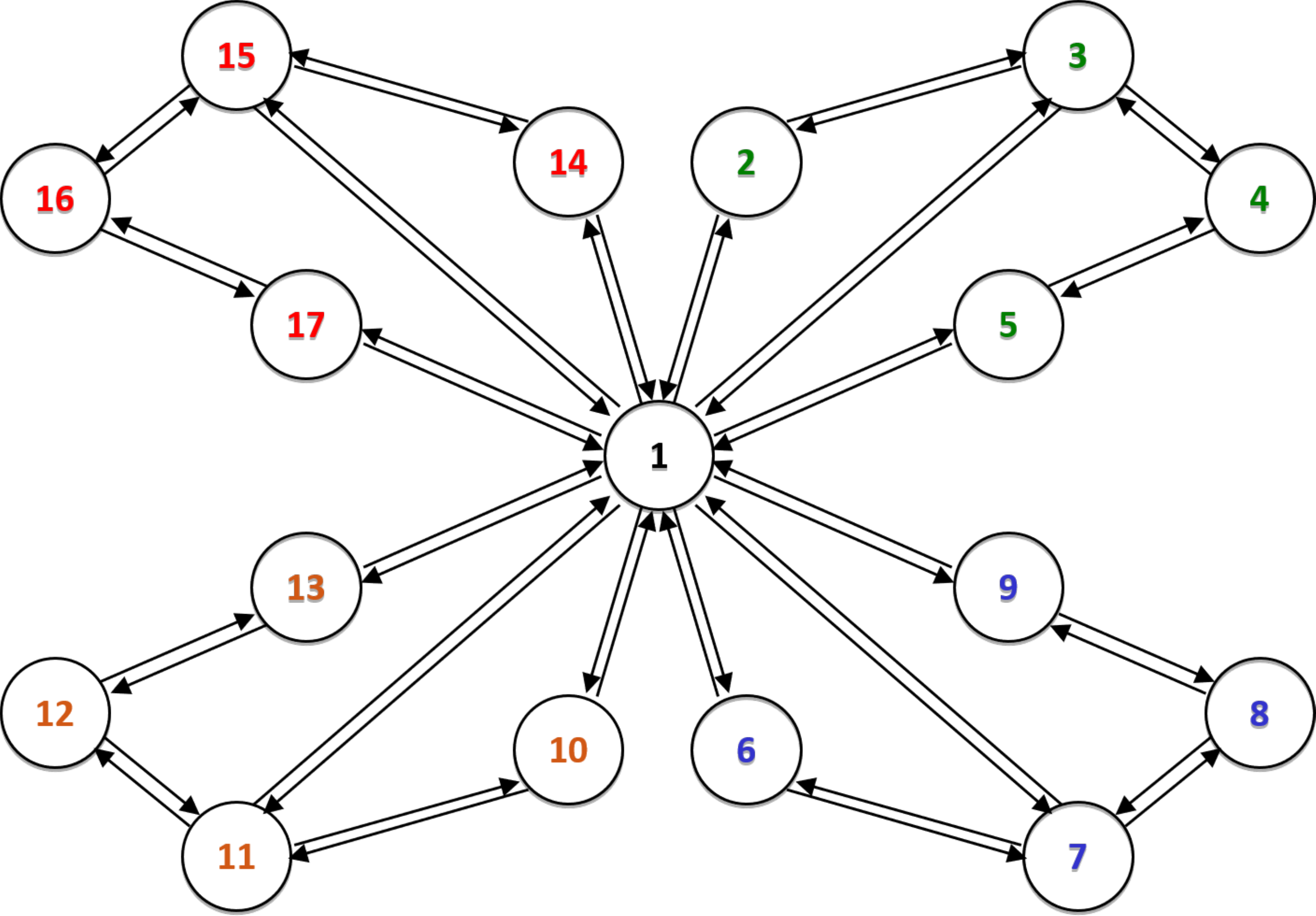}
	\end{center}
	\caption{The kinetic Markov network of the ribosomal elongation cycle. At every codon 
	position, the ribosome undergoes changes between different conformations that are labelled 
	by $i=1, 2, 3,...,17$. The $k_{ij}$ are the transition rates to move 
	from conformation $i$ to conformation $j$. Notice that there are multiple pathways that 
	the ribosome can follow.}
\label{fig4:detailednetwork}
\end{figure}

In reality, the incoming aa-tRNA need not be cognate to the codon in the ribosomal A site (correct aa-tRNA). Instead, it may be a mischarged cognate (wrongly charged), or a near cognate (one of the three nucleotides of the aa-tRNA anti-codon does not match the three nucleotides in the codon) or a non-cognate (none of the nucleotides on the aa-tRNA match) aa-tRNA. Therefore, the overall model of the elongation cycle consists of four subnetworks, 
as shown in Fig.~\ref{fig4:detailednetwork}, each of which looks identical to the 5-state network of 
Fig.\ref{fig1:translation_pict}. The binding of the four possible aa-tRNA molecules (each as a distinct ternary complex formed with GTP and EF-Tu) cause transitions to their respective subnetworks from the state 1.

High fidelity of translation beyond the level guaranteed by thermodynamics is known to arise from kinetic proofreading (transition $3 \to 1$ in Fig.\ref{fig1:translation_pict}) whereby an aa-tRNA is rejected. The overall network depicted in Fig.~\ref{fig4:detailednetwork} implies that even a correctly charged cognate tRNA may get rejected by kinetic proofreading, albeit with a low probablity, in spite of perfect codon-anticodon matching. In contrast, a mischarged cognate aa-tRNA may escape detection by the same quality control mechanism leading to an eventual translational error by incorporating a wrong amino acid in the elongating protein.

It may be noted that in the original version of the model reported earlier by Dutta and 
Chowdhury \cite{dutta_generalized_2017}, some of the transitions were assumed to irreversible because the reverse transitions were not observed in any experiments. In the extended version adopted here all the irreversible transitions are replaced by reversible transitions where the transitions that have 
not been observed experimentally are treated as highly improbable  (Fig.~\ref{fig1:translation_pict}) by assigning a  hypothetical small rate $10^{-5}$ s$^{-1}$.
This weak reversibility condition \cite{Harr07}, which is based on the law of mass action, allows for a discussion of the entropy production in the process. Moreover, the cycle `1-14-15-16-17-1' which corresponds to the incorporation of non cognate tRNA is very improbable. Therefore the probability of a ribosome being in the conformational states 14, 15, 16, 17 must be extremely low. To ensure that, we need to have very low forward rates and very high rejection rates. To elaborate upon this point, let us consider the transition of a ribosome from 1 to 14. While the rate of transition from state 1 to state 14 is very low, the ribosome, if it somehow reaches the state 14, gets trapped in that state if the rejection rate is also low. This would lead to a non negligible probability of occurrence of state 14. To avoid this anomaly, a high rate $10^{5}$ s$^{-1}$ is assigned to rejection rates that have not been observed in any experiments so far.

\subsection{Stochastic reaction kinetics} 

During the course of an elongation cycle, the ribosome is fed by energy from EF-Tu and EF-G 
mediated hydrolysis of GTP to GDP and inorganic phosphate $\mathrm{P_i}$ and ATP hydrolysis during aminoacylation for ribosome catalyzed transpeptidation. These free energies, $\Delta G$ in each hydrolysis, 
are consumed for (a) decoding the genetic information encoded in the codon sequences of the mRNA into the amino acid sequence forming the polypeptide, and (b) also for translocation along the mRNA track. 

\begin{figure}
	\begin{center}
		\includegraphics[width=0.7\columnwidth,angle=-90]{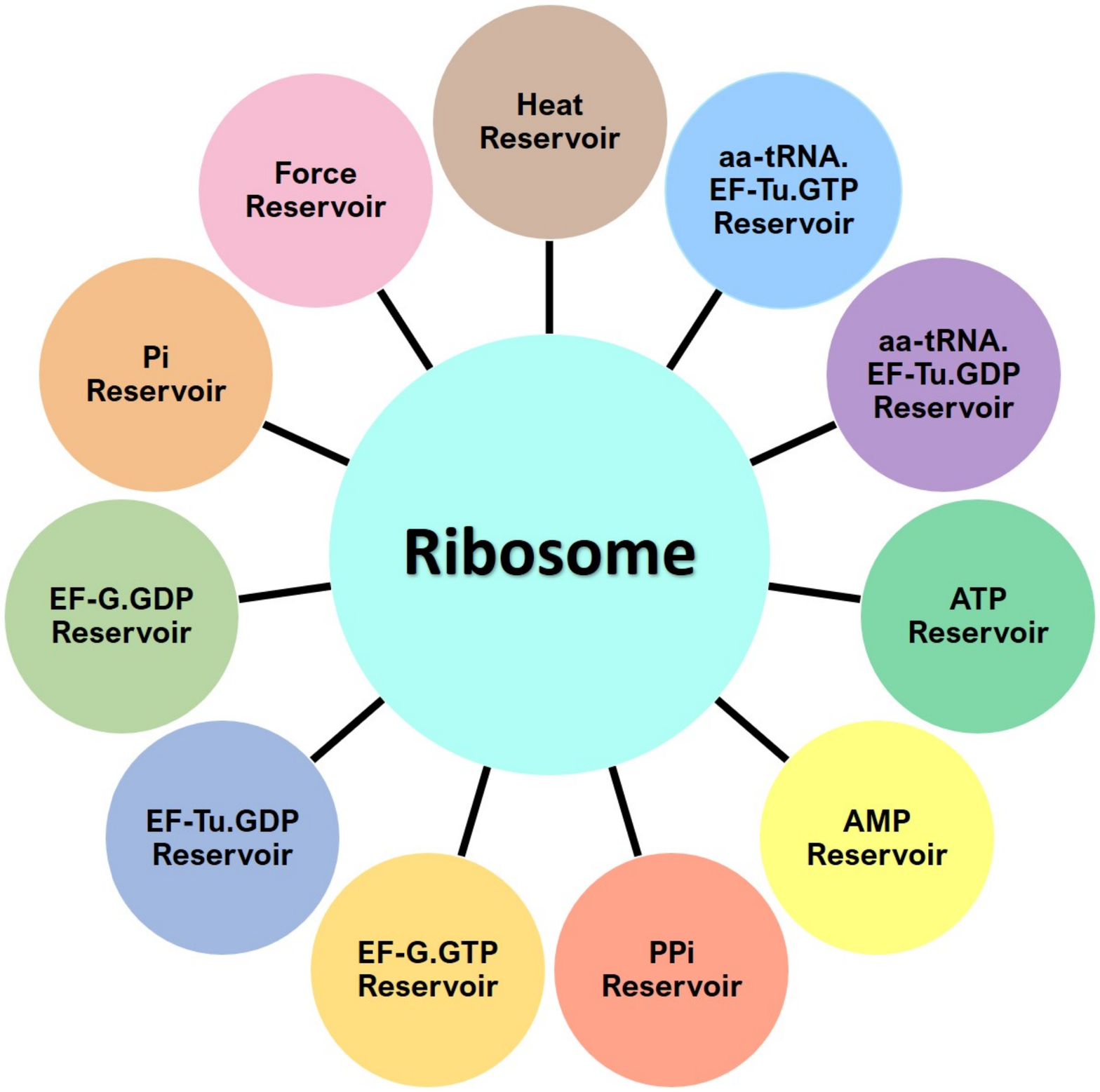}
	\end{center}
	{\caption{Overview of the different reservoirs which are in contact with the Ribosome 
		machinery, viz. (i) Heat Reservoir characterized by temperature $T$; 
		(ii) Particle Reservoirs for the chemical species aa-tRNA.EF-Tu.GTP, aa-tRNA.EF-Tu.GDP, EF-Tu.GDP, 
		EF-G.GTP, EF-G.GDP, and $\mathrm{P_i}$ characterized by their respective chemical potentials; 
		and (iii) Force reservoir characterized by the force $F_{ext}$.}}
	\label{fig3:reservoir}
\end{figure}

The complex ribosome machine can thus be viewed as a small system coupled 
to multiple reservoirs that act as sources and sinks of particles and energy 
for the system (Fig.~\ref{fig3:reservoir}). Under isothermal conditions
the different reservoirs are: (1) Thermal reservoir at temperature $T$; 
(2) Particle reservoirs for the chemical species aa-tRNA.EF-Tu.GTP, aa-tRNA.EF-Tu.GDP, EF-Tu.GDP, 
EF-G.GTP, EF-G.GDP, $Pi$, ATP, AMP and $PPi$ characterized by the chemical potentials 
$\mu_\text{aa-tRNA.EF-Tu.GTP}$, $\mu_\text{aa-tRNA.EF-Tu.GDP}$, $\mu_\text{EF-Tu.GDP}$, $\mu_\text{EF-G.GTP}$, 
$\mu_\text{EF-G.GDP}$, $\mu_\text{Pi}$, $\mu_\text{ATP}$, $\mu_\text{AMP}$, $\mu_\text{PPi}$; 
(3) ``Force reservoir'' that comes into play in the presence of an external force $F_{ext}$ acting on the ribosome, such as a load force opposing the natural forward stepping of the machine. 

The ribosome may remain attached to a membrane and pull the mRNA template through it, translating one codon after another. In contrast, we have assumed the mRNA template to be static along which the ribosome steps forward unidirectionally by one codon at a time. The difference between the two scenarios is merely the difference in the choice of the frame of reference, as pointed out explicitly earlier in a paper by Cozzarelli et al. \cite{cozzarelli06}. The experimental set up that faithfully captures the scenario envisaged in our theoretical modeling is that developed a few years ago by Bustamante and collaborators \cite{bustamante2014}.

The extra work that is done in the mechanical movement on the mRNA track in order to
overcome the load force is also supplied by the free energy released in hydrolysis. A load
force leads to a reduction in the speed of the ribosome, unless it is compensated (or overcompensated)
by an increase of the chemical potential differences that contribute to the free energy.
Analogously, an external force may act in the same direction as the natural motion of the ribosome,
thus either enhancing its speed or reducing the free energy required from hydrolysis.

Thus the functioning of the ribosome machine depends on the thermodynamic forces generated by 
chemical potential differences of the particle reservoirs and on the external force $F_{ext}$
applied to the machine. The chemical potential differences of the particle reservoirs 
for the ribosome apparatus arising from hydrolysis are given by
\begin{eqnarray} 
\label{eq:chem_pot_diffTu}
\Delta \mu_\text{Tu,1} 
& = & \mu_\text{\text{aa-tRNA}.EF-Tu.GTP} - \mu_\text{EF-Tu.GDP} - \mu_\text{Pi}  \\
\label{eq:chem_pot_diffTu}
\Delta \mu_\text{Tu,2} 
& = & \mu_\text{\text{aa-tRNA}.EF-Tu.GTP} - \mu_\text{\text{aa-tRNA}.EF-Tu.GDP}- \mu_\text{Pi}  \\
\label{eq:chem_pot_diffG}
\Delta \mu_\text{G} & = & \mu_\text{EF-G.GTP} - \mu_\text{EF-G.GDP} - \mu_\text{Pi} \\
\label{eq:chem_pot_diffG}
\Delta \mu_\text{A} & = & \mu_\text{ATP} - \mu_\text{AMP} - \mu_\text{PPi}
\end{eqnarray}
It should be noted that EF-Tu is a GTPase that catalyzes the selection and binding 
of aa-tRNA with the help of hydrolysis of chemical fuel GTP whereas EF-G is a GTPase 
that catalyzes the translocation step of the ribosome.

The system attains chemical equilibrium, without any average net displacement of the ribosome, when 
$\Delta \mu_\text{Tu,1}=\Delta \mu_\text{Tu,2}=\Delta \mu_\text{G}=\Delta \mu_\text{A}=0$. However, 
when $\mu_\text{aa-tRNA.EF-Tu.GTP} \gg \mu_\text{EF-Tu.GDP}
+ \mu_\text{Pi}$ and $\mu_\text{EF-G.GTP} \gg \mu_\text{EF-G.GDP} + \mu_\text{Pi}$, 
the likelihood that the GTP will bind to the active site for hydrolysis is much higher 
than the binding of GDP for GTP synthesis. Again, $\mu_\text{ATP} \gg \mu_\text{AMP}
+ \mu_\text{PPi}$ increases the likelihood that deacyl tRNA gets aminoacylated with the help of aminoacyl tRNA synthetase. These non-vanishing chemical potential 
differences drive the system out of equilibrium and generate a directed movement of the 
ribosome as indicated above. Moreover, the conversion of chemical 
energy into mechanical energy involves a thermodynamic cost which results in an increase 
of the entropy in the environment.

From this thermodynamic perspective, the biological processes that the ribosome 
undergoes during an elongation cycle can be understood as Markovian transitions 
between the distinct states, driven by thermal fluctuations, external forces, and chemical 
reservoirs that supply the molecules required for the transitions to take place. 
In our approach, the transition rates between the states 
are assumed to be independent of the spatial position of the ribosome on the mRNA track. 
This allows us to study the elongation cycle just in terms of the  internal 
states of the ribosome, without reference to its location on the mRNA template.

Thus the model reduces to a multi-pathway process as shown in Fig.~\ref{fig4:detailednetwork}, 
with the transition rates from a state $i$ to some other state $j$ denoted by $k_{ij}$. 
The experimental values of the rate constants used in our model are shown in 
Table~\ref{tab:experimental_values}. The rates depend on the concentrations of the complexes that bind to the ribosome in the following manner:\\
$k_{12} = \omega^{0}_{12} ~[\text{aa-tRNA.EF-Tu.GTP}]_{co}$,\\ 
$k_{16} = \omega^{0}_{16} ~[\text{aa-tRNA.EF-Tu.GTP}]_{mc}$,\\ 
$k_{1,10} = \omega^{0}_{1,10} ~[\text{aa-tRNA.EF-Tu.GTP}]_{nr}$,\\ 
$k_{1,14} = \omega^{0}_{1,14} ~[\text{aa-tRNA.EF-Tu.GTP}]_{no}$,\\ 
$k_{45} = \omega^{0}_{45} ~[\text{EF-G.GTP}]$,\\ 
$k_{89} = \omega^{0}_{89} ~[\text{EF-G.GTP}]$,\\ 
$k_{12,13} = \omega^{0}_{12,13} ~[\text{EF-G.GTP}]$,\\ 
$k_{14,15} = \omega^{0}_{14,15} ~[\text{EF-G.GTP}]$,\\ 
$k_{13} = \omega^{0}_{13} ~[\text{aa-tRNA.EF-Tu.GDP}] ~[\text{Pi}]$,\\ 
$k_{17} = \omega^{0}_{17} ~[\text{aa-tRNA.EF-Tu.GDP}] ~[\text{Pi}]$,\\ 
$k_{1,11} = \omega^{0}_{1,11} ~[\text{aa-tRNA.EF-Tu.GDP}] ~[\text{Pi}]$,\\ 
$k_{1,15} = \omega^{0}_{1,15} ~[\text{aa-tRNA.EF-Tu.GDP}] ~[\text{Pi}]$,\\ 
$k_{43} = \omega^{0}_{43} ~[\text{EF-Tu.GDP}] ~[Pi]$,\\ 
$k_{12,11} = \omega^{0}_{12,11} ~[\text{EF-Tu.GDP}] ~[\text{Pi}]$,\\ 
$k_{16,15} = \omega^{0}_{16,15} ~[\text{EF-Tu.GDP}] ~[\text{Pi}]$,\\ 
$k_{87} = \omega^{0}_{87} ~[\text{EF-Tu.GDP}] ~[\text{Pi}]$,\\ 
$k_{15} = \omega^{0}_{15} ~[\text{EF-G.GDP}] ~[\text{Pi}]$,\\ 
$k_{17} = \omega^{0}_{17} ~[\text{EF-G.GDP}] ~[\text{Pi}]$. \\
Here the square brackets $[.]$ denotes the concentration of the complexes. The subscript $co$, $mc$, $nr$ and $no$ represent the cognate, mischarged, near cognate and non cognate aa-tRNA respectively. The $\omega_{ij}$ denote the binding rate constant for the complexes that bind to the ribosome.

For our calculations, we have used the concentration of ternary complexes of cognate, near-cognate and non-cognate aa-tRNA to be $50 \, \mu \mathrm{M}$, i.e.,\\ $[\text{aa-tRNA.EF-Tu.GTP}]_{co}=[\text{aa-tRNA.EF-Tu.GTP}]_{nr}\\=[\text{aa-tRNA.EF-Tu.GTP}]_{no}=[\text{EF-G.GTP}]=50 \, \mu \mathrm{M}$.\\ The concentration of the mischarged aa-tRNA ternary complex is taken to be $5 \mu M$. We have taken the concentration of mischraged ternary complex to be low because the relatively rare event of  mischarging of aa-tRNA occurs only when the aminoacylation of tRNA escapes quality control done be by amino acyl tRNA synthetase. This error  generally occurs when the cell is under stress. The concentration for the rest of the complexes were taken to be $[\text{aa-tRNA.EF-Tu.GDP}]=[\text{EF-Tu.GDP}] = [\text{EF-G.GDP}] = [\text{Pi}] = 50 \mu \mathrm{M}$.

We have used the experimental transition rates that were reported by Rodnina et al.\cite{rodnina_ribosome_2017} (steps involving initial binding, accomodation, proofreading and peptide elongation for cognate and near cognate aa-tRNA) and Belardinelli et al \cite{riccardo_belardinelli_choreography_2016} (steps involving translocation). The experiments in ref.\cite{rodnina_ribosome_2017} were conducted at $20^{\circ}$C, whereas in ref.\cite{riccardo_belardinelli_choreography_2016} the experiments were carried out at $37^{\circ}$C. Since the rates in these experiments are very sensitive to temperature, we have estimated the rates at $37^{\circ}$C that correspond to the rates reported in ref.\cite{rodnina_ribosome_2017} at $20^{\circ}$C using the Arrhenius equation following the method used by Rudorf et al. \cite{rudorf_protein_2015}. Moreover, we have assumed that the steps involved in translocation have the same rates for cognate, mischarged and near cognate aa-tRNA \cite{hatzimanikatis2016}, as the movements involved in translocation are practically independent of the extent of codon-anticodon matching. Furthermore, we have also assumed 
the transition rates for the mischarged cognate aa-tRNA to be identical to those of cognate aa-tRNA, because both exhibit identical codon-anticodon base pairing  \cite{moghal_mistranlation_2014}. For the case of non-cognate aa-tRNA, the transition rates are experimentally unavailable but, as expected, the chances of their incorporation is negligibly small.

\begin{table}[t]
\centering
\scriptsize
\renewcommand{\arraystretch}{1.5}
\begin{threeparttable}
\begin{tabular}{{P{0.25\linewidth}P{0.25\linewidth}P{0.25\linewidth}P{0.25\linewidth}}}
	\hline 
	\multicolumn{1}{M{0.25\linewidth}}{\centering {\bf For correctly charged cognate tRNA}}
	& \multicolumn{1}{M{0.25\linewidth}}{\centering {\bf For incorrectly charged cognate tRNA}}
	& \multicolumn{1}{M{0.25\linewidth}}{\centering {\bf For near-cognate tRNA}}
	& \multicolumn{1}{M{0.25\linewidth}}{\centering {\bf For non-cognate tRNA}}\\
	\hline
	$\omega^{0}_{12}=$ $170 \pm 25 \; \mu M^{-1}s^{-1}$  \tnote{[1]}& $\omega^{0}_{16}=$ $170 \pm 25 \; \mu M^{-1}s^{-1}$ &$\omega^{0}_{1,10}=$ $170 \pm 25 \; \mu M^{-1}s^{-1}$ \tnote{[1]} &$\omega^{0}_{1,14}= $ $170 \pm 25 \; \mu M^{-1}s^{-1}$ \\
	$k_{21}=$ $700 \pm 270 \; s^{-1}$ \tnote{[1]}& $k_{61}=$ $700 \pm 270 \; s^{-1}$ &$k_{10,1}=$ $700 \pm 270 \; s^{-1}$ \tnote{[1]}&$k_{14,1}= $ $700 \pm 270 \; s^{-1}$ \\
	$k_{23}=$ $1500 \pm 450 \; s^{-1}$ \tnote{[1]}& $k_{67}=$ $1500 \pm 450 \; s^{-1}$ &$k_{10,11}=$ $1500 \pm 450 \; s^{-1}$ \tnote{[1]}&$k_{14,15}= $ $10^{-5} \; s^{-1}$ \\
	$k_{32}=$ $2 \pm 0.6 \; s^{-1}$ \tnote{[1]}& $k_{76}=$ $2 \pm 0.6 \; s^{-1}$ &$k_{11,10}=$ $1100 \pm 330 \; s^{-1}$ \tnote{[1]}&$k_{15,14}= $ $10^{5} \; s^{-1}$ \\
	$k_{31}=$  $1 \; s^{-1}$ \tnote{[1]}& $k_{71}=$ $1 \; s^{-1}$ &$k_{11,1}=$ $4 \pm 0.7 \; s^{-1}$ \tnote{[1]}&$k_{15,1}= $ $10^{5} \; s^{-1}$ \\
	$\omega^{0}_{13}=$  $4 \cdot 10^{-9} \; \mu M^{-2}s^{-1}$ & $\omega^{0}_{17}=$ $4 \cdot 10^{-7}  \; \mu M^{-2}s^{-1}$ &$\omega^{0}_{1,11}=$ $4 \cdot 10^{-9}  \; \mu M^{-1}s^{-2}$ &$\omega^{0}_{1,15}= $ $4 \cdot 10^{-9}  \; \mu M^{-2}s^{-1}$ \\
	$k_{34}=$  $200 \pm 40 \; s^{-1}$ \tnote{[1]} & $k_{78}=$ $200 \pm 40 \; s^{-1}$ &$k_{11,12}=$ $0.26 \pm 0.04 \; s^{-1}$ \tnote{[1]}&$k_{15,16}= $ $10^{-5} \; s^{-1}$ \\
	$\omega^{0}_{43}=$  $4 \cdot 10^{-9} \; s^{-1}$ & $\omega^{0}_{87}=$ $4 \cdot 10^{-9}   \; \mu M^{-2}s^{-1}$ &$\omega^{0}_{12,11}=$ $4 \cdot 10^{-9}  \; \mu M^{-2}s^{-1}$ &$\omega^{0}_{16,15}= $ $40 \;   \; \mu M^{-2}s^{-1}$ \\
	$\omega^{0}_{45}=$ $55 \pm 6 \; \mu M^{-1}s^{-1}$ \tnote{[2]}& $\omega^{0}_{89}=$ $55 \pm 6 \; \mu M^{-1}s^{-1}$ &$\omega^{0}_{12,13}=$ $55 \pm 6 \; \mu M^{-1}s^{-1}$ &$\omega^{0}_{16,17}= $ $10^{-6} \; \mu M^{-1}s^{-1}$ \\
	$k_{54}=$  $65 \pm 10 \; s^{-1}$ \tnote{[2]}
	 & $k_{98}=$  $65 \pm 10 \; s^{-1}$ &$k_{13,12}=$  $65 \pm 10 \; s^{-1}$ &$k_{17,16}= $  $10^{5} \; s^{-1}$ \\
	$k_{51}=$  $4 \pm 1 \; s^{-1}$ \tnote{[2]} & $k_{91}=$ $4 \pm 1 \; s^{-1}$ &$k_{13,1}=$ $4 \pm 1 \; s^{-1}$ &$k_{17,1}= $ $10^{-5} \; s^{-1}$ \\
	$k_{15}=$ $10^{-5} \; s^{-1}$ & $k_{19}=$ $10^{-5} \; s^{-1}$ &$k_{1,13}=$ $10^{-5} \; s^{-1}$ &$k_{1,17}= $ $10^{-5} \; s^{-1}$ \\
	\hline
	\end{tabular}
\begin{tablenotes}
\item[1] ref.\cite{rodnina17}. 
\item[2] ref.\cite{belardinelli16} 
\end{tablenotes}
\hrule
\end{threeparttable}
\caption{Values of the rate constants used in our model.}
\label{tab:experimental_values}
\end{table}


\section{Results and discussion I: Generalized thermodynamic forces and fluxes}

\subsection{Stationary state of the elongation cycle}

To keep track of the position of the ribosome along the mRNA template we describe
the ribosome by its chemical state $i$ at time $t$ and the number of monomers $n_m$ in the polypeptide chain
that it has polymerized up to time $t$. This number directly yields the displacement $\Delta x = n_m {\ell}$
by time $t$ along the mRNA. 
The stochasticity of the process then leads to a description of the dynamics in terms
of the probability $P_{i}(n_m,t)$ that at time $t$ the 
ribosome is at ``position'' $n_m$ in the ``chemical'' state $i$. 
The full master equation for the probability to find the ribosome at time $t$ in the 
chemical  state $i$ at codon $n_m$ reads

From the full master equation (\ref{eq:full_master_equation_1}) - (\ref{eq:full_master_equation_17}) as shown in Appendix \ref{app:full_master_equation}, 
one obtains the reduced master equation for the probability distribution $P_i(t)$ for the internal states 
by summing over all positions $n_m$ and defining $P_i(t) = \sum_{n_m} P_i(n_m,t)$. This yields
\begin{equation}
\label{eq:master_equation}
\frac{d P_{i}(t)}{dt} = \sum_{j=1}^7 \left[k_{ji} P_{j}(t) - k_{ij} P_{i}(t)\right]
\end{equation} 
for the probability distribution of the chemical states.

In the stationary state the time-dependence drops out from both sides of the reduced master equation (\ref{eq:master_equation}). 
For the steady-state probabilities, the master equation  can be solved numerically very efficiently with standard computer  routines for any choice of the numerical values of the transition rates. However, we are interested in the
exact analytical solution, i.e., in the stationary probabilities as functions of the transition  rates. 

\begin{figure}
	\begin{center}
		\includegraphics[width=\columnwidth]{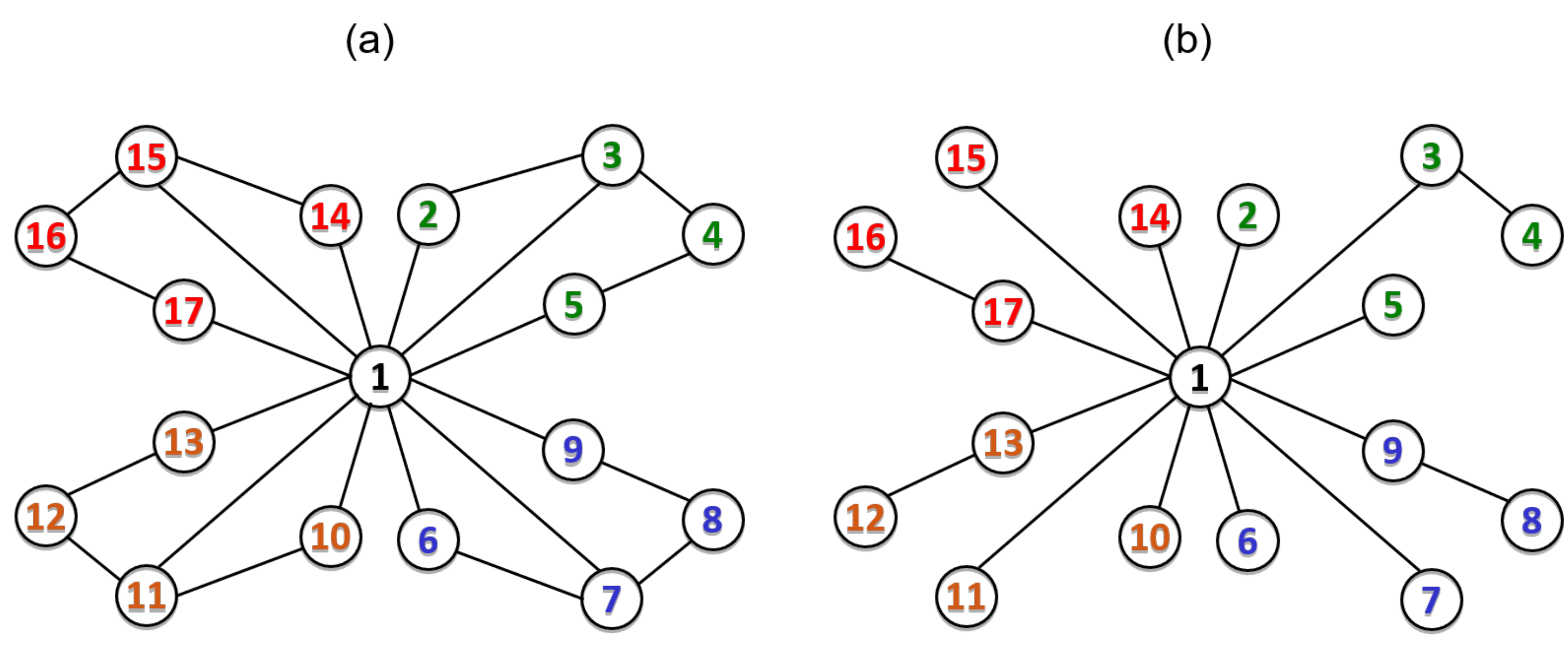}
	\end{center}
	\caption{Graph representation: Figure (a) shows the undirected graph for our network as shown in Fig.~\ref{fig4:detailednetwork} and Figure (b) shows an example of spanning tree for the undirected graph. There are in total 14641 spanning trees for our network.}
	\label{fig5:undirected_cycle_model}
\end{figure}

To this end, we adapt the ideas of \cite{schnakenberg_network_1976} in which the Markov network shown in Fig.~\ref{fig4:detailednetwork} is represented by a graph where each of the vertices represents a distinct state $i$ ($i=1,2,\dots,17$) and the directed edges $i \rightarrow j$ represent the possible transitions from the state $i$ to  the state $j$. 
From the graph an {\it undirected graph} is obtained by replacing the directed edges by undirected edges (see Fig.~\ref{fig5:undirected_cycle_model} (a)). 
A {\it spanning tree} of an undirected graph is a sub-graph which is maximal in the sense that it includes all the vertices of the graph, with minimum possible edges which implies the absence of any loop (see Fig.~\ref{fig5:undirected_cycle_model} (b)). A directed spanning tree, say the $\mu$-th ($\mu=1,2,\dots, M$), $T^{\mu}_{i}(G)$ of a graph $G$ can be obtained by directing all the edges of the undirected spanning tree $T^{\mu}(G)$ towards the vertex $i$. To each of the directed spanning trees $T^{\mu}_{i}(G)$, we assign a numerical value, $A(T^{\mu}_{i})(G)$, which is defined as the product  of the $|V|-1$ transition rates in the tree. The steady state probability distributions are then given by 
\begin{equation} \label{eq:p_st}
P_i=\mathcal{Z}^{-1} \sum^{M}_{\mu=1} A(T^{\mu}_{i})
\end{equation}
where the normalization factor
\begin{equation} \label{eq:part_func}
\mathcal{Z}=\sum^{|V|}_{i=1} \sum^{M}_{\mu=1} A(T^{\mu}_{i}). 
\end{equation}
plays a role similar to that of the partition function.
The detailed step-by-step derivation of $P_{i}$ for a smaller network is given in the Appendix \ref{App:rssp}. We have written a dedicated Matlab code that computes all the steps in the graph theoretic calculation.

Using the experimentally measured values of the rates listed in 
Tab.~\ref{tab:experimental_values} in the analytical expressions for $P_i$ ($i=1,2,\dots,17$) one obtains the numerical values as shown in Tab.~\ref{tab:probabilities}.
\begin{table}[hbt!]
	\centering
	\renewcommand{\arraystretch}{1}
	\begin{tabular}{M{0.3\linewidth}M{0.3\linewidth}}
		\hline
			$P_1 = 0.00059$ & $P_2 = 0.0023$ \\
			$P_3 = 0.017$ & $P_4 =  0.021$ \\
			$P_5 = 0.84$ & $P_6 =  0.00023$ \\
			$P_7 = 0.0017$ & $P_8 =  0.0021$ \\
			$P_9 = 0.084$ & $P_{10} =  0.0071$ \\
			$P_{11} = 0.0096$ & $P_{12} =  0.000016$ \\
			$P_{13} = 0.00062$ & $P_{14} =  0.007$ \\
			$P_{15} = 4.2*10^{-13}$ & $P_{16} =  5.9*10^{-14}$ \\
			$P_{17} = 5.9*10^{-14}$ \\
			\hline
	\end{tabular}
	\caption{Numerical values for the probabilities}	
	\label{tab:probabilities}
\end{table}

By ergodicity, the stationary values $P_i$ represent the fraction of time
the ribosome spends in state $i$. We point out that for the correctly charged cognate aa-tRNA 
the overwhelming amount of
time ($> 80 \%$) is spent in the state 5 (from which translocation, accompanied by
GTP hydrolysis and EF-G.GDP dissociation, takes place), followed by that in the state 3
(from which proofreading and elongation, accompanied by
EF-Tu.GDP dissociation, take place). This is followed by the time spent in the states 9 and 6 for similar reasons in case of mischarged aa-tRNA.  This is again followed by the probability of the states 13 and then 11 for the same reasons in case of near cognate aa-tRNA. The incorporation of near cognate amino acid has much lower probability than mischarged aa-tRNA as the aa-tRNAs go through the quality control process of  the ribosome. The probability of the ribosome incorporating a non-cognate amino acid is negligibly small.

\subsection{Transition flux, cycle flux and their relations}

The right hand side of the master equation ((\ref{eq:full_master_equation_1}) - (\ref{eq:full_master_equation_17})) (as shown in Appendix \ref{app:full_master_equation}) is the
negative sum of the net {\it probability currents}
\begin{equation} 
\label{eq:prob_current}
J_{ij}(t) = k_{ij} P_{i}(t) - k_{ji} P_{j}(t)
\end{equation} 
from state $i$ to state $j$. Notice that  all pairs $i$ and $j$ that contribute to the master equation are neighbours in the network graph Fig.~\ref{fig4:detailednetwork}. For the steady state solution $P_{i}$, the associated stationary probability currents $k_{ij} P_{i} - k_{ji} P_{j}$ are denoted by $J_{ij}$, without argument $t$. They are related to the {\it macroscopic mean transition fluxes} 
\begin{equation}
J^\ast_{ij}:= k_{ij}N_i - k_{ji}N_j 
\end{equation}
between states $i$ and $j$ in the direction $i\rightarrow j$, where in an ensemble of $N$ identical ribosomes translating an mRNA an  average of $N_i$ are in state $i$. Since in the stationary 
state one has $N_i = N P_{i}$ 
one finds 
\begin{equation}
J^\ast_{ij} = N(k_{ij} P_{i} - k_{ji} P_{j}) = N J_{ij}.
\end{equation}
Therefore, we shall refer to the stationary probability currents $J_{ij}=-J_{ji}$ as {\it transition fluxes}. If the ribosome would be in full thermal equilibrium (i.e., not only with respect to the temperature), the process would satisfy detailed balance and, consequently, all transition fluxes would vanish. However, as pointed out above, this is not the case because of the non-equilibrium nature of the steady state of the model under investigation here.

Next we define 
the auxiliary variables $Q_i(t) := \sum_{n_m} n_m P_i(n_m,t)$.
With the step length $\ell$ of a translocation from one codon to the next, one obtains from the $Q_i(t)$ the
mean position $X(t) = \ell \sum_i Q_i(t)$ and the mean velocity $v(t) = \dot{X}(t)$ of the ribosome.
From the full master equation (\ref{eq:full_master_equation_1}) - (\ref{eq:full_master_equation_7}) 
one obtains (by shifting the summation index in the sum over $n_m$ for the terms involving site $n_m\pm1$) the
simple expression
\begin{equation}
\label{eq:stat_velo}
v(t)=  \ell \left[J_{51}(t) + J_{91}(t) + J_{13,1}(t) + J_{17,1}(t)\right]
\end{equation}
for the average velocity for the ribosome.

In a similar fashion one obtains the mean hydrolysis rate. One introduces as stochastic variable the net number
$m(t)$ of GTP hydrolysis and effective GTP synthesis events. This number is incremented by $+1$ for say $2\to 3$, $5\to 1$ takes place and incremented by $-1$ when the
reverse transitions take place. This yields a master equation for the joint probability $P_i(n_m,m,t)$ similar to 
(\ref{eq:full_master_equation_1}) - (\ref{eq:full_master_equation_7}), but with terms 
like $k_{51} P_{5}(n_m-1,m-1,t)$, $k_{32} P_{3}(n_m,m+1,t)$ and so on instead of 
$k_{51} P_{5}(n_m-1,t)$, $k_{32} P_{3}(n_m,t)$ (and so on). The net hydrolysis up to time $t$ is then
given by $H(t) = \sum_i \sum_{n_m} \sum_m m P_i(n_m,m,t)$ and the master equation yields the exact expression
\begin{equation}
\label{eq:hydrolysis}
h(t)=J_{23}(t) +J_{67}(t) +J_{10,11}(t) +J_{14,15}(t) +J_{51}(t) + J_{91}(t) + J_{13,1}(t) + J_{17,1}(t) 
\end{equation}
for mean hydrolysis rate $h(t) = dH(t)/dt$. 
In the steady-state, the time-dependence drops out of both sides of (\ref{eq:stat_velo}) as well as those of 
(\ref{eq:hydrolysis}).

In order to get deeper insight into the non-equilibrium nature of the chemo-mechanical cycle of the  ribosome we next express the transition fluxes in terms of cycle fluxes.
Six different single loops (cycles regardless of their orientation), labeled by $\kappa \in$ \{ (a), (b), (c), (d), (e), (f), (g), 
(h), (i), (j), (k), (l) \} in Fig.~\ref{fig5:cycle1} and Fig.~\ref{fig6:cycle2}, can arise in the network model shown in \ref{fig4:detailednetwork}. It will transpire that transition cycles allow for a deeper understanding of the kinetic activity of the network than just by focusing on the transition fluxes.

\begin{figure}[h!]
\begin{center}
\includegraphics[width=0.9\columnwidth]{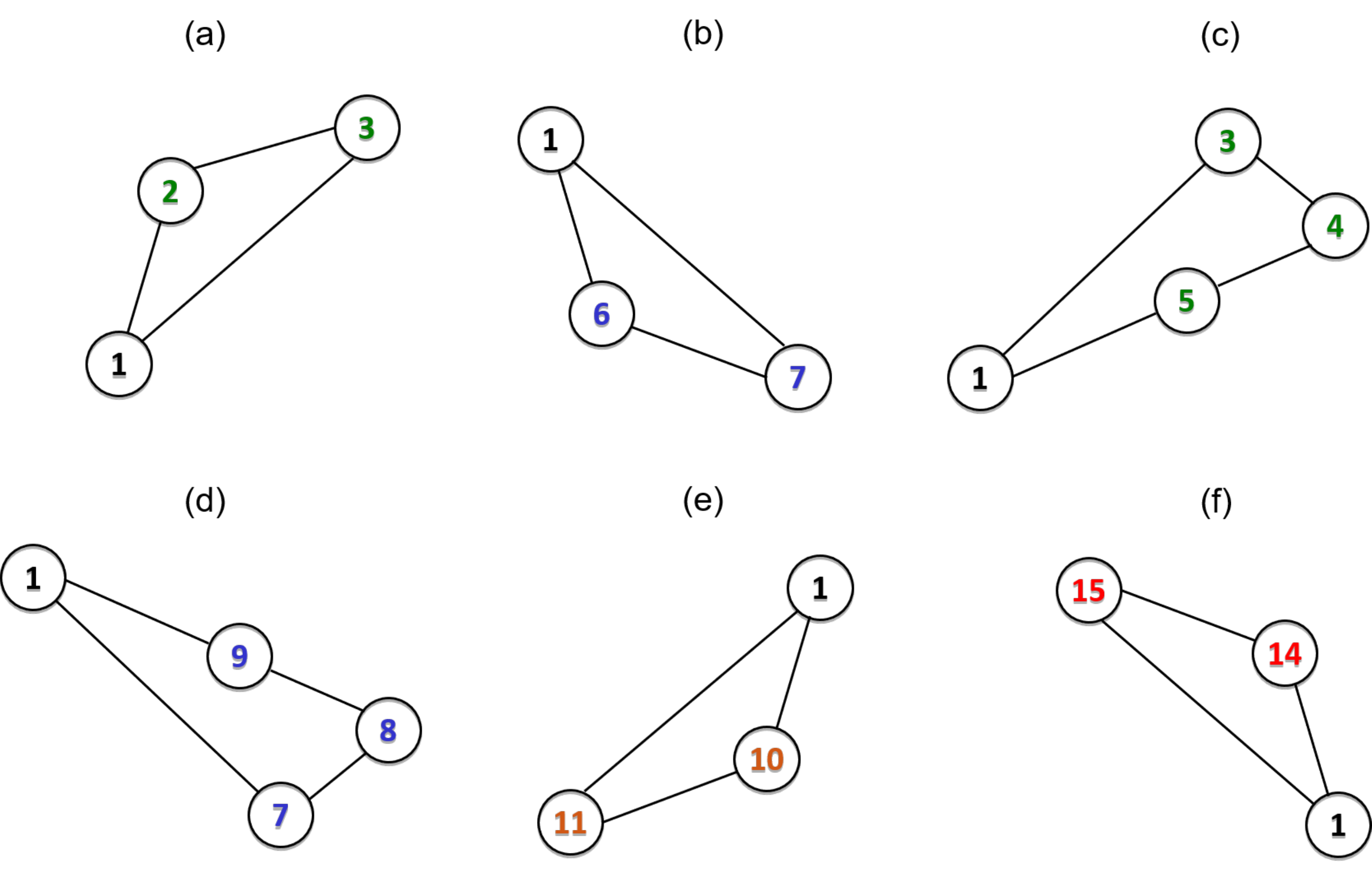}
\end{center}
\caption{The graphs (a)-(f) show the six different loops (unoriented cycles) labelled by $\kappa = a,b,\dots,f$ and indicated by the solid lines that are present in the Fig.~\ref{fig5:undirected_cycle_model}.}
\label{fig5:cycle1}
\end{figure}

\begin{figure}[h!]
	\begin{center}
		\includegraphics[width=0.9\columnwidth]{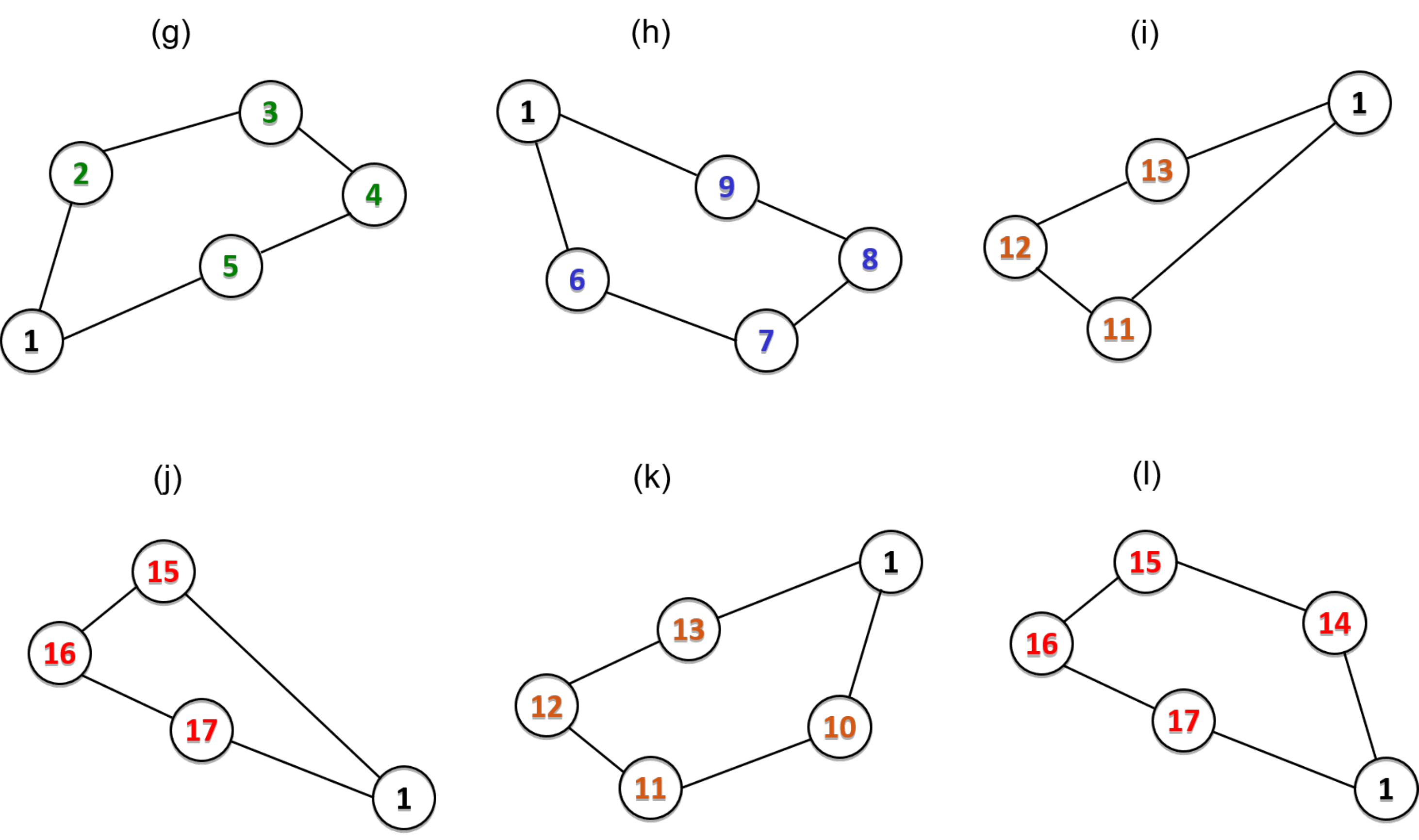}
	\end{center}
	\caption{The graphs (g)-(l) show the six different loops (unoriented cycles) labelled by $\kappa = g,h,\dots,l$ and indicated by the solid lines that are present in the Fig.~\ref{fig5:undirected_cycle_model}}.
	\label{fig6:cycle2}
\end{figure}}

To make the role of cycles quantitative, we first define the {\it cycle products} 
\begin{equation}
\Pi_{\kappa\pm} := \prod_{(i,j) \in (\kappa\pm)} k_{ij} 
\label{eq-cycleProduct}
\end{equation}
as the products of rates $k_{ij}$ of the edges contributing to cycle $\kappa$ in clockwise 
($+$) or anticlockwise ($-$) orientations, respectively.

Next we summarize a graph theoretic approach for deriving the analytical expressions for the cycle fluxes. 
Following Hill \cite{hill_free_1989}, we associate to a given oriented cycle $(\kappa\pm)$ the cycle rate constants 
\begin{equation}
J_{\kappa,\pm} = \sum_{\mu, (i,j)} k_{ij} A(T^{\mu}_{i})
\end{equation}
over those directed trees $\mu$ and edges $(i,j)$ that contribute to all graphs with cycle $(\kappa\pm)$. Notice that 
adding a new directed edge oriented from $i$ to $j$ to a directed spanning tree $T^{\mu}_{i}$ converts the latter into a graph $T^{\mu}_{ij}$ that has a single oriented cycle. Thus, from the undirected spanning trees one can construct the {\it flux diagrams} each of which has a single undirected cycle and directed branches that feed into this cycle. Then, it follows that
\begin{equation}
J_{\kappa,\pm} = \Pi_{\kappa,\pm} R_{\kappa}/ \mathcal{Z}
\label{eq-Jkappa}
\end{equation}
where 
\begin{equation}
R_{\kappa} =  \sum_{\substack{\text{all the flux diagrams}\\ \text{containing cycle } \kappa}}  \biggl( \prod_{\substack{\text{(i, j) are the directed edges}\\ \text{feeding into cycle }\kappa \text{ in the flux diagram}}}  k_{ij} \biggr). 
\label{eq-Rkappa}
\end{equation} 
The expression (\ref{eq-Jkappa}) for $J_{\kappa,\pm}$ can be interpreted as the cycle product multiplied by the weight factor $R_{\kappa}/\mathcal{Z}$ where the latter accounts for the flux from the rest of the network into the cycle. 
Using the expressions above, we arrive at the expression for the cycle fluxes
\begin{equation}
J_{\kappa} = J_{\kappa,+} - J_{\kappa,-} = (\Pi_{\kappa,+}-\Pi_{\kappa,-})R_{\kappa}/\mathcal{Z}.
\label{eq-Jkappa2}
\end{equation}
where $R_{\kappa}$ is given by (\ref{eq-Rkappa}). Both $J_{\kappa,+}$ and $J_{\kappa,-}$ are positive quantities while $J_{\kappa}$ can be positive or negative.

\begin{table}[hbt!]
	\centering
	\renewcommand{\arraystretch}{1}
	\begin{tabular}{llll}
		\hline
		$\kappa \quad $ & $\Pi_{\kappa,+}$ &  $\Pi_{\kappa,-}$ & $J_{\kappa}$ \\
		\hline
		(a) & $1.3*10^{7}$ & $0.014$ & $0.017$  \\
		(b) & $0.014$ & $1.3*10^{6}$ & $-0.0017$  \\
		(c) & $22$ & $6.5*10^{-9}$ & $5.8*10^{-9}$  \\
		(d) &  $6.5*10^{-9}$ & $22$ & $-5.8*10^{-9}$  \\
		(e) & $5.1*10^{7}$ & $7.7$ & $0.0385$  \\
		(f) & $700$ & $8500$ & $-3.3*10^{-8}$  \\
		(g) & $2.8*10^{13}$ & $9.1*10^{-6}$ & $3.38$  \\
		(h) & $9.1*10^{-6}$ &$2.8*10^{12}$ & $-0.34$  \\
		(i) & $0.029$ & $2.6*10^{-8}$ & $4.3*10^{-12}$  \\
		(j) & $1*10^{10}$ & $1*10^{-20}$ & $2.9*10^{-9}$  \\
		(k) & $3.6*10^{10}$ & $0.0050$ & $0.0025$  \\
		(l) & $7*10^{12}$ & $8.5*10^{-17}$ & $2.9*10^{-9}$  \\
		\hline
	\end{tabular}
	\caption{Numerical values for $\Pi_{\kappa,\pm}$ and the cycle fluxes 
		$J_{\kappa,\pm}$,
		obtained from the rates of Table~\ref{tab:experimental_values}. 
		Since $J_{\kappa,+} \gg J_{\kappa,-}$ for all cycles $\kappa$, one has $J_\kappa \approx J_{\kappa,+}$.
		The units of the cycle products $\Pi_{\kappa,\pm}$ depend on the loop $\kappa$. 
		The units for $J_{\kappa,+}$, $J_{\kappa,-}$ and $J_{\kappa}$ are s$^{-1}$.}
	\label{tab:pi_values}
\end{table}

For clear understanding, as an example, all the flux diagrams corresponding to the smaller network are displayed in the Appendix \ref{app-flux_diag_Ribo} along with the outline of the related calculations. Each transition flux through the oriented edge $(i,j)$ can be decomposed into the sum of cycle fluxes.  This decomposition for the 17-state model yields

\begin{eqnarray}
\label{JtJc12}
& & J_{12} = J_{23} = J_a + J_g \\
\label{JtJc13}
& & J_{13}  = - J_a + J_c \\
\label{JtJc34}
& & J_{34}=J_{45}=J_{51}=J_c+J_g \\
\label{JtJc16}
& & J_{16} = J_{67} = -J_b - J_h \\
\label{JtJc17}
& & J_{17}  =  J_b - J_d \\
\label{JtJc78}
& & J_{78}=J_{89}=J_{91}=-J_d-J_h \\
\label{JtJc110}
& & J_{1,10} = J_{10,11} = J_e + J_k \\
\label{JtJc111}
& & J_{1,11}  = - J_e + J_i \\
\label{JtJc1112}
& & J_{11,12}=J_{12,13}=J_{13,1}=J_i+J_k \\
\label{JtJc114}
& & J_{1,14} = J_{14,15} = -J_f - J_l \\
\label{JtJc115}
& & J_{1,15}  = - J_j + J_f \\
\label{JtJc1516}
& & J_{15,16}=J_{16,17}=J_{17,1}=-J_j-J_l \\
\end{eqnarray}
Note that this list of decompositions is complete. All transition fluxes are uniquely expressed in terms of cycle fluxes. 
However, it is not possible to invert this and express the cycle fluxes in terms of transition fluxes because only four transition fluxes are linearly independent while the number of cycle fluxes is six.
The numerical values of the cycle fluxes are given in Table~\ref{tab:pi_values}. 
The largest flux goes through cycle (g,+), which represents correct translation
with correctly charged cognate aa-tRNA. The second largest flux passes through cycle (h,-), 
corresponding to incorporation of wrong amino acid due mischarged cognate aa-tRNA binding. The third largest flux passes through cycle (e,+), 
corresponding to successful error correction when near cognate aa-tRNA binds. The cycle (k,+), (unsuccessful error correction leading up to missense error) has a small flux.

It is worth pointing out here that our decomposition into cycles is complete and carried out by inspection. A formal prescription for identifying the independent thermodynamic forces (affinities) has been proposed in recent years by Esposito and collaborators \cite{polettini16,rao18}. However, the latter formal approach is not required for analysing the entropy production of the specific machine (ribosome) and specific process (translation) under our consideration in this paper. Nevertheless, we hope to report the alternative analysis, based on Esposito's prescription, for the kinetics of template-directed polymerization, in a future publication.

\subsection{Energy balance in steady-state}

The two ratios 
\begin{equation} 
\label{eq:ratio_flux}
\frac{J_{\kappa,+}}{J_{\kappa,-}} = \frac{\Pi_{\kappa,+}}{\Pi_{\kappa,-}}
=: \mathrm{e}^{\frac{X_{\kappa}}{k_{B}T}}
\end{equation}
depend only on the rate constants around the cycle $\kappa$ itself. 
They define the so-called generalized thermodynamic forces $X_{\kappa}$
\cite{schnakenberg_network_1976,hill_free_1989} generated in a non-equilibrium system 
when coupled to different reservoirs. In a stationary state, which does not involve any 
change in internal energy, these forces are equivalent to
the heat exchange $\Delta Q_{\kappa}$ that arises from the transitions 
through a loop $\kappa$ and proportional to the corresponding entropy change 
\begin{equation}
\label{SXQ}
\Delta S_{\kappa} = X_\kappa/T = \Delta Q_{\kappa}/T,
\end{equation}
thus exposing the entropic nature
of the generalized thermodynamic forces in the stationary regime.

Thus the generalized thermodynamic forces highlight the connection 
between the irreversibility of a non-equilibrium process and its heat dissipation since 
(\ref{eq:ratio_flux}) implies
\begin{equation}
\label{eq:flux_force_rel2} 
J_\kappa =J_{\kappa,+} \left(1-\mathrm{e}^{-\frac{\Delta Q_{\kappa}}{k_B T}}\right) = J_{\kappa,+} \left(1-\mathrm{e}^{-\frac{X_{\kappa}}{k_B T}}\right) 
\end{equation}
which demonstrates (a) that the heat dissipation and the cycle flux 
vanish simultaneously (corresponding to thermal equilibrium),
and (b) that otherwise (i.e., out of equilibrium) they always have the same sign. 
In other words, the generalized thermodynamic forces express the direction of the cycle fluxes.

The ratios (\ref{eq:ratio_flux}) also provide a link between the kinetics and thermodynamics 
of the network which can be further developed as follows. During the transition from 
state $i$ to state $j$, the internal energy $U_i$ of the system in state $i$ can change 
due to three factors:
(i) a chemical potential change $\Delta \mu_{ij} =\mu_i - \mu_j$ arising from the 
coupling of the states to the particle reservoirs with the chemical potentials 
$\mu_\text{aa-tRNA.EF-Tu.GTP}$, $\mu_\text{EF-Tu.GDP}$, $\mu_\text{EF-G.GTP}$, 
$\mu_\text{EF-G.GDP}$, $\mu_\text{ATP}$, $\mu_\text{AMP}$, $\mu_\text{PPi}$ and $\mu_\text{Pi}$ introduced above, (ii) the mechanical work
$W_{ij}$ which the machine performs in the transition $i\to j$ to overcome the 
external force $F_{ext}$,
and (iii) the heat exchange $Q_{ij}$ with the surrounding medium. 
Conservation of energy for the transition from $i$ to $j$ therefore reads
\begin{equation} 
\label{eq:energy_cons_single_trans}
U_i - U_j = \Delta \mu_{ij} - W_{ij} - Q_{ij}.
\end{equation}

Next we apply conservation of energy to a cycle.
Since the internal energy is a state function, the total change in the internal 
energy for a cycle must be zero in the steady state. The heat exchange 
$\Delta Q_{\kappa} = T \Delta S_{\kappa}$ in a cycle $(\kappa,+)$ 
is given by (\ref{eq:ratio_flux}) in terms of the reaction rates.
For a cycle $(\kappa,+)$, energy conservation (\ref{eq:energy_cons_single_trans}) thus
reads 
\begin{equation} 
\label{eq:energy_cons_cycle}
\Delta \mu_{\kappa,+} - W_{\kappa,+} - \Delta Q_{\kappa} = 0
\end{equation}
where $W_{\kappa,+}$ is the work done against the external force in the complete cycle in positive
direction and 
$\Delta \mu_{\kappa,+}$ denotes the net chemical potential difference in the complete cycle.

Using (\ref{eq:ratio_flux}) and (\ref{SXQ}) in (\ref{eq:energy_cons_cycle}) we arrive at the equivalent
steady state balance condition 
\begin{equation}
\label{eq:steady_bal}
\Delta \mu_{\kappa,+} - W_{\kappa,+}  = X_\kappa
= k_B T \ln \frac{\Pi_{\kappa,+}}{\Pi_{\kappa,-}}
\end{equation}
which yields the relation between the transition rates (through $\Pi_{\kappa,\pm}$), the chemical energy input and the mechanical work for any cycle 
in the network, conveniently expressed through the stationary generalized thermodynamic 
forces.
This relation will be used below to determine the range of normal operation of the 
ribosome, i.e., with positive mean velocity of the ribosome along the mRNA and positive rate of hydrolysis.

\section{Results and discussion II: Modes of operation, mechanics and stochastic thermodynamics} 

\subsection{Modes of operation}

The analysis carried out in this subsection for the ribosome is very similar to that reported 
earlier \cite{liepelt09} for the cytoskeletal molecular motor kinesin. As elaborated above, the elongation factors EF-Tu and EF-G catalyze the hydrolysis of the 
fuel molecules, i.e., GTP,  into GDP and $\mathrm{P_i}$. Also, ATP is hydrolyzed into AMP and $\mathrm{PP_i}$ during aminoacylation with help of aminoacyl synthetase. The energy released in these reactions 
acts as the chemical energy input for the ribosome machine in each elongation cycle. Thus the 
operation of the ribosome depends on the respective chemical potential differences
and the external force $F_{ext}$. Its modes of operation are characterized in terms of the 
average velocity and the average rate of hydrolysis both of which are positive in the normal 
mode of operation. In the following, we identify the different modes of operation of the 
ribosome, normal as well as abnormal, exploiting the cyclic energy balance relations
(\ref{eq:energy_cons_cycle}).

\subsubsection{Sign of cycle fluxes and modes of operation}

Substituting (\ref{JtJc34}), (\ref{JtJc78}), (\ref{JtJc1112}) and (\ref{JtJc1516}) into the time-independent (\ref{eq:stat_velo}) we get 
\begin{equation} 
\label{eq:vel}
v=  \ell \left[J_{51} + J_{91} + J_{13,1} + J_{17,1}\right]  = {\ell} (J_c - J_d + J_g - J_h + J_i - J_j + J_k - J_l) 
\end{equation}

for the average velocity of the ribosome in terms of the cycle fluxes;  which yields the numerical value 
$3.7 \,  \mathrm{nm} \,\mathrm{s}^{-1}$ after substitution of the values of parameters listed in 
Tab.~\ref{tab:pi_values}. Our result is of the same order of the experimentally observed elongation rates \cite{proshkin10}. Note that while our result is in nm/s, the experimentally observed rate is given in aa/s and recall that for every addition of amino acid, the ribosome takes one codon step which is approximately equal to 1 nm.

Likewise, substituting the expressions (\ref{JtJc12}), (\ref{JtJc16}), (\ref{JtJc110}), (\ref{JtJc114}), (\ref{JtJc34}), (\ref{JtJc78}), (\ref{JtJc1112}) and (\ref{JtJc1516}) 
into (\ref{eq:hydrolysis}) in the steady state we get the rate of GTP hydrolysis 

\begin{eqnarray} 
\label{eq:hydro_rate}
h &=&J_{23} +J_{67} +J_{10,11} +J_{14,15} +J_{51} + J_{91} + J_{13,1} + J_{17,1} \nonumber \\
&=& J_a - J_b + J_c - J_d + J_e - J_f + J_i - J_j + 2 (J_g - J_h + J_k - J_l)
\end{eqnarray}
in the network model in terms of the cycle fluxes; it predicts the numerical value 
$h=7.5 \,\mathrm{s}^{-1}$ after substitution of the parameter values listed in Tab.~\ref{tab:pi_values}.

To investigate under which external conditions the ribosome functions in normal mode, 
we make a simplification of the calculation by showing only the GTP molecules explicitly. According to 
this simplification, we have two GTP molecules hydrolyzed for every mechanical step of the 
ribosome between the codons, one in the transition $2\to 3$ and one in the transition $5\to 1$
(similarly for the other pathways). The hydrolysis is then driven by a chemical
potential difference $\Delta \mu$ that comes into play in these transitions. Since the sign of the transition 
fluxes does not depend in a straightforward fashion on the chemical potential difference and the 
external force, we study the process in terms of the cycle fluxes.

For the twelve individual cycles labeled by $\kappa$, as displayed in Fig.~\ref{fig5:cycle1} and Fig.~\ref{fig6:cycle2}, 
the analysis of the process discussed in detail in Sec.~\ref{Sec:rcn} yields
\begin{eqnarray}
\Delta \mu_{a,+} & = & \Delta \mu_{e,+} \, = \, \Delta \mu \\
\Delta \mu_{b,+} & = & \Delta \mu_{f,+} \, = \, - \Delta \mu \\
\Delta \mu_{c,+} & = & \Delta \mu_{i,+} \, = \,  2 \Delta \mu \\
\Delta \mu_{d,+} & = & \Delta \mu_{j,+} \, = \, -2 \Delta \mu \\
\Delta \mu_{g,+} & = & \Delta \mu_{k,+} \, = \, 3 \Delta \mu \\
\Delta \mu_{h,+} & = & \Delta \mu_{l,+} \, = \, -3 \Delta \mu \\ \nonumber
\end{eqnarray}
with the effective chemical potential difference $\Delta \mu_{\kappa,+} = 
\mu^\text{GTP}_{\kappa,+} - \mu^\text{GDP}_{\kappa,+} - \mu^{\mathrm{P_i}}_{\kappa,+}$ or $\Delta \mu_{\kappa,+} = 
\mu^\text{ATP}_{\kappa,+} - \mu^\text{AMP}_{\kappa,+} - \mu^{\mathrm{PP_i}}_{\kappa,+}$.

Recalling that $W = {\ell} F_{ext}$ is the work performed against the external force in one translocation step, 
one also derives from the description of the six cycles 
\begin{eqnarray}
W_{a,+} & = & W_{b,+} \, = \, W_{e,+}  \, = \, W_{f,+} \, = \, 0\\
W_{d,+} & = & W_{h,+} \, = \, W_{j,+}  \, = \, W_{l,+} \, = \, W\\
W_{c,+} & = & W_{g,+} \, = \, W_{i,+}  \, = \, W_{k,+} \, = \, -W\\ \nonumber
\end{eqnarray}
Thus one gets from (\ref{eq:steady_bal})
\begin{eqnarray}
X_a &=& X_e \, = \, \Delta \mu \\
X_b &=& X_f \, = \, -\Delta \mu \\
X_c &=& X_i \, = \, 2\Delta \mu -  W \\
X_d &=& X_j \, = \,  W - 2\Delta \mu \\
X_g &=& X_k \, = \, (3\Delta \mu) -  W \\
X_h &=& X_l \, = \,  W - (3\Delta \mu) \\ \nonumber
\end{eqnarray}

\begin{figure}[h]
\begin{center}
\includegraphics[width=0.7\columnwidth]{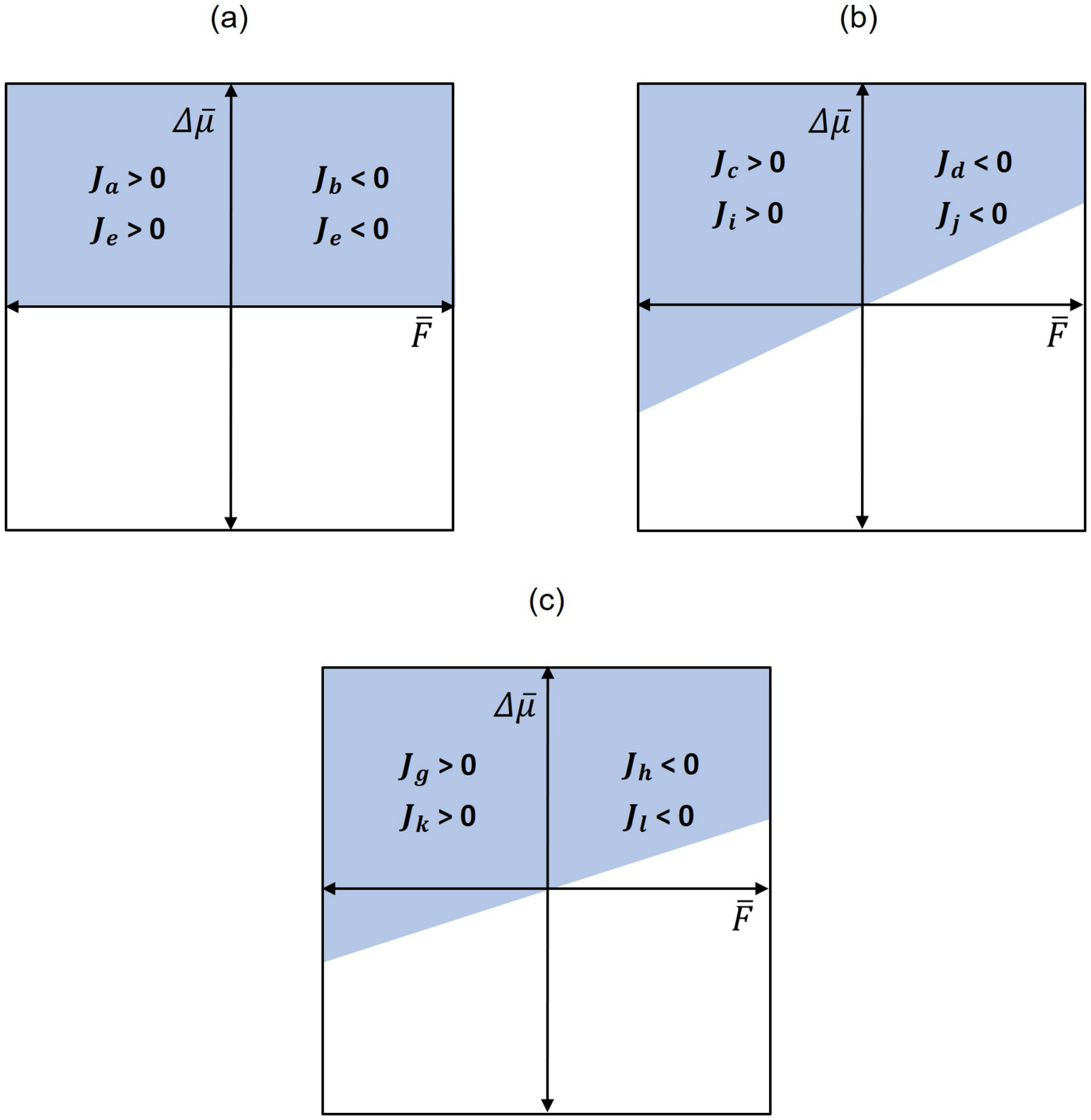}
\end{center}
\caption{The figure shows the variation of sign of cycle fluxes with the variation of the chemical potential difference $\Delta\mu$ and the applied force $F_{ext}$. The shaded region corresponds to the conditions written on it.}
\label{fig8:operat_diag1}
\end{figure}

\begin{figure}[h]
\begin{center}
\includegraphics[width=0.7\columnwidth,angle=-90]{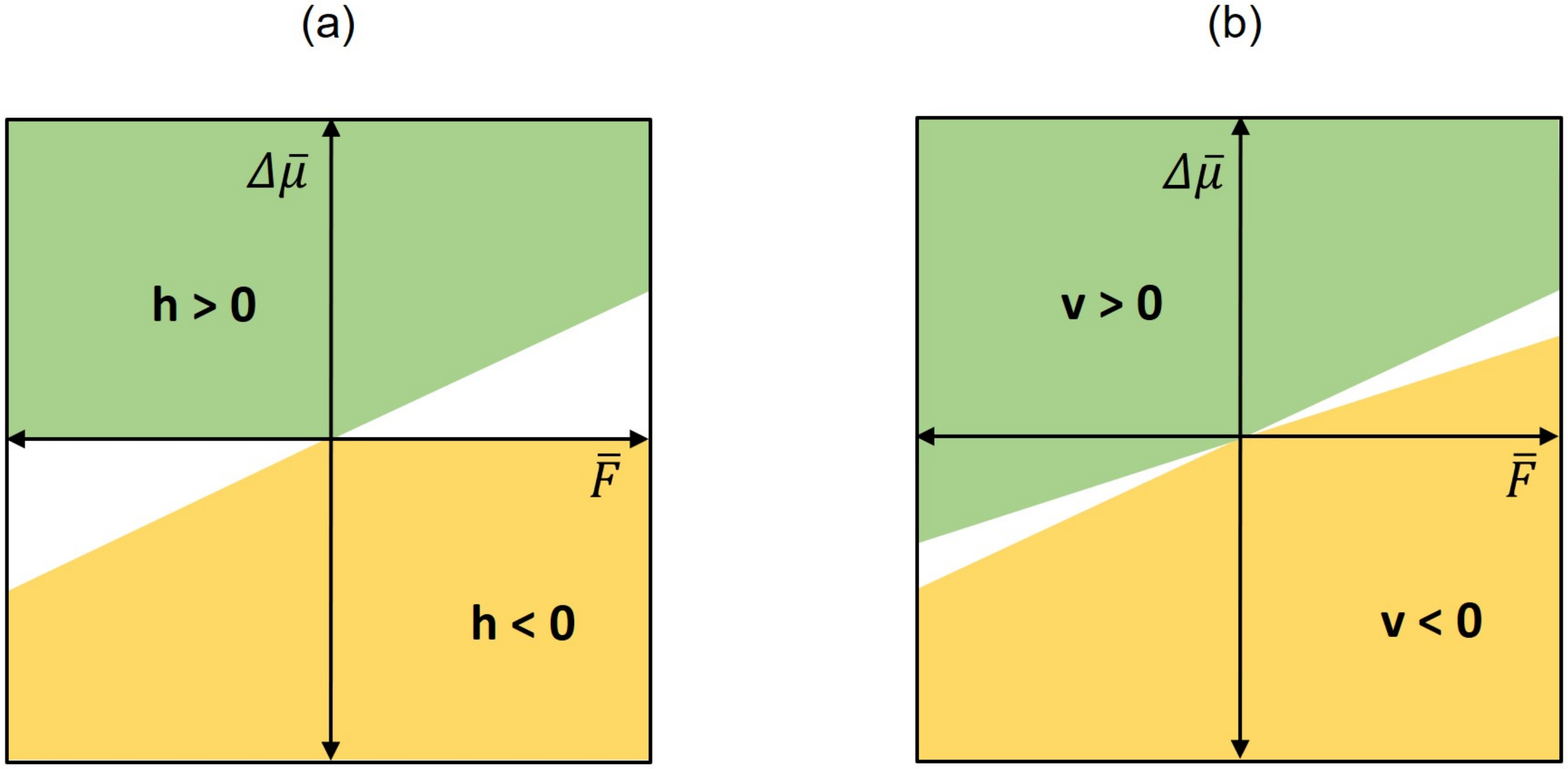}
\end{center}
\caption{Modes of operation of the ribosome as function
of scaled chemical potential difference $\Delta \bar{\mu}$ and the scaled force 
$\bar{F}$. Inside the green shaded regions (upper left) 
the motor velocity $v$ and the hydrolysis rate $h$ are  positive 
for all values of $\Delta \bar{\mu}$ and  
$\bar{F}$. Inside the pink shaded regions (lower right) these
quantities are negative for all values of $\Delta \bar{\mu}$ and  
$\bar{F}$. The change of sign 
takes place inside the respectve white regions.}
\label{fig9:operat_diag2}
\end{figure}

Since the cycle flux and the generalized thermodynamic force always have the same sign and must vanish simultaneously,
\begin{equation}
\label{eq:flux_force_rel2} 
J_\kappa =J_{\kappa,+} \left(1-\mathrm{e}^{-\frac{\Delta Q_{\kappa}}{k_B T}}\right) = J_{\kappa,+} \left(1-\mathrm{e}^{-\frac{X_{\kappa}}{k_B T}}\right) 
\end{equation}
 The conditions that must be satisfied are shown in in Fig.\ref{fig8:operat_diag1}  on the $F_{ext}-\Delta \mu$-plane. Combining the information displayed in these figures, we identify the various modes of operation of the ribosome, from (\ref{eq:vel}) and (\ref{eq:hydro_rate}), on the $\bar{F}-\Delta \bar{\mu}$-plane in terms of $v$ and $h$,  as shown in Fig.~\ref{fig9:operat_diag2}. Here, $\bar{F}=\frac{F \ell}{k_{B}T}$  and $\Delta \bar{\mu}=\frac{\Delta \mu}{k_{B}T}$ are the scaled force and chemical potential .

The lines where velocity and hydrolysis rate change sign are not accessible
to this analysis. 
At the equilibrium point $W=0=\Delta\mu$ both $v$ and $h$ vanish.

\subsubsection{Stall force and balanced potential}

For a given concentration for [GTP],[GDP],[P], the stall force $F_s$ is given by
the condition
\begin{equation} 
\label{eq:F_s}
v(F_{ext}\text{,[GTP],[GDP],[P])})=0 \mbox{ for } F_{ext} = F_s  .
\end{equation}

After substituting the values of the rate constants and $\ell=1 \; \mathrm{nm}$ in (\ref{eq:vel}), the expression for velocity in terms of $\Delta \mu$ and $F_{ext}$ is
	\begin{eqnarray}
	v&=&
	[-2.9*10^{-9} (1-e^{\frac{2 \Delta \mu-F_{ext}}{k_BT}})-2.9*10^{-9} (1-e^{\frac{3 \Delta \mu-F_{ext}}{k_BT}}) \nonumber \\ 
	&+& 3.4 (1-e^{\frac{-3 \Delta \mu+F_{ext}}{k_BT}})+5.8*10^{-9} (1-e^{\frac{-2 \Delta \mu+F_{ext}}{k_BT}})]
	\mathrm{nm} \, \mathrm{s}^{-1}
	\label{eq:v_exp}
	\end{eqnarray}
Using (\ref{eq:F_s}) and (\ref{eq:v_exp}), 
we obtain the stall force $F_s(\Delta \mu)$ as a function of $\Delta \mu$.

\begin{figure}[h!]
\begin{center}
\includegraphics[width=0.7\columnwidth,angle=-90]{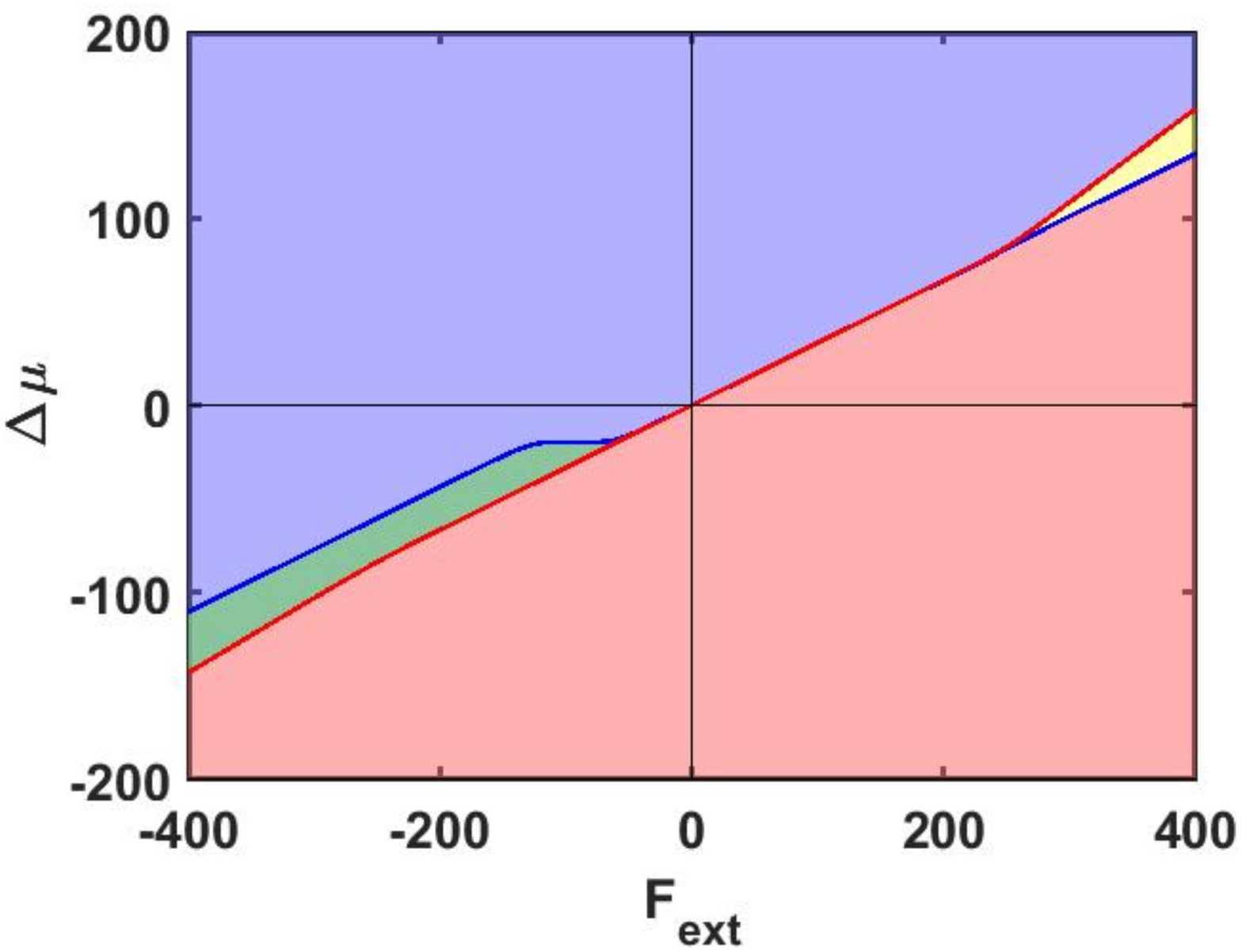}
\end{center}
\caption{Operational diagram for the ribosome. The stall force $F_{s}$ as a function of $\Delta \mu$ is shown by the red line. The balanced potential $\Delta \mu_{b}$ as a function of $F_{ext}$ is shown by the blue line. These lines divide the ($F_{ext}$,$\Delta \mu$) plane into four different regions HF (hydrolysis-forward) (blue), HB (hydrolysis-backward) (yellow), SB (synthesis-backward) (red) and SF (synthesis-forward) (green). The lines were computed using the transition rates given in Table I. $F_{ext}$ is in units of pN and $\Delta \mu$ is in units of pN.nm}
\label{fig1:operat_diag}
\end{figure}

Following a similar route, the zeros of the hydrolysis rate give us the balanced concentrations of GTP, GDP, P, i.e, which in turn gives us the balanced potential $\Delta \mu_b$. 
The condition for balanced potential $\Delta \mu_b$ is given by
\begin{equation} 
\label{eq:mu_b}
h(F_{ext}\text{,[GTP],[GDP],[P])})=0  \mbox{ for }  \Delta \mu =\Delta \mu_{b}  .
\end{equation}

After substituting the values of the rate constants in (\ref{eq:hydro_rate}), the expression for hydrolysis rate in terms of $\Delta \mu$ and $F_{ext}$ is:
\begin{eqnarray}
h&=& \left[6.8-0.055 e^{-(\Delta \mu/k_BT)}+3.0*10^{-9} e^{\Delta \mu/k_BT} \right. \nonumber \\
&+&2.9*10^{-9} e^{(2 \Delta \mu-F)/k_BT}+5.9*10^{-9} e^{(3 \Delta \mu-F)/k_BT} \nonumber \\
&-& \left. 6.8 e^{(-3 \Delta \mu+F)/k_BT}-5.8*10^{-9} e^{(-2 \Delta \mu+F)/k_BT} \right] \mathrm{s}^{-1}
\label{eq:hyd}
\end{eqnarray}
Using Eqs. (\ref{eq:hyd}) and  (\ref{eq:mu_b}), we obtain the expression for the balanced potential
$\Delta \mu_{b}(F_{ext})$ as function of $F_{ext}$. 

As shown in Fig. \ref{fig1:operat_diag}, the conditions of vanishing velocity (red line) and vanishing hydrolysis rate (blue line) divides the (F, $\Delta \mu$)-plane into four different regions. In operation mode HF (blue), ribosome couples GTP hydrolysis to forward mechanical steps, while in the operation mode HB (yellow), the ribosome couples GTP hydrolysis to backward steps. In the operation mode SB (red), the ribosome couples GTP synthesis to backward steps, while in the operation mode SF (green), the ribosome couples GTP synthesis to forward steps. The SF region appears only for negative external force (corresponding to a force {\it along} the natural direction of motion) when this force is strong enough.

The two functions $F_{s}(\Delta \mu)$ and $\Delta \mu_b(F)$ intersect when there is both mechanical and chemical equilibrium i.e., $F_{ext}, \Delta \mu=0$. We note that the stall force and the balanced potential approach the straight line $F_{ext} =3 \Delta \mu$ as we move closer to the chemo-mechanical equilibrium, i.e., $F_{ext}=\Delta \mu=0$. This represents the ideal operating curve which directly follows from the linear response theory. Near the chemo-mechanical equilibrium, the ribosome works with 100\% efficiency i.e., $F_{ext} \ell= 3 \Delta \mu$.  The efficiency is given by the ratio of the mechanical work performed by ribosome against the external force and the chemical energy consumption.

The validity of our prediction for GTP synthesis under strong enough load force can be verified only after systematic experimental studies are carried out using a set up of the type developed by Liu et al. \cite{bustamante2014}. Can a ribosome really synthesize GTP under the conditions described above? A similar question on the possibility of ATP synthesis by cytoskeletal motors under externally applied load force has been raised earlier in the literature in the context of the modes of their operation. In spite of some indirect indication in support, the possibulity of ATP synthesis during load-induced backstepping still remains an open question \cite{hyeon10}. We believe that the possibility of GTP synthesis during load-induced backward stepping is allowed by the principles of stochastic kinetics and thermodynamics. However, some constraints arising from the structures of the ribosome and the accessory proteins involved during the elongation cycle may make the probability current of the process in reality negligibly small.

\subsection{Ribosome velocity}

We study in more detail the ribosome velocity under normal operation
as a function of the rates $k_{23}$ and $k_{51}$ for the processes that involve
GTP hydrolysis, with all other rates kept at their experimental
or hypothetical values.  
From the exact expression (\ref{eq:vel}) one concludes that
the velocity as a function of any two rates $k$, $k'$, with all other
rates kept fixed, is of the form
\begin{equation}
\label{vtworates}
v(k,k') =  \frac{\alpha_1 + \alpha_2  k + \alpha_3 k' 
+ \alpha_4 k k' }
{\beta_1  + \beta_2 k + \beta_3 k'  
+ \beta_4 k k' + \beta_5 k^2 + \beta_6 {k'}^2} \mathrm{nm} \, \mathrm{s}^{-1}
\end{equation}
with coefficients $\alpha_l$, $\beta_l$ that depend on the choice of rates
$k$, $k'$.

As a function of the rates $k=k_{51}$ and $k'=k_{23}$ one obtains from the exact stationary 
distribution the coefficients 
$\alpha_1 = 1.23\cdot10^{-8}$, $\alpha_2 = 3.53  \mathrm{s}$, 
$\alpha_3 = 5.87\cdot10^{-12} \mathrm{s}$, 
$\alpha_4 = 7.78\cdot10^{-2} \mathrm{s}^2$ and $\beta_1 = 6.69\cdot10^{-8}$, 
$\beta_2 = 1.31 \mathrm{s}$, 
$\beta_3 = 7.46\cdot10^{-2} \mathrm{s}$, $\beta_4 = 2.14 \cdot10^{-3} \mathrm{s}^2$, 
$\beta_5 = 4.36 \cdot10^{-9} \mathrm{s}^2$, $\beta_6 = 0$. Neglecting numerical prefactors  of order
$10^{-6}$ and smaller and introducing the dimensionless unit rates $\tilde{k}_{ij} := {k}_{ij} \, \mathrm{s}$ 
(i.e., the rates expressed in units of a second) one obtains
\begin{eqnarray}
\label{v2351}
v & = &  \frac{3.53 \tilde{k}_{51} + 7.78\cdot10^{-2}  \tilde{k}_{51} \tilde{k}_{23}  }
{1.31 \tilde{k}_{51} + 7.46\cdot 10^{-2} \tilde{k}_{23} + 2.14 \cdot10^{-3} \tilde{k}_{51} \tilde{k}_{23}}
\mathrm{nm} \, \mathrm{s}^{-1} .
\end{eqnarray}

\begin{figure}[h]
\begin{center}
\includegraphics[width=0.7\columnwidth]{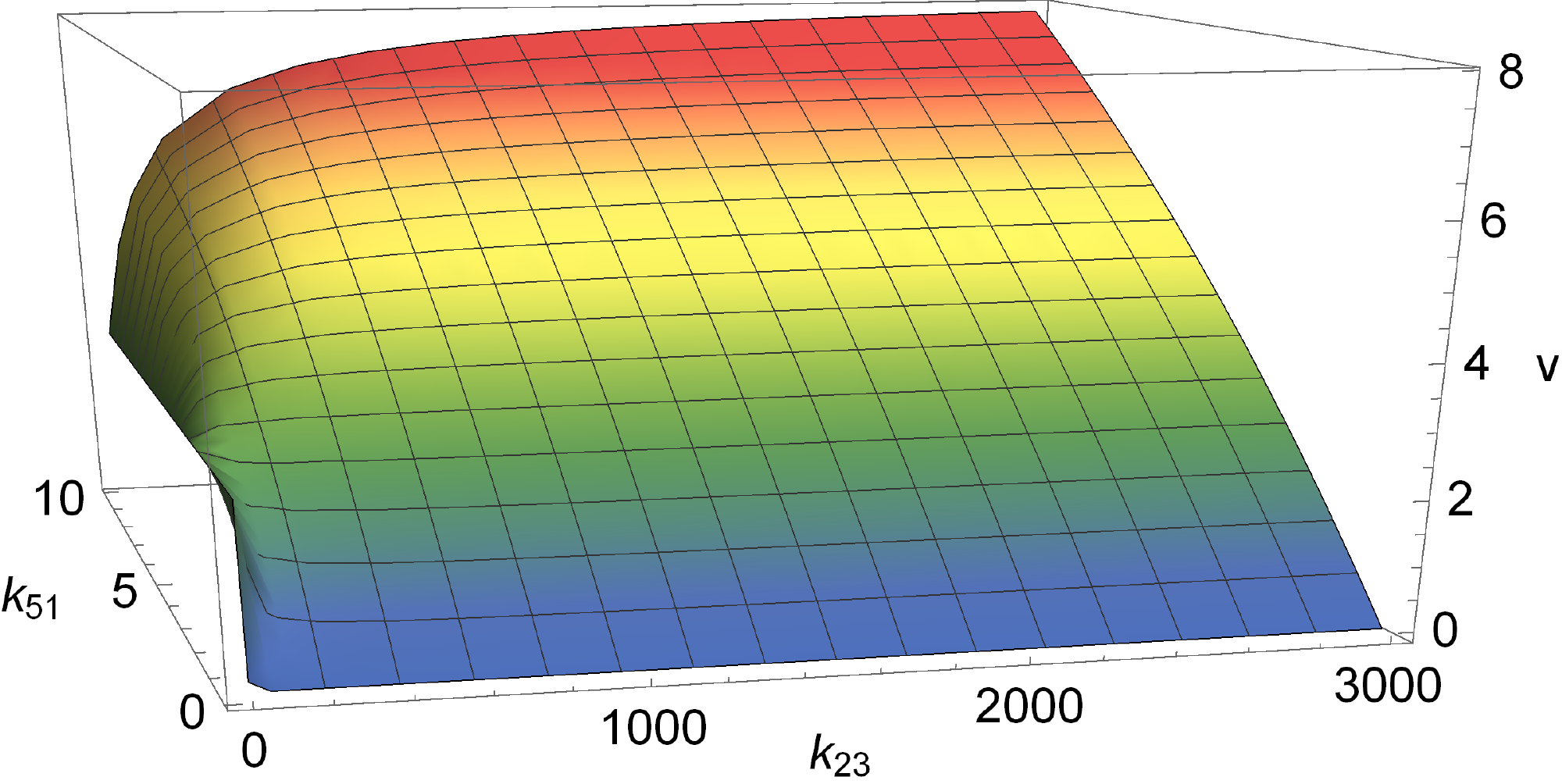}
\end{center}
\caption{Velocity $v$ in $\mathrm{nm}\,\mathrm{s}^{-1}$ as a function of the rates
$k_{51}$ and $k_{23}$ in $\mathrm{s}^{-1}$.}
\label{fig:velocity2351}
\end{figure}

Notice that the velocity becomes a constant even when the rates $k_{51}$ and $k_{23}$
become large and saturates. 
This saturation effect arises because even if hydrolysis and translocation would
be instantaneous, the velocity would still be limited by the rate of the other
processes in the elongation cycle. As seen in Fig.~\ref{fig:velocity2351}, the dependence on the rate $k_{23}$
is very weak in the experimentally relevant range around $k_{23} = 1500 \,\mathrm{s}^{-1}$ and 
saturates to a value that is approximately proportional to $k_{51}$ in the experimentally 
relevant range around $k_{51} = 4 \,\mathrm{s}^{-1}$. 
This saturation effect indicates that the main limiting factor for the
velocity is the second hydrolysis in the elongation cycle.  

\begin{figure}[h]
\begin{center}
\includegraphics[width=0.7\columnwidth]{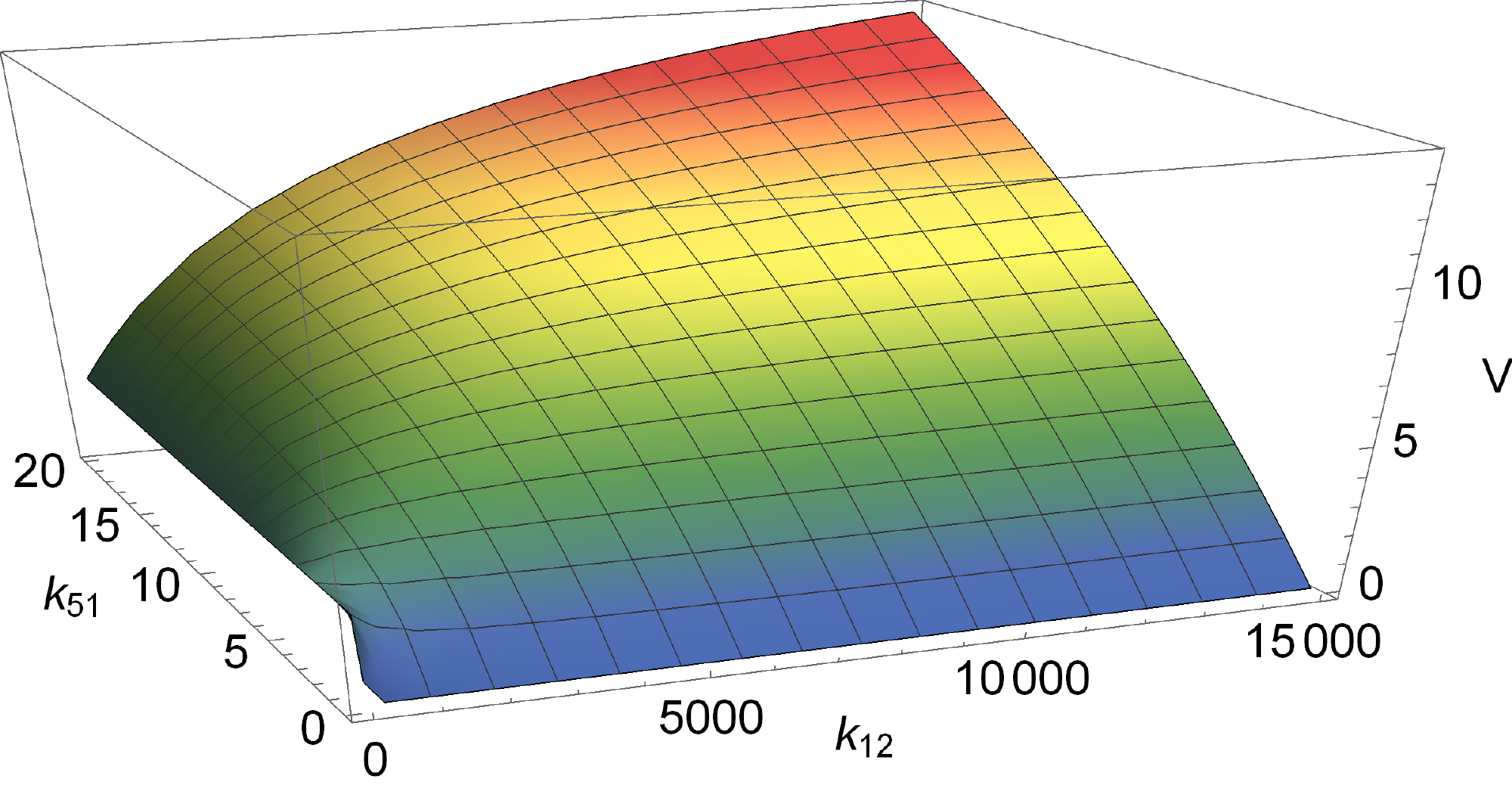}
\end{center}
\caption{Velocity $v$ in $\mathrm{nm}\,\mathrm{s}^{-1}$ as a function of the rates
$k_{51}$ and $k_{12}$ in $\mathrm{s}^{-1}$.}
\label{fig:velocity1251}
\end{figure}

To explore the velocity saturation due to the second GTP hydrolysis further we consider the
velocity as a function of $k=k_{51}$ and $k'=k_{12}$, the latter being the rate at which
GTP for the first hydrolysis is supplied by the ternary complex \text{EF-Tu.GTP.aa-tRNA}. 
As a function of these two rates the
constants entering (\ref{vtworates}) are given by
$\alpha_1 = 2.1\cdot10^{-9}$, $\alpha_2 = 1.1 \mathrm{s}$, $\alpha_3 = 0$,
$\alpha_4 = 1.3\cdot10^{-3} \mathrm{s}^2$ and $\beta_1 = 3.9\cdot10^{-8}$, 
$\beta_2 = 3.8\cdot10^{-1} \mathrm{s}$, 
$\beta_3 = 1.3\cdot10^{-3} \mathrm{s}$, $\beta_4 = 7.8 \cdot10^{-6} \mathrm{s}^2$, 
$\beta_5 = 1.3 \cdot10^{-20} \mathrm{s}^2$, $\beta_6=0 $ which yields the
velocity plot shown in Fig.~\ref{fig:velocity1251}. Neglecting numerical prefactors of order
$10^{-6}$ and smaller one finds
\begin{eqnarray}
\label{v1251}
v & = & \frac{1.1 \tilde{k}_{51}+1.3\cdot10^{-3} \tilde{k}_{51} \tilde{k}_{12}}
{1.3\cdot10^{-3} \tilde{k}_{12} + 0.38 \tilde{k}_{51}} 
\mathrm{nm} \, \mathrm{s}^{-1}.
\end{eqnarray}

The dependence of the velocity on the rate $k_{12}$ is very weak in the experimentally 
relevant range of the order $10^3$ - $10^4$. As $k_{12}$ increases (high concentration of aa-tRNA), the 
velocity becomes limited by the rate $k_{51}$ which is not affected by a high concentration of 
aa-tRNA.  

\subsection{Hydrolysis rate}

\begin{figure}[h]
\begin{center}
\includegraphics[width=0.7\columnwidth]{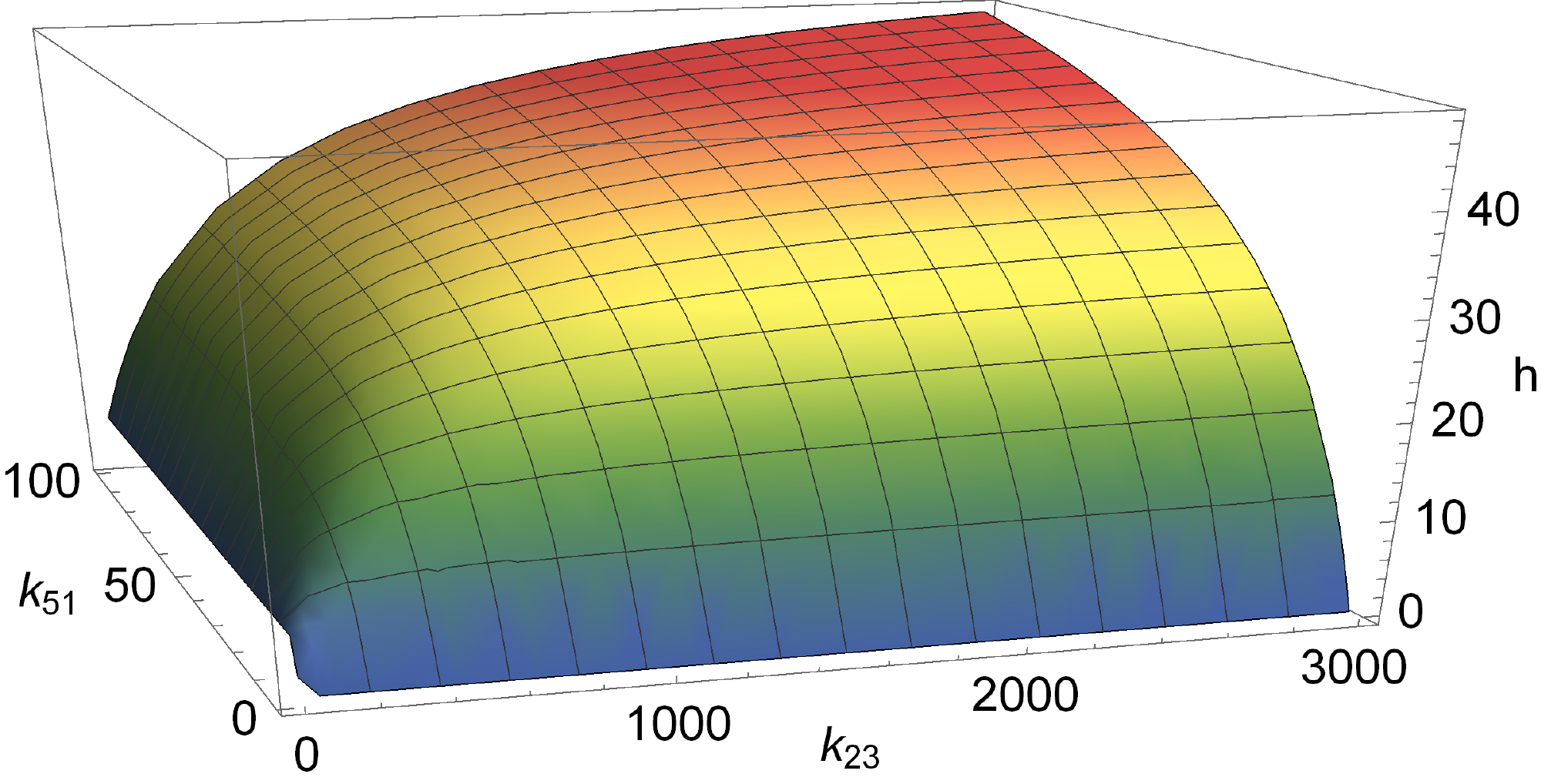}
\end{center}
\caption{Hydrolysis rate $h$ in units of $\mathrm{s}^{-1}$
plotted against the rates $k_{23}$ and $k_{51}$
in units of \red{$\mathrm{s}^{-1}$}.}
\label{fig14:h_k23_k51}
\end{figure}

The transitions $2\to 3$  and $5\to 1$ involve GTP hydrolysis. Therefore, from the Eq.(\ref{eq:hydro_rate}), we get the expression 
\begin{equation}
\label{htworates}
h = \frac{\tilde{\alpha}_1 + \tilde{\alpha}_2 k_{23} + \tilde{\alpha}_3 k_{51}+\tilde{\alpha}_4 k_{23}k_{51}}
{\tilde{\beta}_1  + \tilde{\beta}_2 k_{23} + \tilde{\beta}_3 k_{51} + \tilde{\beta}_4 k_{23} k_{51}+\tilde{\beta}_5 k^{2}_{51}} \mathrm{s}^{-1}
\end{equation}
for the rate of GTP hydrolysis as a function of the rates $k_{23}$ and $k_{51}$, 
where the coefficients $\alpha_l$, $\beta_l$, just like in the case of the velocity, depend on the choice of 
the rates. For  $k=k_{23}$, $k'=k_{51}$ one obtains the coefficients $\tilde{\alpha}_1=2.6\cdot 10^{-8}$, 
$\tilde{\alpha}_2=9.8\cdot 10^{-11} \mathrm{s}$, $\tilde{\alpha}_3=7.5 \mathrm{s}$, 
$\tilde{\alpha}_4=1.6\cdot 10^{-1} \mathrm{s}^2$, 
$\tilde{\beta}_1=1.3\cdot 10^{-7} $, $\tilde{\beta}_2=7.4\cdot 10^{-2} \mathrm{s}$, 
$\tilde{\beta}_3=1.3 \mathrm{s}$, 
$\tilde{\beta}_4=2.1\cdot 10^{-3} \mathrm{s}^2$, $\tilde{\beta}_5=2.2\cdot 10^{-9} \mathrm{s}^2$. 
As a function of $k_{51}$, there is little variation in the rate for 
$k_{23}>1500 \, \mathrm{s}^{-1}$, which is in the experimental relevant range $k_{23} \approx 1500 \, \mathrm{s}^{-1}$,
thus indicating robustness of hydrolysis w.r.t. this rate. On the other hand, in the region of the experimental
value $k_{51} \approx 4 \, \mathrm{s}^{-1}$ for the transition $5\to 1$ (where hydrolysis is accompanied
by translocation), the hydrolysis rate depends strongly on $k_{51}$. The rate of hydrolysis under normal operation is shown in Fig.~\ref{fig14:h_k23_k51}.

\subsection{Entropy Production}

In order to determine how the entropy change $\Delta S_{\kappa,\pm} := \pm \Delta S_{\kappa}$ 
associated with the completion 
of an oriented cycle $(\kappa,\pm)$ contributes to the total entropy production of the process we recall that
the entropy of a system described by Markovian stochastic dynamics is the usual Gibbs entropy
\begin{equation} \label{eq:entropy}
S^{sys}(t)=-k_{B} \sum_{i} P_i(t) \ln(P_i(t)).
\end{equation}
Following Schnakenberg \cite{schnakenberg_network_1976} we split the time evolution 
of the system entropy into two parts
\begin{equation} 
\label{eq:sigma_pf}
\dfrac{d}{dt}S^{sys}(t) =\sigma^{tot}(t)+\sigma^{env}(t)
\end{equation}
with the {\it total entropy production}
\begin{equation} \label{eq:entropy_prod_rate}
\sigma^{tot}(t) := \dfrac{1}{2} k_{B} \sum_{ij} J_{ij}(t) \ln\bigg(\dfrac{P_i(t) k_{ij}}{P_j(t) k_{ji}}\bigg)
\end{equation}
and the {\it entropy flux}
\begin{equation} \label{eq:entropy_flux_rate}
\sigma^{env}(t) := -\dfrac{1}{2}k_{B} \sum_{ij} J_{ij}(t) \ln\bigg(\dfrac{k_{ij}}{k_{ji}}\bigg).
\end{equation}
The entropy flux can be interpreted as the entropy production of the environment 
\cite{lebowitz_gallavotticohen-type_1999,Harr07}. Using the master equation 
(\ref{eq:master_equation}) it is straightforwardly verified that the
time-derivatives satisfy $\dot{S}(t) = \sigma^{tot}(t) + \sigma^{env}(t)$.

In the steady state, the system entropy does not change which implies
\begin{equation}\label{eq:ss_entropy1}
\sigma^{tot} = - \dfrac{1}{2} k_{B} \sum_{ij} J_{ij} \ln{\left(\dfrac{k_{ij}}{k_{ji}}\right)} = - \sigma^{env}.
\end{equation}
In terms of cycle fluxes with the entropy change
\begin{equation}
\label{eq:ss_entropy_c2}
\Delta S_{\kappa,\pm} = k_B \ln \frac{\Pi_{\kappa,\pm}}{\Pi_{\kappa,\mp}}
\end{equation}
along a circle in clockwise or anticlockwise direction one gets the decomposition
\begin{equation}\label{eq:ss_entropy_c1}
\sigma^{tot} = \sum_{\kappa} \left(J_{\kappa,+} \Delta S_{\kappa,+} 
+ J_{\kappa,-} \Delta S_{\kappa,-} \right).
\end{equation}
of the entropy production in terms of cycles (see appendix \ref{app-EntropyProd} for detailed derivation).
Using $\Delta S_{\kappa,-} = - \Delta S_{\kappa,+}$,
which indicates that the cycle $(\kappa,+)$ is the time reversed  trajectory of the cycle $(\kappa,-)$,
we arrive at the entropy production
\begin{equation}\label{eq:ss_entropy_c3}
\sigma^{tot}_{\kappa} = J_{\kappa} \Delta S_{\kappa} 
\end{equation}
for cycle $\kappa$ and at 
\begin{equation}\label{eq:ss_entropy_c4}
\sigma^{tot} = \sum_{\kappa} \sigma^{tot}_{\kappa}
\end{equation}
for the total entropy production. From Tab.~\ref{tab:pi_values} it is readily seen that cycle (b) 
has the overwhelmingly largest contribution to the total entropy production, followed by cycle (c)
and then cycle (a).

As $k_{12}$ is proportional to the concentration of cognate EF-Tu.GTP.aa-tRNA, the increase of $k_{12}$ is easily implemented by the increase of the concentration of cognate EF-Tu.GTP.aa-tRNA.  
Fig.~\ref{fig:entropyprod_vs_TC}(a) shows how the entropy production increases with the increase of the concentration of the cognate ternary complex EF-Tu.GTP.aa-tRNA. We find that the effect is particularly pronounced at small concentration
and gradually flattens out somewhat at higher concentration where the entropy production diverges logarithmically with the further increase of the EF-Tu.GTP.aa-tRNA concentration. A similar trend of variation of the entropy production is observed also in the variation of, for example $k_{23}$ (see Fig.~\ref{fig:entropyprod_vs_TC}(b)) as well as with the variation of $k_{51}$ (Fig.~\ref{fig:entropyprod_vs_TC}(c)). Note that $k_{23}$ is the rate of a step that involves the hydrolysis of a molecule of GTP whereas $k_{51}$ is associated with translocation.

\begin{figure}[h]
\begin{center}
\label{fig:entropyprod_vs_TC}
(a) \\
\includegraphics[width=0.6\columnwidth]{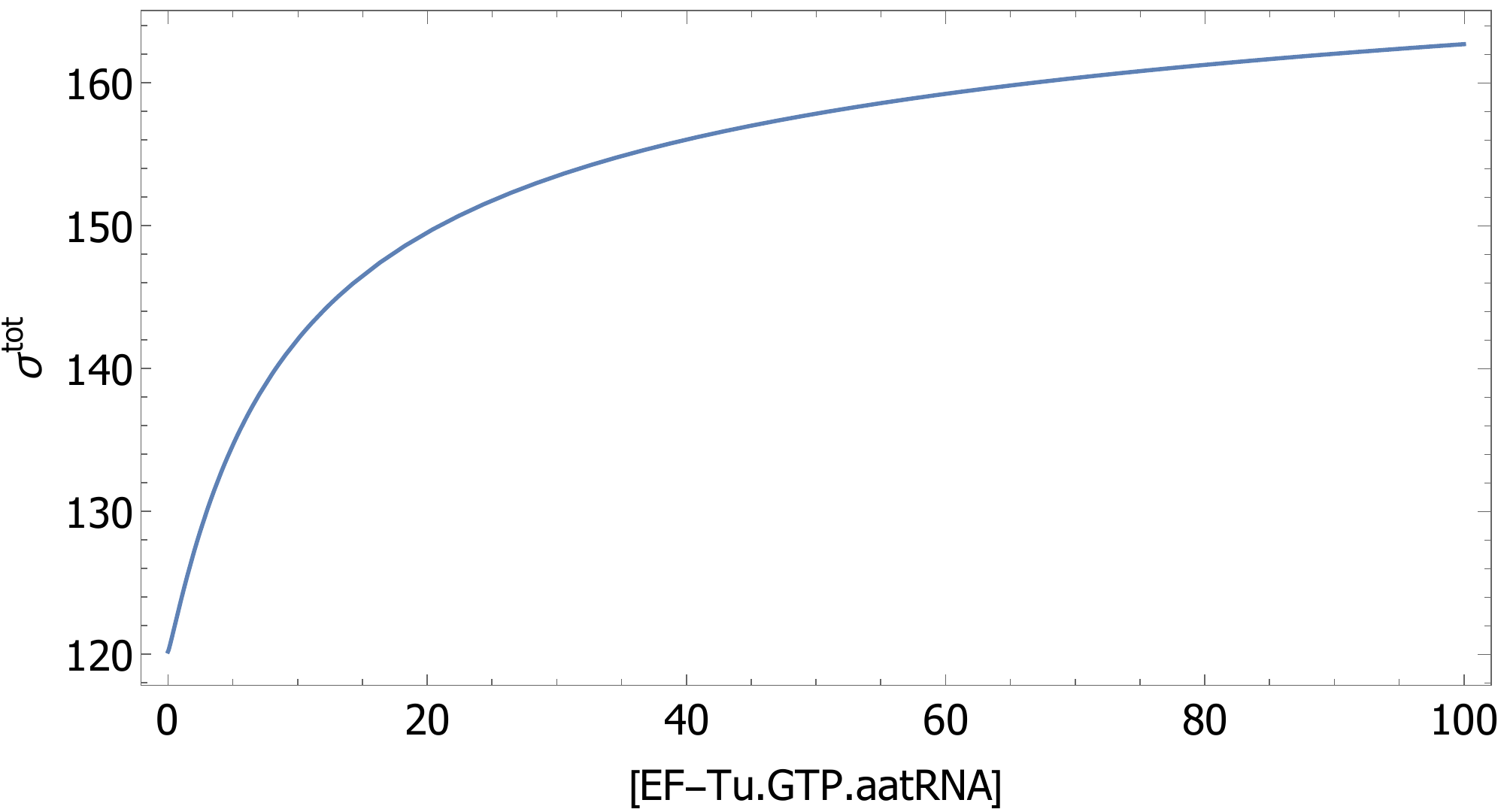} \\
(b)\\
\includegraphics[width=0.6\columnwidth]{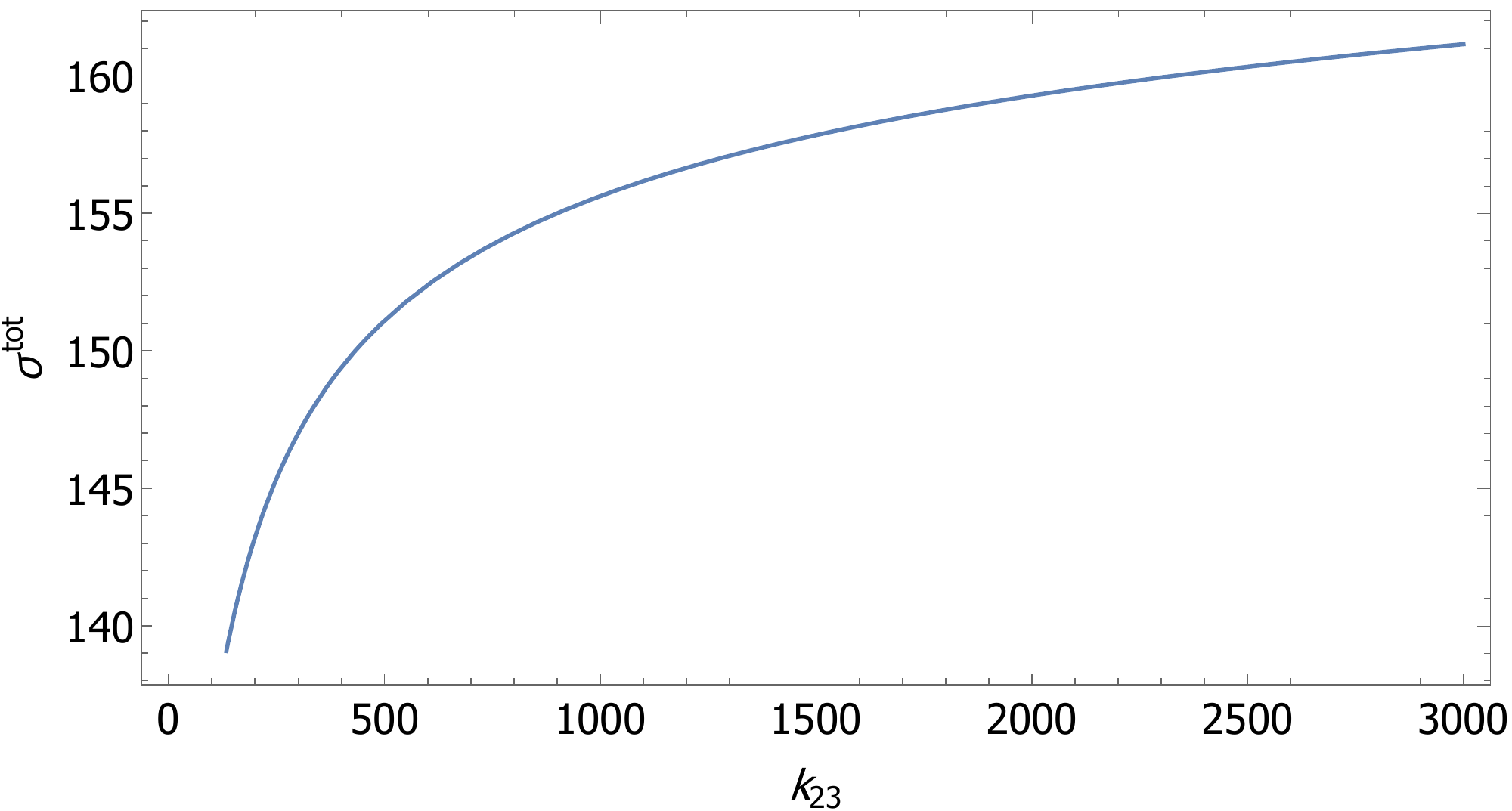} \\
(c)\\
\includegraphics[width=0.6\columnwidth]{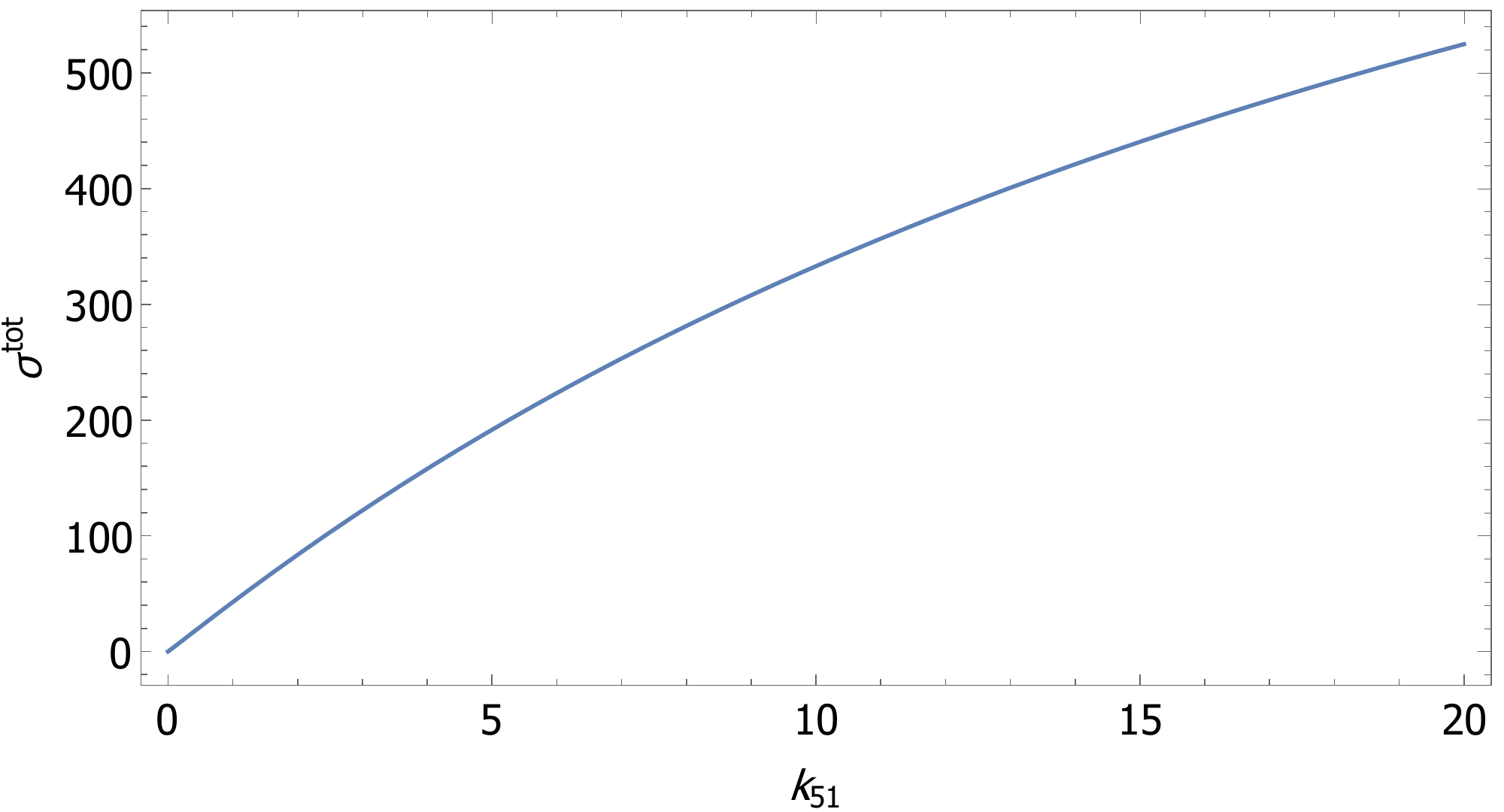} \\
\end{center}
\caption{Entropy production rate $\sigma^{tot} $ in units of seconds plotted against (a) the concentration 
of the cognate ternary complex EF-Tu.GTP.aa-tRNA (in units of $\mu \mathrm{mol}$ which is proportional to the rate 
$k_{12}$), (b) the rate $k_{23}$ of the transition $2 \to 3$ that  involves hydrolysis of GTP, (c) the rate $k_{51}$ of the transition $5 \to 1$ that is associated with translocation. The
remaining rates are kept at their experimental values.}
\label{fig:entropyprod_vs_TC}
\end{figure}

In order to understand the causes and consequences of this common trend of variation of the entropy production with the rates of interstate transitions, it is instructive to consider also the behaviour of the entropy production as a function of the ribosome velocity. As $k_{12}$ increases (i.e., effectively the concentration of cognate EF-Tu.GTP.aa-tRNA increases), the velocity $v(k_{12})$ (\ref{v1251}) of the ribosome, expressed as a function of $k_{12}$, saturates to a value $v^\ast = 3.7 \,  \mathrm{nm} \, \mathrm{s}^{-1}$ as discussed above since in that parameter regime it is limited by the rates along the cycles that involve translocation. These rates are not affected by a high concentration of 
cognate EF-Tu.GTP.aa-tRNA. On the other hand, the contribution to the entropy production from the cycles (a), (b) and (c) -- which involve the ratio $\ln{(k_{12} /k_{21})}$ from the reaction $1\rightleftharpoons 2$ --
keeps increasing and diverges at $v^\ast$ since the inverse function $k_{12}(v)$ diverges at $v^\ast$. A similar divergence appears when any of the transition rates $k_{ij}$ becomes large since the entropy production diverges as $\ln{(k_{ij})}$ while the velocity saturates for large $k_{ij}$ (see Figs.~\ref{fig:entropyprod_vs_TC}(b) and (c)) .

\begin{figure}[h!]
\begin{center}
\includegraphics[width=0.7\columnwidth]{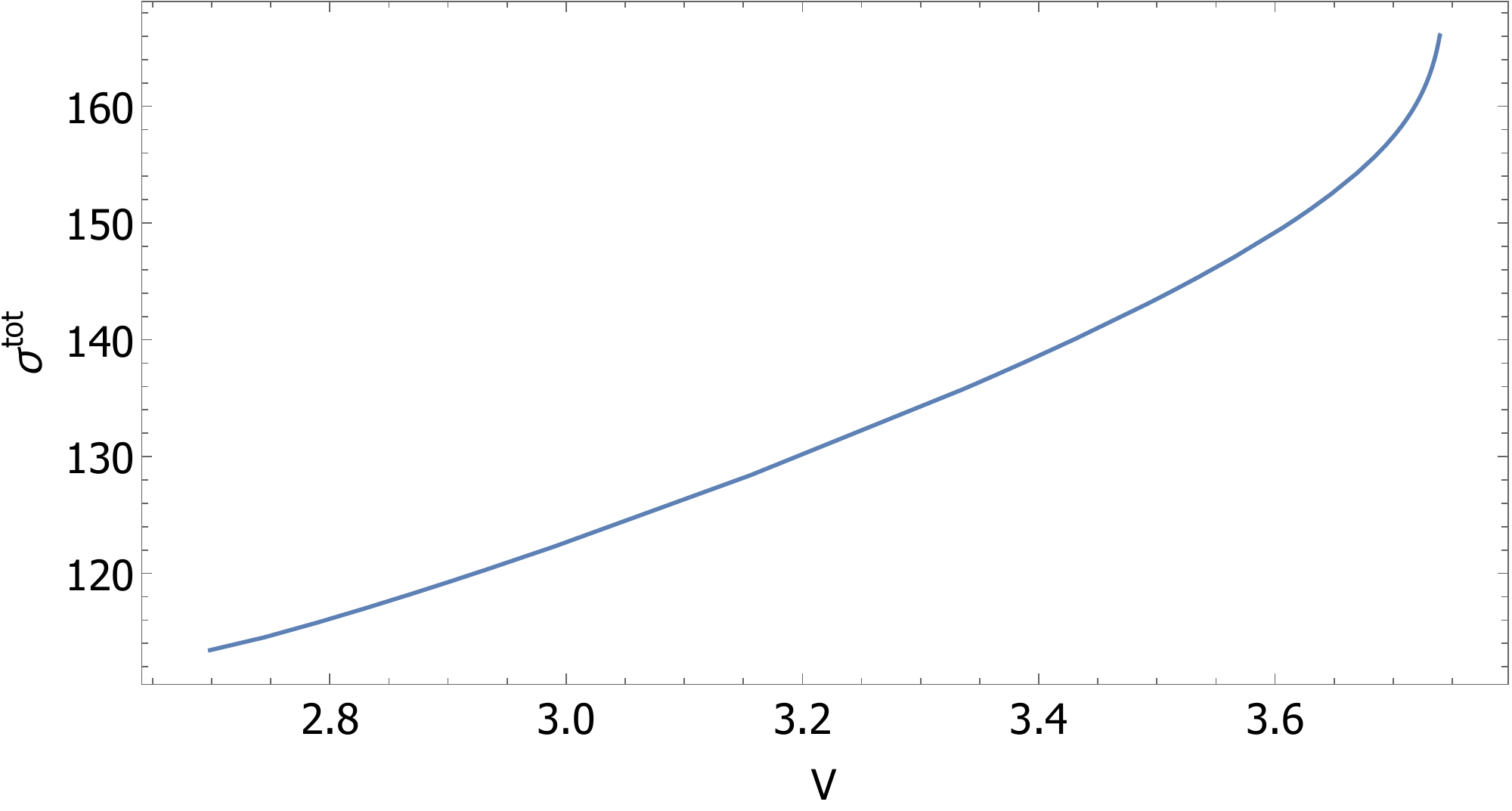}
\end{center}
\caption{ The entropy production rate ($P$) per second plotted against the velocity of translation $v$ in $\mathrm{nm}/\mathrm{s}$,
obtained by varying  the rate $k_{12}$ from 0 to $150000 \mathrm{s}^{-1}$.}
\label{fig8:entropyprod_vel}
\end{figure}

\subsection{Accuracy of translation and its thermodynamic cost}

As pointed out earlier, proof reading and hence rejection of non- or near cognate aa-tRNA or mischarged aa-tRNA, may be faulty,
leading to a transition $7\to 8$, $11\to 12$ and $15\to 16$. Unless the reverse transitions $8\to7$, $12\to 11$ and $16\to 15 $ take place before elongation and
translocation, this process leads to a production of missense error in polypeptide chains
with a total net production rate
\begin{equation}
\label{eq:e6}
e_6 = J_{78}+J_{11,12}+J_{15,16}.
\end{equation}
On the other hand, by a similar argument, correct production occurs with a rate 
\begin{equation}
\label{eq:e4}
e_4 = J_{34}.
\end{equation}
Thus the accuracy of translation, defined by \cite{dutta_generalized_2017} 
\begin{equation}
\phi := \frac{e_4}{e_4+e_6} = \frac{J_c+J_g}{J_c+J_g-J_d-J_h+J_i+J_k-J_l-J_j},
\end{equation}
yields the fraction of proper polypeptide chains in the total production of the ribosome.

\begin{figure}[h]
	\begin{center}
		\includegraphics[width=0.7\columnwidth]{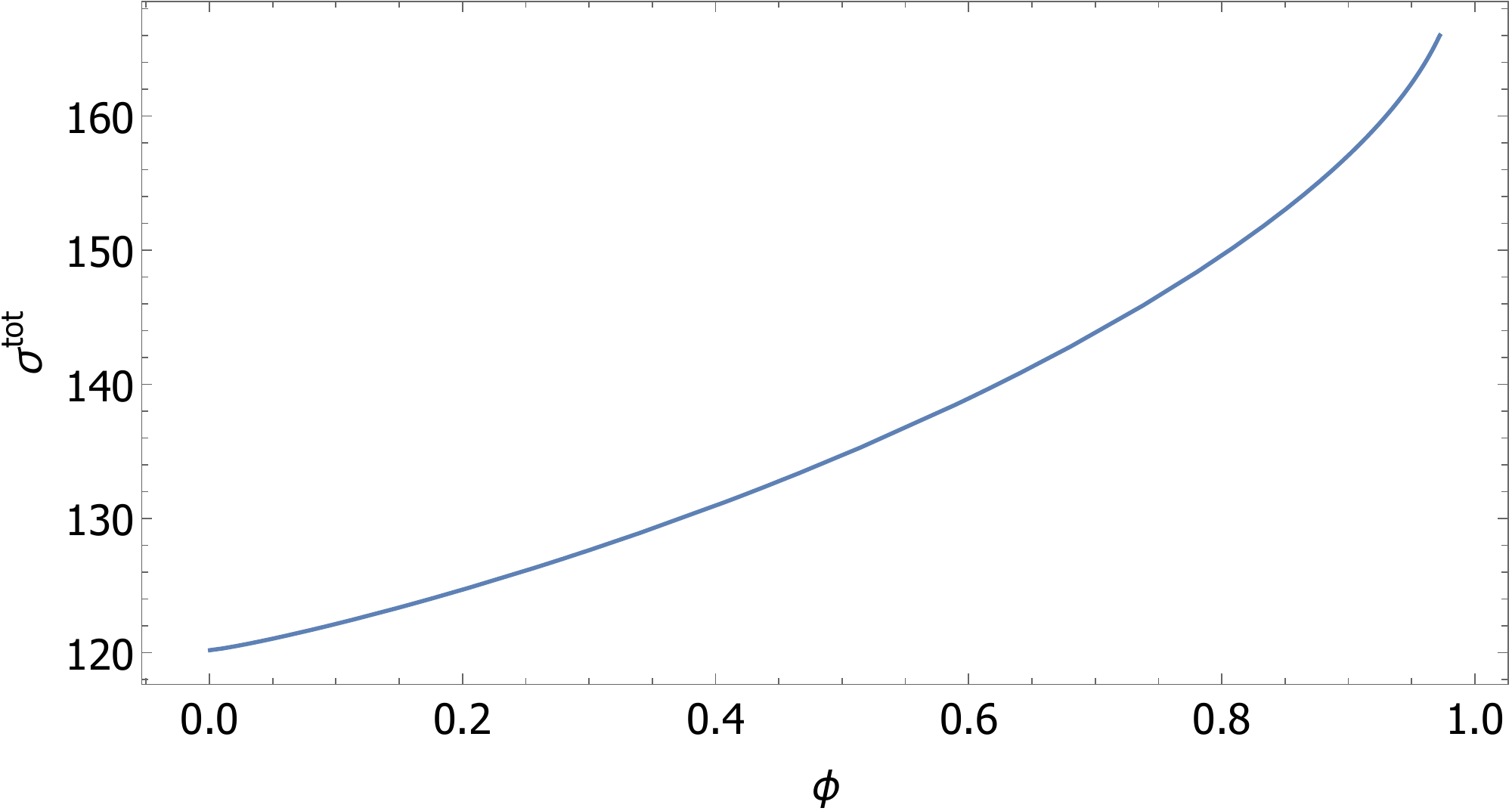}
	\end{center}
	\caption{Entropy production rate $\sigma^{tot} $ is plotted against accuracy of translation $\phi$ by varying the rate $k_{23}$ from 0 to 30000 $s^{-1}$}
	\label{fig13:phi_k34_k36}
\end{figure}

In Fig.~\ref{fig13:phi_k34_k36}, total entropy production rate $\sigma^{tot}$ is plotted against accuracy of translation. The plot shows that as we increase the accuracy of translation, the energetic cost of translation also increases.

\begin{figure}[h]
	\begin{center}
		(a) \\
		\qquad
		\includegraphics[width=0.7\columnwidth]{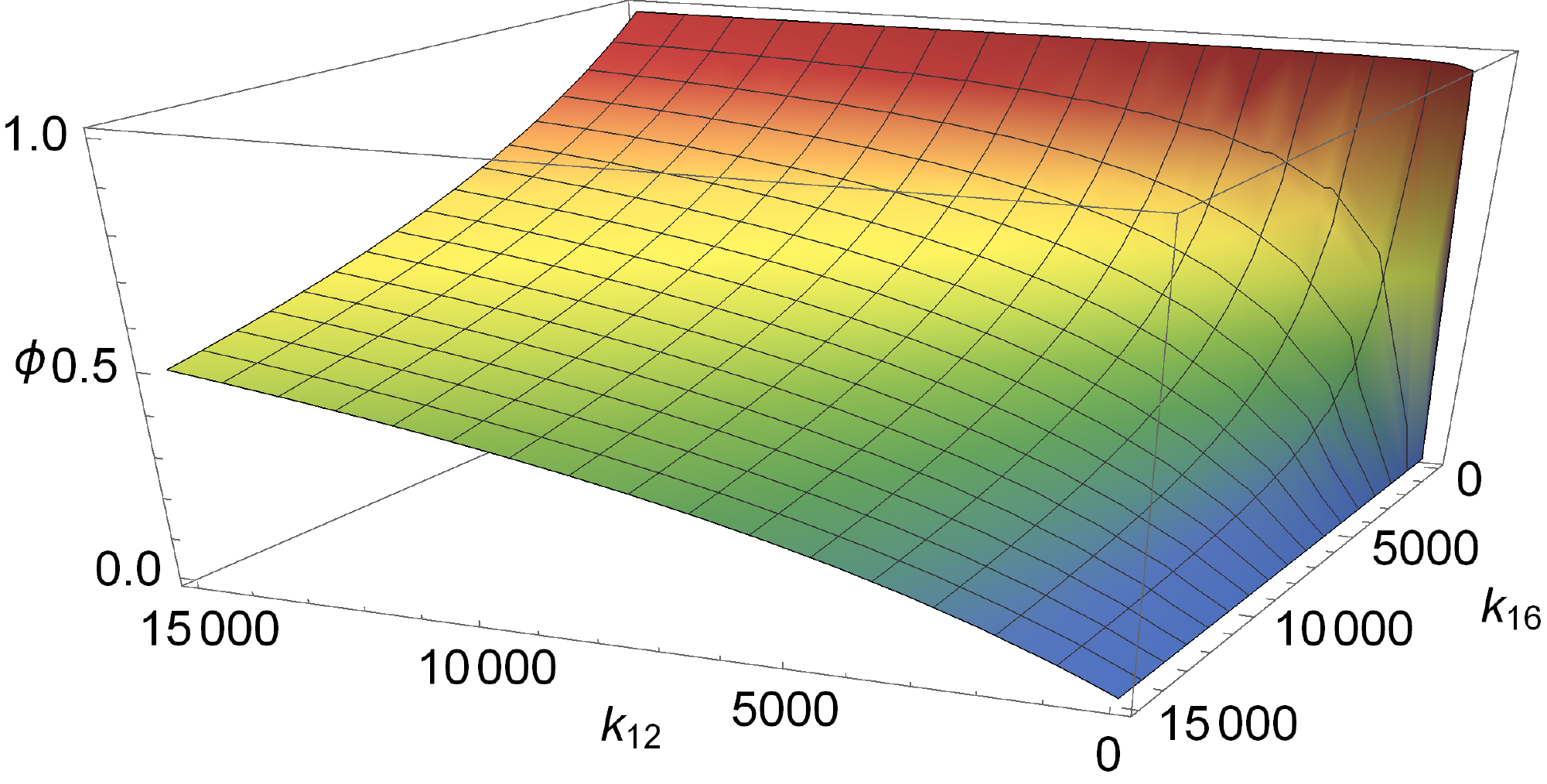} \\
		\qquad
		(b)\\
		\qquad
		\includegraphics[width=0.7\columnwidth]{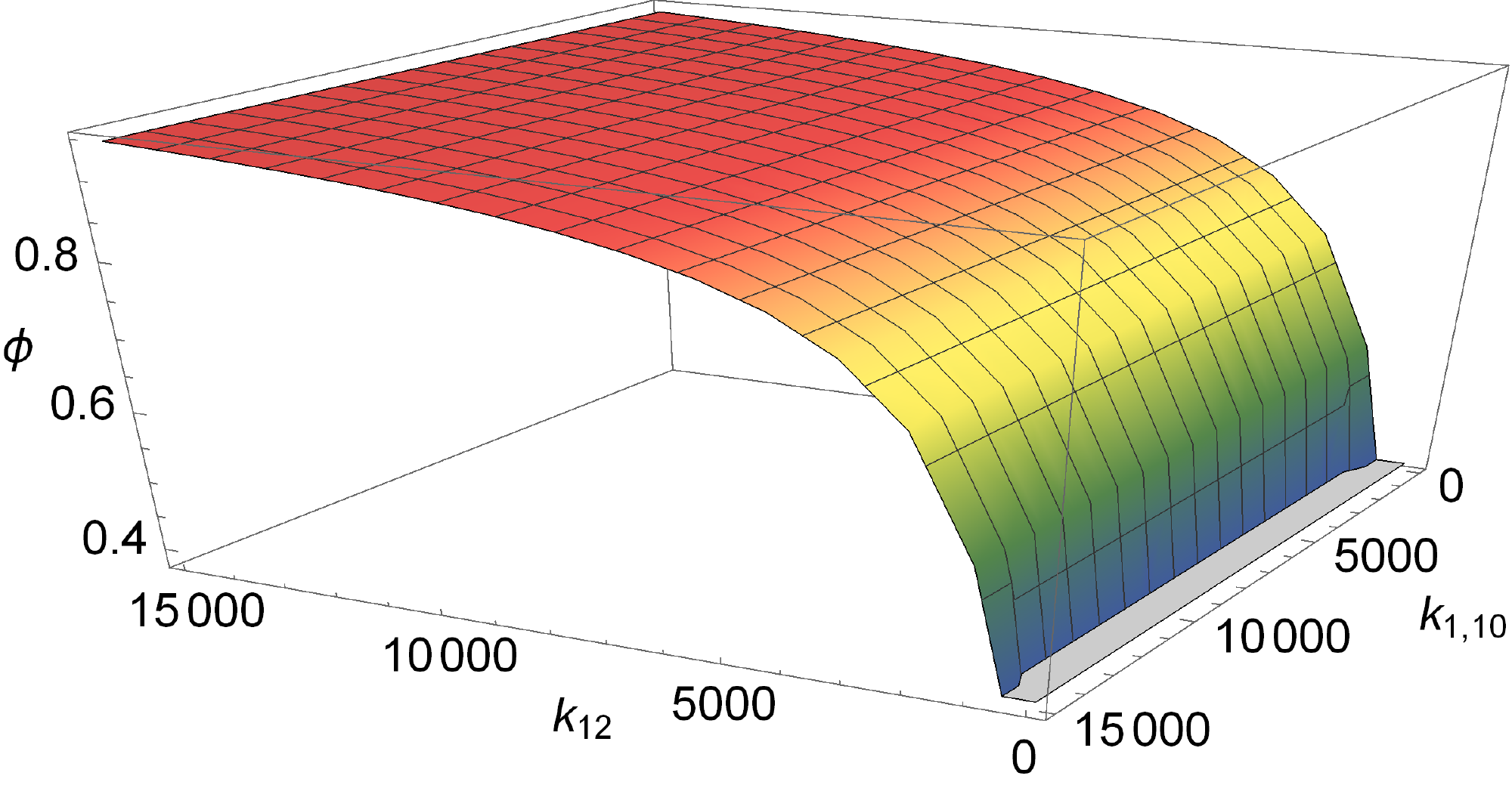} \\	\end{center}
	\caption{Accuracy of translation $\phi$ plotted against (a) the concentration 
		of the cognate ternary complex binding rate $k_{12}$ and mischarged ternary complex binding rate $k_{16}$, (b) the concentration 
		of the cognate ternary complex binding rate $k_{12}$ and near cognate ternary complex binding rate $k_{1,10}$. The
		remaining rates are kept at their experimental values.}
	\label{fig:accuracy_vs_TC}
\end{figure}

In Fig.~\ref{fig:accuracy_vs_TC}, we observe how the accauracy of translation vary with the variation in concentration of cognate and mischarged ternary complexes Fig.~\ref{fig:accuracy_vs_TC}(a) and with the variation of concentration of cognate and near cognate ternary complexes Fig.~\ref{fig:accuracy_vs_TC}(b). This gives us n insight of the competition between different types of ternary complexes. We see that the accuracy varies considerably with the variation in concentration of cognate and mischarged because the mischarged tRNA escapes the proofreading as it has the correct codon-anticodon base pairing. Therefore, if we increase the concentration of mischarged aa-tRNA in the surrounding, it may conisderably affect the accuracy of translation. For the Fig.~Fig.~\ref{fig:accuracy_vs_TC} (b), the accuracy is almost insensitive to the near cognate tRNA concentration. This is because the ribosome ensures rejection of near cognate tRNA through stringent proofreading.

\begin{figure}[h]
	\begin{center}
		(a) \\
		\qquad
		\includegraphics[width=0.7\columnwidth]{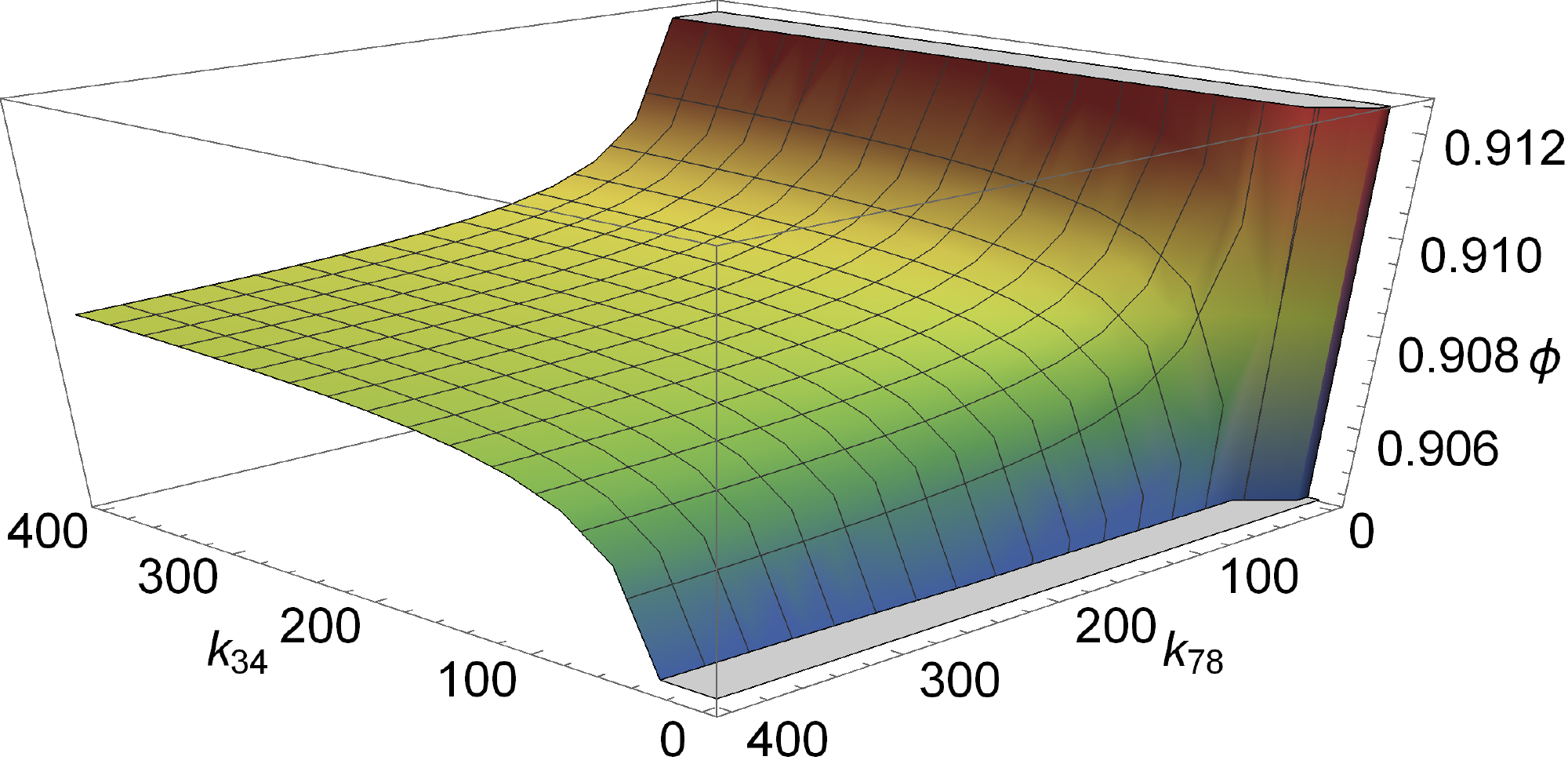} \\
		\qquad
		(b)\\
		\qquad
		\includegraphics[width=0.7\columnwidth]{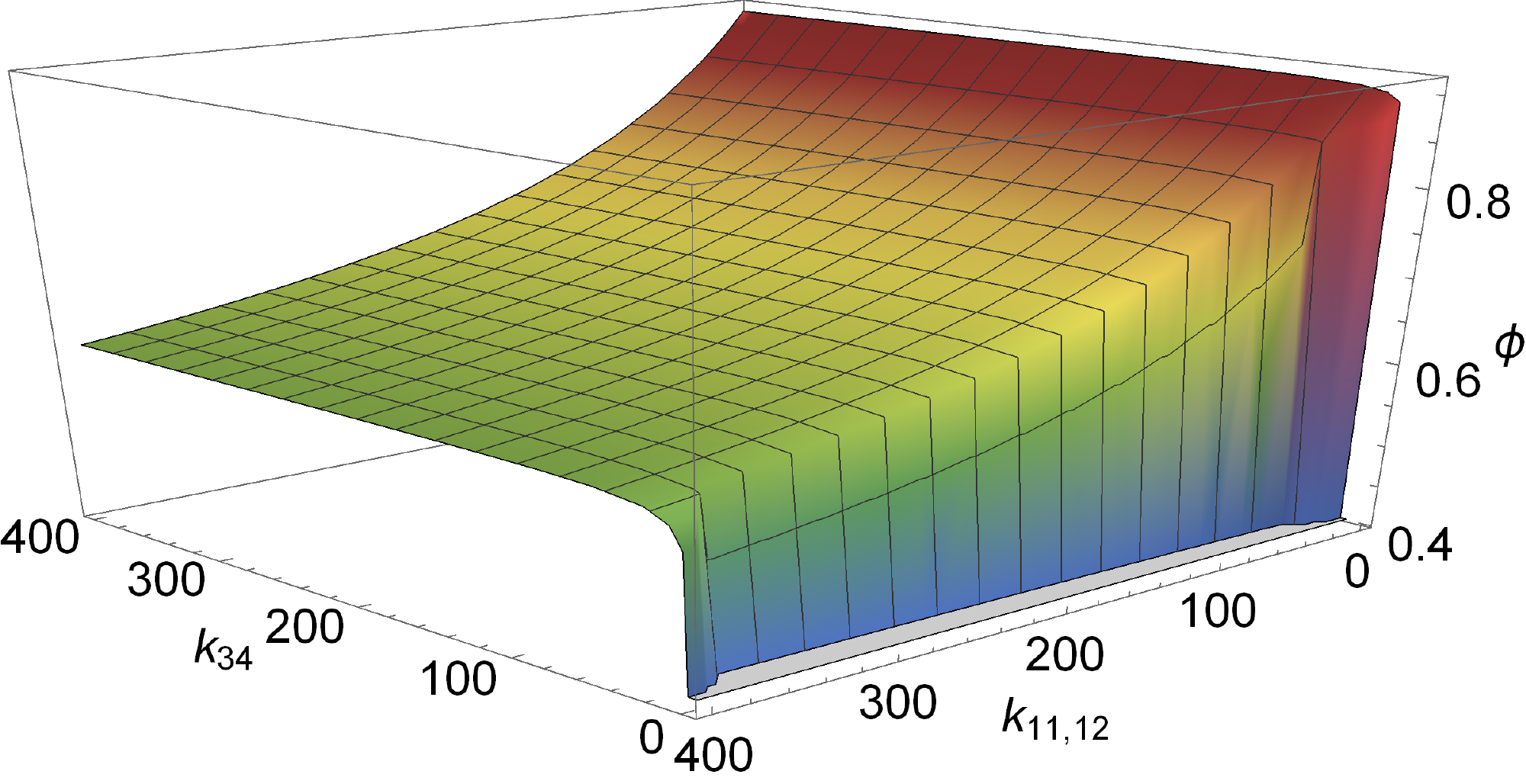} \\
	\end{center}
	\caption{Accuracy of translation $\phi$ plotted against (a) the cognate cycle peptide bond formation rate $k_{34}$ and mischarged cycle peptide bond formation rate $k_{78}$, (b) the cognate cycle peptide bond formation rate $k_{34}$ and mischarged cycle peptide bond formation rate $k_{11,12}$. The
		remaining rates are kept at their experimental values.}
	\label{fig:accuracy_vs_pbond}
\end{figure}

In Fig.~\ref{fig:accuracy_vs_pbond}, we show the variation of accuracy of translation $\phi$ with the cognate cycle peptide bond formation rate $k_{34}$ and mischarged cycle peptide bond formation rate $k_{78}$ (Fig.~\ref{fig:accuracy_vs_pbond} (a)) and the cognate cycle peptide bond formation rate $k_{34}$ and mischarged cycle peptide bond formation rate $k_{11,12}$ in Fig.~\ref{fig:accuracy_vs_pbond} (b).

\section{Conclusions}

The
ribosome is one of the largest multi-component molecular machines. It performs a crucially important biological function called translation (of genetic code) that results in the synthesis of proteins as directed by a mRNA template.  
Although the structure and kinetics of ribosomes have been studied extensively in the past, the stochastic thermodynamics has not received attention so far. Most of the results reported in this paper constitute, to our knowledge, essentially the first step in that direction.

Using a network approach we solved exactly the stationary master equation for a seven-states model of the kinetics  of a ribosome during the elongation stage of translation.  This solution is used for a detailed description of stationary properties arising from the stochasticity of the chemo-mechanical cycle of the ribosome. We have identified the various modes of operation of this machine in terms of its average velocity and the mean rate of GTP hydrolysis. Similar analysis have been reported earlier in the literature for cytoskeletal motor proteins. To our knowledge, this paper reports the first analysis, from the perspective of stochastic thermodynamics, of a molecular machine that carries out template-directed polymerization. 

Our quantitative predictions can be used as benchmarks for simpler models and thus allow for judging the adequacy of such reduced models that incorporate fewer or other internal states of the ribosome.  Moreover, since we used rates obtained from experiments, the comparison of the analytical results with other experimental data allows for a detailed quantitative understanding of the microscopic processes underlying translation, particularly those during the elongation stage.

Finally, we would like to point out that the knowledge of {\it exact} stationary distribution allows for 
the construction of exactly solvable models of many {\it interacting} ribosomes, as has 
been demonstrated recently in a mathematically similar setting for a two-states description 
of transcription elongation by RNA polymerase \cite{BELITSKY2019370}. This approach
can be adapted to more internal states and exact stationary single-motor results open 
up the path to obtaining exact quantitative results for the elongation kinetics of many 
simultaneously transcribing or translating molecular motors.

\section{Acknowledgements}
G.M.S. thanks the Institute of Mathematics and Statistics, where part of this
work was done for kind hospitality. This work was financed in part by Coordena\c c\~ao 
de Aperfei\c coamento de Pessoal de N\'ivel Superior -- Brazil (CAPES) -- Finance Code 001,  by the grants 
2017/20696-0, 2017/10555-0 of the S\~ao Paulo Research Foundation (FAPESP), 
and by the grant 309239/2017-6 of the Conselho Nacional de Desenvolvimento Cient\'ifico e 
Tecnol\'ogico (CNPq). This work was also supported partly by SERB (India) through a J.C. Bose National Fellowship (D.C.). D.C. also thanks Frank J\"ulicher and the Visitors Program at the MPI-PKS for hospitality in Dresden where a part of the manuscript was completed during the final stages of this work. 

\appendix

\section{Master Equations}
\label{app:full_master_equation}

The full master equation for the probability to find the ribosome at time $t$ in the chemical  state $i$ at codon $n_m$ reads
{\small \begin{eqnarray}
\label{eq:full_master_equation_1}
\frac{d}{dt} P_{1}(n_m,t) & = & k_{21} P_{2}(n_m,t) - k_{12} P_{1}(n_m,t) + k_{31} P_{3}(n_m,t) - k_{13} P_{1}(n_m,t) \nonumber \\
& & + k_{51} P_{5}(n_m-1,t) - k_{15} P_{1}(n_m,t) + k_{91} P_{9}(n_m-1,t) - k_{19} P_{1}(n_m,t) \nonumber \\
& & + k_{71} P_{7}(n_m,t) - k_{17} P_{1}(n_m,t) + k_{61} P_{6}(n_m,t) - k_{16} P_{1}(n_m,t) \nonumber \\
& & + k_{101} P_{10}(n_m,t) - k_{110} P_{1}(n_m,t) + k_{111} P_{11}(n_m,t) - k_{111} P_{1}(n_m,t) \nonumber \\
& & + k_{131} P_{13}(n_m-1,t) - k_{113} P_{1}(n_m,t) + k_{171} P_{17}(n_m-1,t) - k_{117} P_{1}(n_m,t) \nonumber \\
& & + k_{151} P_{15}(n_m,t) - k_{115} P_{1}(n_m,t) + k_{141} P_{14}(n_m,t) - k_{114} P_{1}(n_m,t) \\
\label{eq:full_master_equation_2}
\frac{d}{dt} P_{2}(n_m,t)
& = & k_{12} P_{1}(n_m,t) - k_{21} P_{2}(n_m,t) + k_{32} P_{3}(n_m,t) - k_{23} P_{2}(n_m,t) \\
\label{eq:full_master_equation_3}
\frac{d}{dt} P_{3}(n_m,t) 
& = & k_{13} P_{1}(n_m,t) - k_{31} P_{3}(n_m,t) + k_{23} P_{2}(n_m,t) - k_{32} P_{3}(n_m,t) \nonumber \\
& & + k_{43} P_{4}(n_m,t) - k_{34} P_{3}(n_m,t) \\
\label{eq:full_master_equation_4}
\frac{d}{dt} P_{4}(k,t)
& = & k_{34} P_{3}(n_m,t) - k_{43} P_{4}(n_m,t) + k_{54} P_{5}(n_m,t) - k_{45} P_{4}(n_m,t) \\
\label{eq:full_master_equation_5}
\frac{d}{dt} P_{5}(n_m,t)
& = & k_{15} P_{1}(n_m+1,t) - k_{51} P_{5}(n_m,t) + k_{45} P_{4}(n_m,t) - k_{54} P_{5}(n_m,t) \\
\label{eq:full_master_equation_6}
\frac{d}{dt} P_{6}(n_m,t)
& = & k_{16} P_{1}(n_m,t) - k_{61} P_{6}(n_m,t) + k_{76} P_{7}(n_m,t) - k_{67} P_{6}(n_m,t) \\
\label{eq:full_master_equation_7}
\frac{d}{dt} P_{7}(n_m,t)
& = & k_{17} P_{1}(n_m,t) - k_{71} P_{7}(n_m,t) + k_{67} P_{6}(n_m,t) - k_{76} P_{7}(n_m,t)\nonumber \\
& & + k_{87} P_{8}(n_m,t) - k_{78} P_{7}(n_m,t) \\
\label{eq:full_master_equation_8}
\frac{d}{dt} P_{8}(n_m,t)
& = & k_{78} P_{7}(n_m,t) - k_{87} P_{8}(n_m,t) + k_{98} P_{9}(n_m,t) - k_{89} P_{8}(n_m,t)  \\
\label{eq:full_master_equation_9}
\frac{d}{dt} P_{9}(n_m,t)
& = & k_{19} P_{1}(n_m+1,t) - k_{91} P_{9}(n_m,t) + k_{89} P_{8}(n_m,t) - k_{98} P_{9}(n_m,t)  \\
\label{eq:full_master_equation_10}
\frac{d}{dt} P_{10}(n_m,t)
& = & k_{110} P_{1}(n_m,t) - k_{101} P_{10}(n_m,t) + k_{1110} P_{11}(n_m,t) - k_{1011} P_{10}(n_m,t) \\ \nonumber
\end{eqnarray}}

\small{\begin{eqnarray}
\label{eq:full_master_equation_11}
\frac{d}{dt} P_{11}(n_m,t)
& = & k_{111} P_{1}(n_m,t) - k_{111} P_{11}(n_m,t) + k_{1011} P_{10}(n_m,t) - k_{1110} P_{11}(n_m,t)  \nonumber \\
& + & k_{1211} P_{12}(n_m,t) - k_{1112} P_{11}(n_m,t) \\
\label{eq:full_master_equation_12}
\frac{d}{dt} P_{12}(n_m,t)
& = & k_{1112} P_{11}(n_m,t) - k_{1211} P_{12}(n_m,t) + k_{1312} P_{13}(n_m,t) - k_{1213} P_{12}(n_m,t) \\
\label{eq:full_master_equation_13}
\frac{d}{dt} P_{13}(n_m,t)
& = & k_{1213} P_{12}(n_m,t) - k_{1312} P_{13}(n_m,t) + k_{113} P_{1}(n_m+1,t) - k_{131} P_{13}(n_m,t) \\
\label{eq:full_master_equation_14}
\frac{d}{dt} P_{14}(n_m,t)
& = & k_{114} P_{1}(n_m,t) - k_{141} P_{14}(n_m,t) + k_{1514} P_{15}(n_m,t) - k_{1415} P_{14}(n_m,t) \\
\label{eq:full_master_equation_15}
\frac{d}{dt} P_{15}(n_m,t)
& = & k_{1415} P_{14}(n_m,t) - k_{1514} P_{15}(n_m,t) \\
\label{eq:full_master_equation_16}
\frac{d}{dt} P_{16}(n_m,t)
& = & k_{1516} P_{15}(n_m,t) - k_{1615} P_{16}(n_m,t) + k_{1716} P_{17}(n_m,t) - k_{1617} P_{16}(n_m,t) \\
\label{eq:full_master_equation_17}
\frac{d}{dt} P_{17}(n_m,t)
& = & k_{1617} P_{16}(n_m,t) - k_{1716} P_{17}(n_m,t) + k_{117} P_{1}(n_m+1,t) - k_{171} P_{17}(n_m,t)  \\ \nonumber
\end{eqnarray}}

The normalization condition is
\begin{equation}
\label{normalization}
\sum_{i=1}^{17} \sum_{n_m} P_{i}(n_m,t)=1 .
\end{equation}

\section{Graph theoretic solution of master equations}
\label{App:rssp}

In order to find the stationary solution of the master equation (\ref{eq:master_equation}) 
we follow \cite{schnakenberg_network_1976}. We demonstrate the solution explicitly
for a simplified effective version of the model with only seven states. The solution of the full model
proceeds along completely analogous lines.

\subsection{Step 1: Constructing Graph and Undirected Graph}
\begin{figure}[t]
	\begin{center}
		\includegraphics[width=0.5\textwidth]{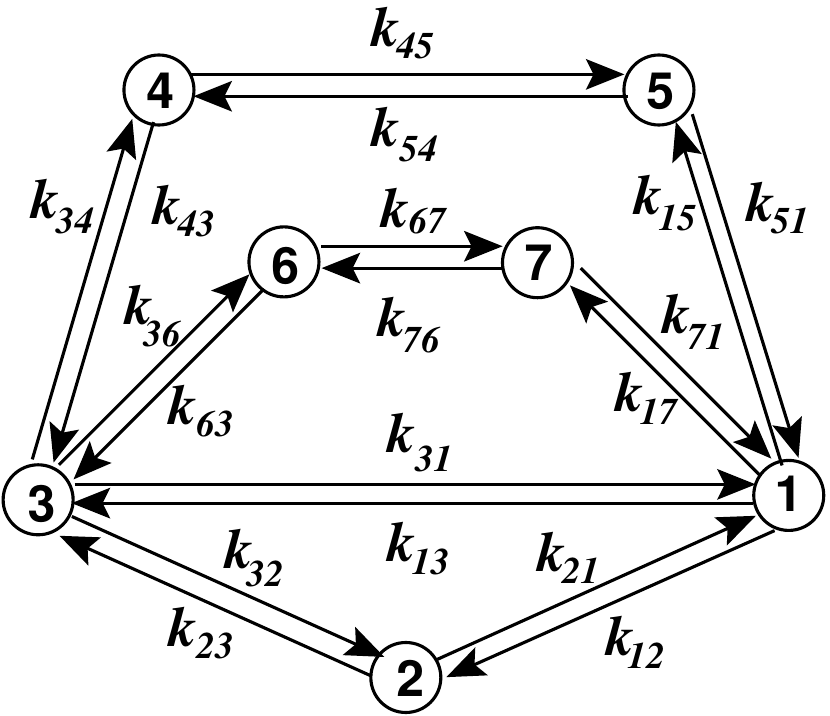}
	\end{center}
	\caption{The kinetic Markov network of the ribosomal elongation cycle. At every codon 
		position, the ribosome undergoes different conformation changes that are labelled 
		by $i=1, 2, 3, 4, 5, 6, 7$. The $k_{ij}$ are the transition rates to move 
		from conformation $i$ to conformation $j$. Notice that there are multiple pathways that 
		the ribosome can follow.}
	\label{fig:small_network}
\end{figure}

\begin{figure}
\begin{center}
\includegraphics[width=.5\columnwidth]{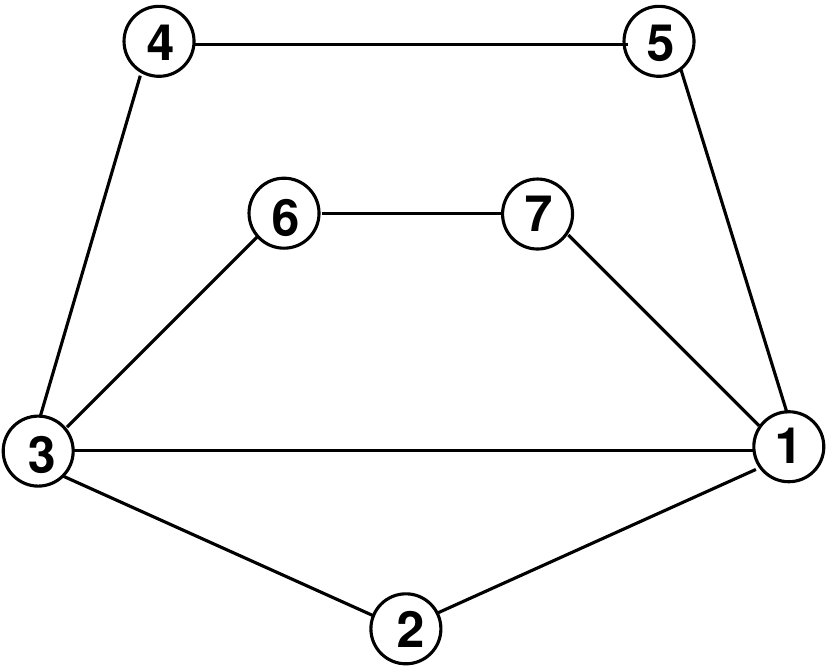}
\end{center}
\caption{Undirected graph representation of the network of 7-states.}
\label{fig:graph_undirected_st}
\end{figure}

 From the graph an {\it undirected graph} is obtained by replacing the directed edges by undirected edges. 
The undirected graph for the 7-state model is displayed in Fig.\ref{fig:graph_undirected_st}. This graph
is fully determined by the vertex set $V=\{1,2,3,4,5,6,7\}$ and the edge set
$E=\{(1,2),(1,3),(1,5),(1,7),(2,3),(3,4),(3,6),(4,5),(6,7)\}$. Here the edges are
not directed, i.e., edge $(i,j)$ is the same as the edge $(j,i)$, as opposed to
oriented edges $\vec{(i,j)}$ displayed below by an arrow pointing from $i$ to $j$.

\subsection{Step 2:  Constructing undirected spanning trees from the undirected graph}
Recall that a {\it spanning tree} of an undirected graph is a sub-graph which is a maximal tree that includes all the vertices of the graph, with minimum possible edges. All possible spanning trees of graph $G$ have the same number of edges and vertices. It doesn't contain any cycle. Adding just one edge will create a cycle and removing one edge will make the graph disconnected. Let $T^{\mu}(G)$ ($\mu=1,2,\dots, M$) represent the $\mu$-th undirected spanning tree of graph $G$.

\begin{figure}[h]
\includegraphics[width=0.9\columnwidth]{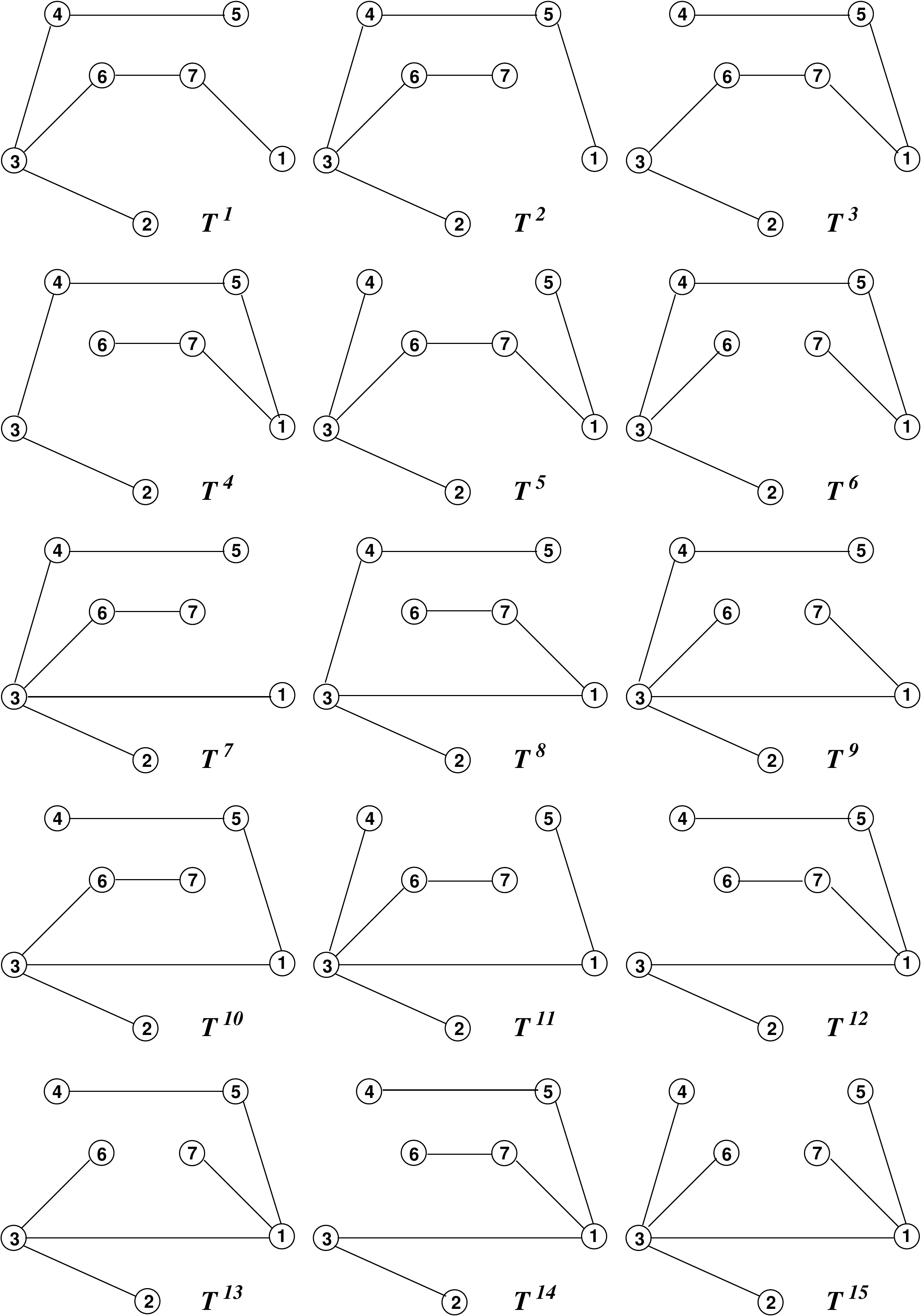}
\caption{All spanning trees $T^{\mu}$ without edge $(1,2)$.}
\label{fig:st12}
\end{figure}

\begin{figure}[h]
\includegraphics[width=0.9\columnwidth]{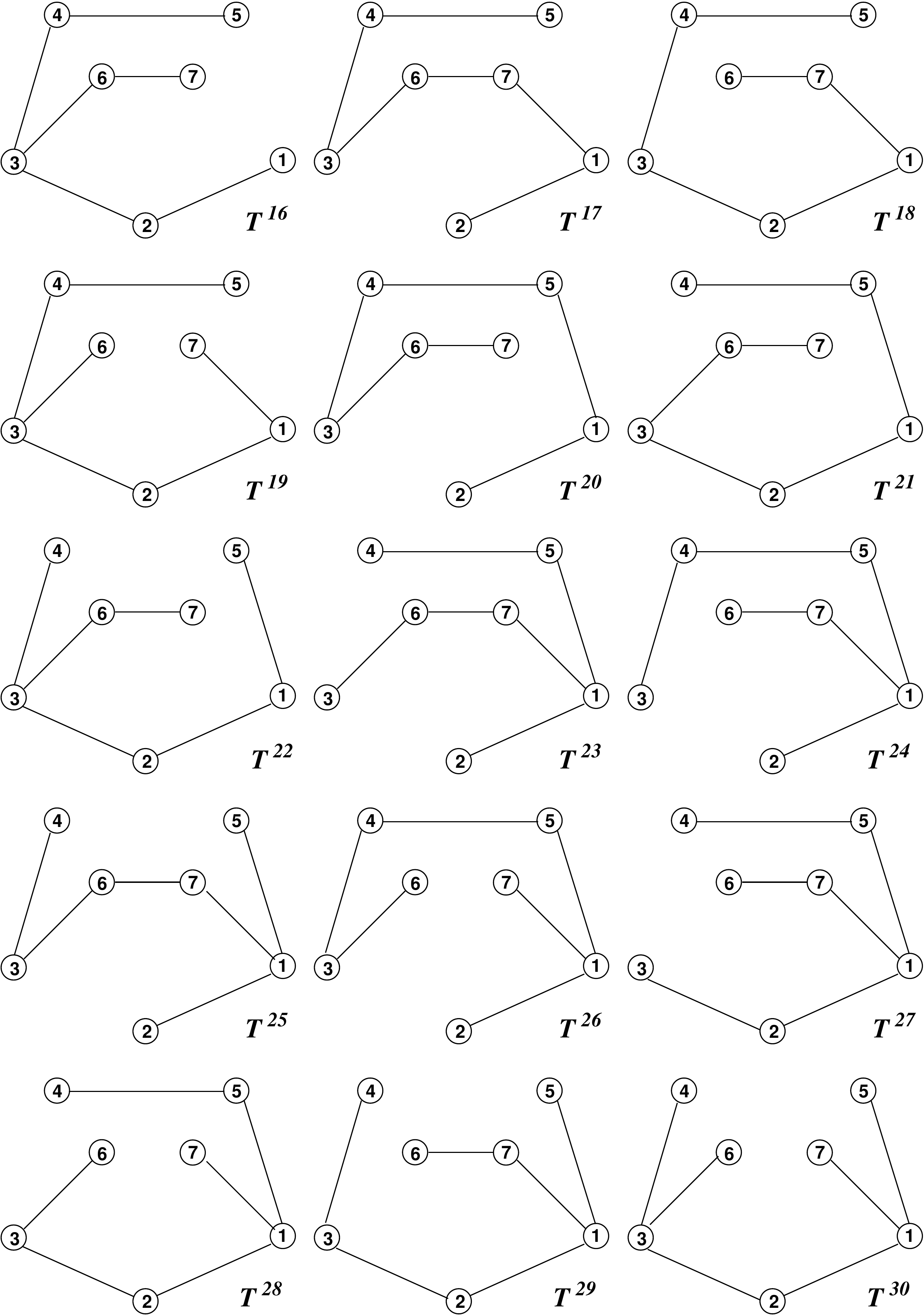}
\caption{All spanning trees $T^{\mu}$ that have edge $(1,2)$ but not edge $(1,3)$.}
\label{fig:st13}
\end{figure}

\begin{figure}[ht]
\includegraphics[width=0.9\columnwidth]{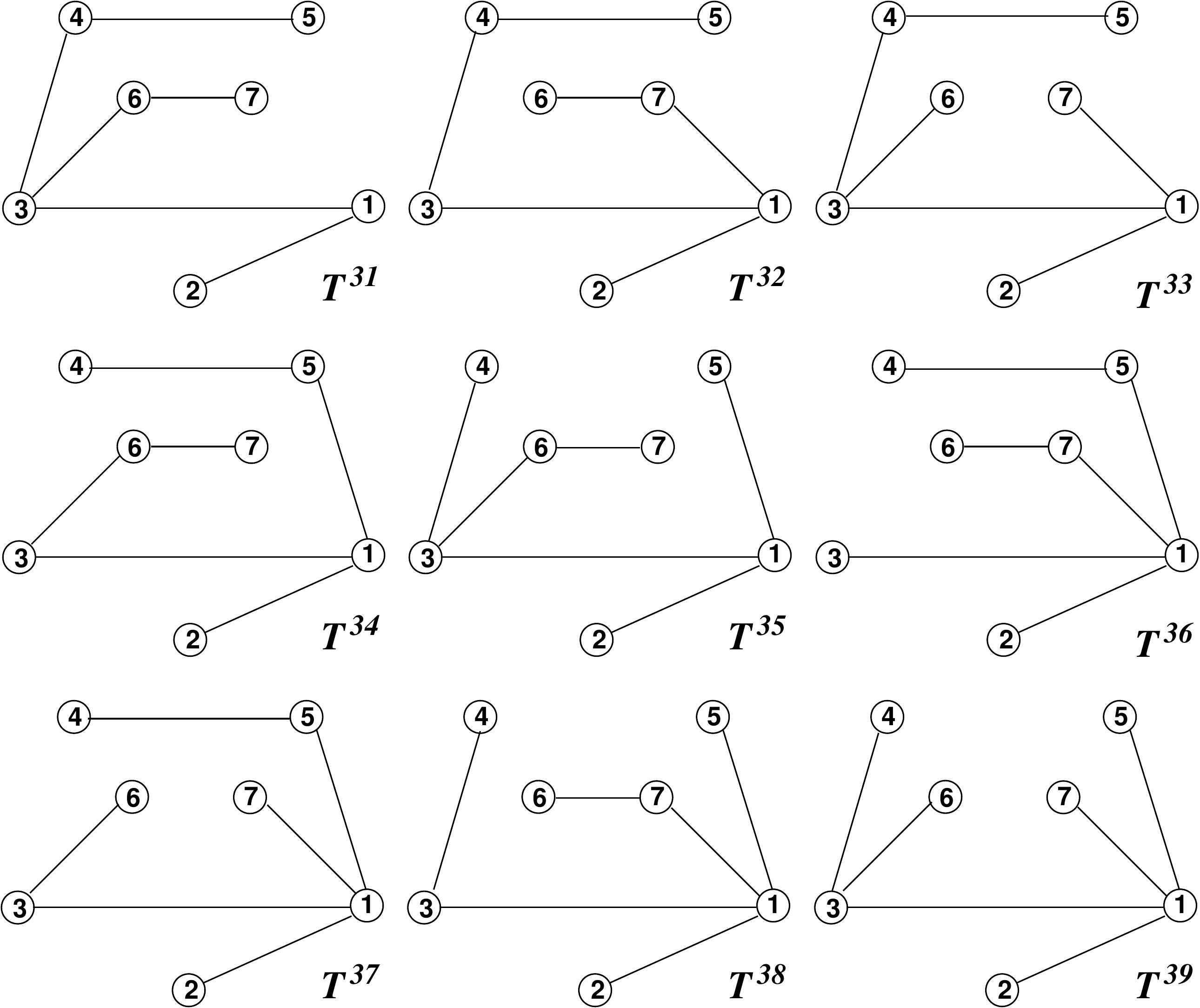}
\caption{All spanning trees $T^{\mu}$ that have edge $(1,2)$ and edge $(1,3)$ but not edge $(2,3)$.}
\label{fig:st23}
\end{figure}

One can construct the spanning trees by removing $|E|-|V|+1$ edges (for our graph, 9-7+1=3 edges) 
from the graph, where $|E|$ is the number of edges and $|V|$ is the number of vertices. This
yields $M=39$ distinct undirected spanning trees. For a systematic construction we group the
spanning trees into three classes: (I) All spanning trees without edge $(1,2)$, (II) All spanning trees that have edge 
$(1,2)$ but not edge $(1,3)$, and (III) All spanning trees that have edge $(1,2)$ and edge $(1,3)$ 
but not edge $(2,3)$. In total there are 39 spanning trees, see Figs.~\ref{fig:st12} - \ref{fig:st23}.

\subsection{Step 3:  Constructing directed spanning trees from undirected spanning trees}

\begin{figure}[h]
\begin{center}
\includegraphics[width=0.8\columnwidth]{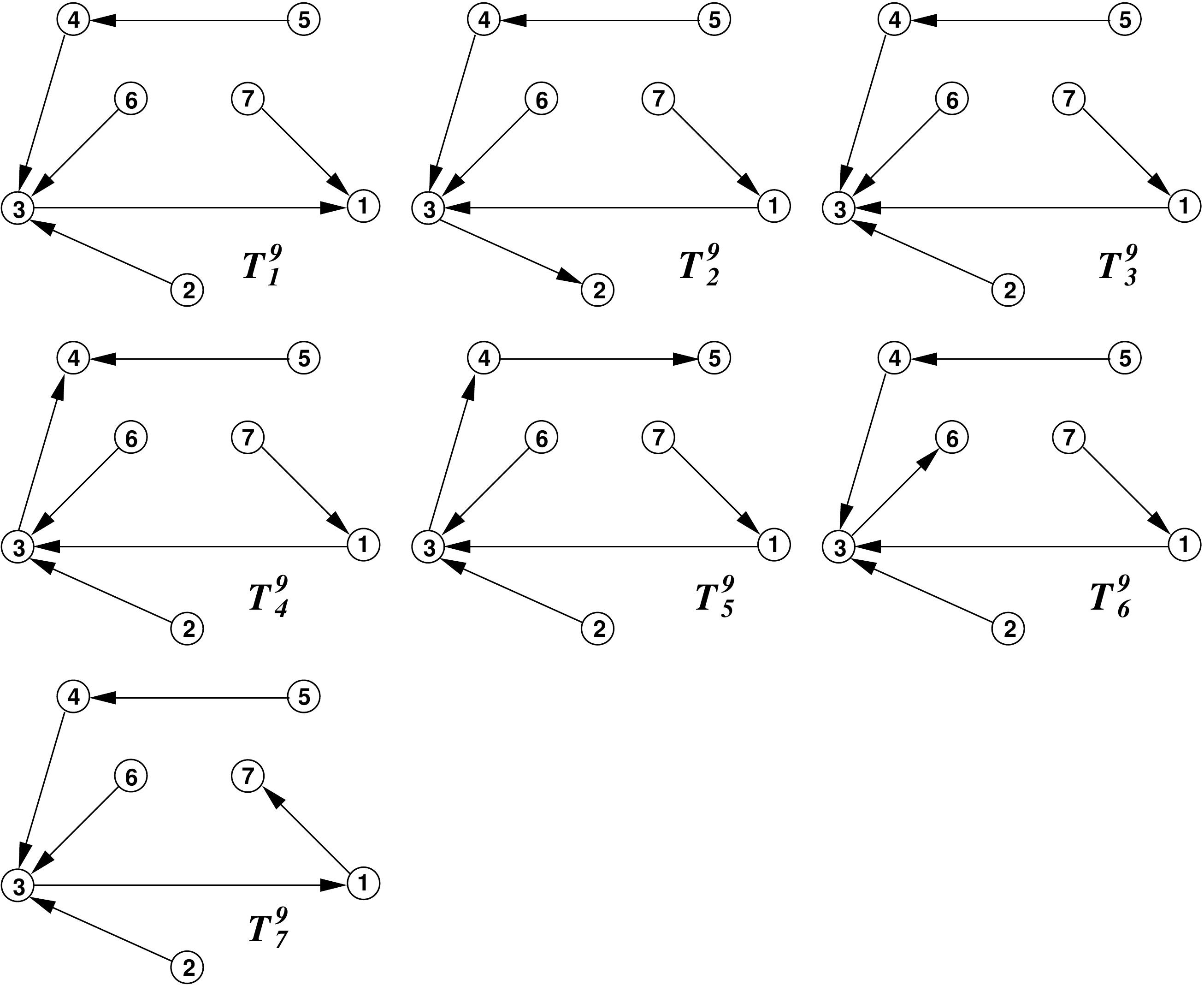}
\end{center}
\caption{All the seven directed spanning trees $T^{9}_i$, obtained from the undirected  spanning tree $T^{9}$ of Fig.~\ref{fig:st12} by directing all the edges towards the root vertex $i$.}
\label{fig:directed_st}
\end{figure}

Recall that a  directed spanning tree $T^{\mu}_{i}(G)$ can be obtained by directing all the edges of the undirected spanning tree $T^{\mu}(G)$ towards the vertex $i$. Thus, for each undirected spanning tree $T^{\mu}$ displayed in Figs.~\ref{fig:st12} - \ref{fig:st23}, the six {\it directed} spanning trees $T^{\mu}_{i}$, $i \in \{1,\dots,7\}$, are obtained by directing all the edges of $T^{\mu}$ towards the vertex $i$. This construction is illustrated  in 
Fig. \ref{fig:directed_st} for the undirected spanning tree $T^{9}$.  Since, for every undirected spanning tree $T^{\mu}(G)$ and  for a particular root vertex $i$, there is exactly one directed spanning tree,  this construction yields a total of $7 \times 39= 273$ directed spanning trees $T^{\mu}_{i}$ ($\mu=1,2,\dots, 7$).

\subsection{Step 4:  Steady-state solution in terms of contributions from directed spanning trees}
To each of the directed spanning trees 
$T^{\mu}_{i}$, we assign a numerical value, $A(T^{\mu}_{i})$, which is defined as the product 
of the $|V|-1=6$ transition rates in the tree, with transitions $i\to j$ defined along the 
orientation $\vec{(i,j)}$.
The steady state probability distributions are then given by (\ref{eq:p_st}) with the
normalization factor (\ref{eq:part_func}).

As a result of this construction, all unnormalized steady state probabilities 
$\tilde{P}_i = \sum^{M}_{\mu=1} A(T^{\mu}_{i})$ for the 7-state model 
are a sum of 39 monomials. Each monomial is a product of seven rates $k_{ij}$ such that each 
edge $(i,j)$ is represented exactly once. The monomials thus differ only in the orientation
in which an edge appears. So, finally, for the 7-state model one finds 
\begin{equation}
P_{i} = \tilde{P_{i}}/\mathcal{Z},
\end{equation} 
with
\begin{equation} 
\label{eq:part_func2}
\mathcal{Z}=\sum^{7}_{i=1} \tilde{P}_i, 
\end{equation}
 where
\begin{eqnarray}
\tilde{P}_1 &=& k_{21} k_{31} k_{43} k_{51} k_{71} k_{67}+k_{21} k_{31} k_{51} k_{45} k_{71} k_{67}\nonumber \\ 
& & + k_{31} k_{23} k_{51} k_{45} k_{71} k_{67}+k_{31} k_{23} k_{43} k_{51} k_{71} k_{67}\nonumber \\
& & + k_{51} k_{45} k_{34} k_{23} k_{71} k_{67}+k_{21} k_{32} k_{43} k_{51} k_{71} k_{67}\nonumber \\ 
& & + k_{51} k_{71} k_{67} k_{36} k_{23} k_{43}+k_{51} k_{45} k_{71} k_{67} k_{36} k_{23}\nonumber \\
& & + k_{21} k_{32} k_{51} k_{45} k_{71} k_{67}+k_{21} k_{51} k_{45} k_{34} k_{71} k_{67}\nonumber \\ 
& & + k_{21} k_{51} k_{45} k_{71} k_{67} k_{36}+k_{21} k_{51} k_{71} k_{67} k_{36} k_{43}\nonumber \\
& & + k_{21} k_{71} k_{67} k_{36} k_{43} k_{54}+k_{21} k_{31} k_{43} k_{54} k_{71} k_{67}\nonumber \\ 
& & +k_{21} k_{32} k_{43} k_{54} k_{71} k_{67}+k_{31} k_{23} k_{43} k_{54} k_{71} k_{67}\nonumber \\
& & + k_{71} k_{67} k_{36} k_{23} k_{43} k_{54}+k_{21} k_{32} k_{43} k_{54} k_{63} k_{76}\nonumber \\ 
& & +k_{51} k_{45} k_{34} k_{23} k_{63} k_{76}+k_{31} k_{23} k_{43} k_{54} k_{63} k_{76}\nonumber \\
& & + k_{21} k_{31} k_{43} k_{54} k_{63} k_{76}+k_{21} k_{51} k_{45} k_{34} k_{63} k_{76}\nonumber \\ 
& & +k_{21} k_{32} k_{43} k_{63} k_{76} k_{51}+k_{21} k_{31} k_{43} k_{63} k_{76} k_{51}\nonumber \\
& & + k_{31} k_{23} k_{43} k_{63} k_{76} k_{51}+k_{31} k_{23} k_{63} k_{76} k_{51} k_{45}\nonumber \\ 
& & +k_{21} k_{31} k_{63} k_{76} k_{51} k_{45}+k_{21} k_{32} k_{63} k_{76} k_{51} k_{45}\nonumber \\
& & + k_{21} k_{32} k_{63} k_{51} k_{45} k_{71}+k_{21} k_{51} k_{45} k_{34} k_{63} k_{71}\nonumber \\ 
& & +k_{21} k_{31} k_{63} k_{51} k_{45} k_{71}+k_{31} k_{23} k_{63} k_{51} k_{45} k_{71}\nonumber \\
& & + k_{51} k_{45} k_{34} k_{23} k_{63} k_{71}+k_{21} k_{32} k_{43} k_{63} k_{51} k_{71}\nonumber \\ 
& & +k_{31} k_{23} k_{43} k_{63} k_{51} k_{71}+k_{21} k_{31} k_{43} k_{63} k_{51} k_{71}\nonumber \\
& & + k_{21} k_{31} k_{43} k_{54} k_{63} k_{71}+k_{31} k_{23} k_{43} k_{54} k_{63} k_{71}\nonumber \\ 
& & +k_{21} k_{32} k_{43} k_{54} k_{63} k_{71},
\end{eqnarray}

\begin{eqnarray}
\tilde{P}_2 &=& k_{12} k_{31} k_{43} k_{51} k_{71} k_{67}+k_{12} k_{31} k_{51} k_{45} k_{71} k_{67}\nonumber \\ 
& & +k_{32} k_{13} k_{51} k_{45} k_{71} k_{67}+k_{32} k_{13} k_{51} k_{71} k_{67} k_{43}\nonumber \\
& & + k_{32} k_{43} k_{54} k_{15} k_{71} k_{67}+k_{12} k_{51} k_{71} k_{67} k_{32} k_{43}\nonumber \\ 
& & +k_{32} k_{43} k_{63} k_{76} k_{17} k_{51}+k_{32} k_{63} k_{76} k_{17} k_{51} k_{45}\nonumber \\
& & + k_{12} k_{51} k_{45} k_{71} k_{67} k_{32}+k_{12} k_{51} k_{45} k_{34} k_{71} k_{67}\nonumber \\ 
& & +k_{12} k_{51} k_{45} k_{71} k_{67} k_{36}+k_{12} k_{51} k_{71} k_{67} k_{36} k_{43}\nonumber \\
& & + k_{12} k_{71} k_{67} k_{36} k_{43} k_{54}+k_{12} k_{31} k_{43} k_{54} k_{71} k_{67}\nonumber \\ 
& & +k_{12} k_{71} k_{67} k_{32} k_{43} k_{54}+k_{32} k_{13} k_{71} k_{67} k_{43} k_{54}\nonumber \\
& & + k_{32} k_{43} k_{54} k_{63} k_{76} k_{17}+k_{12} k_{32} k_{43} k_{54} k_{63} k_{76}\nonumber \\ 
& & +k_{32} k_{43} k_{54} k_{15} k_{63} k_{76}+k_{32} k_{13} k_{43} k_{54} k_{63} k_{76}\nonumber \\
& & + k_{12} k_{31} k_{43} k_{54} k_{63} k_{76}+k_{12} k_{51} k_{45} k_{34} k_{63} k_{76}\nonumber \\ 
& & +k_{12} k_{51} k_{32} k_{43} k_{63} k_{76}+k_{12} k_{31} k_{43} k_{63} k_{76} k_{51}\nonumber \\
& & + k_{32} k_{13} k_{51} k_{43} k_{63} k_{76}+k_{32} k_{13} k_{51} k_{45} k_{63} k_{76}\nonumber \\ 
& & +k_{12} k_{31} k_{63} k_{76} k_{51} k_{45}+k_{12} k_{51} k_{45} k_{32} k_{63} k_{76}\nonumber \\
& & + k_{12} k_{51} k_{45} k_{71} k_{32} k_{63}+k_{12} k_{51} k_{45} k_{34} k_{63} k_{71}\nonumber \\ 
& & +k_{12} k_{31} k_{63} k_{51} k_{45} k_{71}+k_{32} k_{13} k_{51} k_{45} k_{71} k_{63}\nonumber \\
& & + k_{32} k_{43} k_{54} k_{15} k_{71} k_{63}+k_{12} k_{51} k_{71} k_{32} k_{43} k_{63}\nonumber \\ 
& & +k_{32} k_{13} k_{51} k_{71} k_{43} k_{63}+k_{12} k_{31} k_{43} k_{63} k_{51} k_{71}\nonumber \\
& & + k_{12} k_{31} k_{43} k_{54} k_{63} k_{71}+k_{32} k_{13} k_{71} k_{43} k_{54} k_{63}\nonumber \\ 
& & +k_{21} k_{32} k_{43} k_{54} k_{63} k_{71},
\end{eqnarray}

\begin{eqnarray}
\tilde{P}_3 &=& k_{13} k_{21} k_{51} k_{71} k_{67} k_{43}+k_{13} k_{21} k_{51} k_{45} k_{71} k_{67}\nonumber \\ 
& & +k_{13} k_{51} k_{45} k_{71} k_{67} k_{23}+k_{13} k_{51} k_{71} k_{67} k_{23} k_{43}\nonumber \\
& & + k_{23} k_{43} k_{54} k_{15} k_{71} k_{67}+k_{23} k_{12} k_{51} k_{71} k_{67} k_{43}\nonumber \\ 
& & +k_{23} k_{43} k_{63} k_{76} k_{17} k_{51}+k_{23} k_{63} k_{76} k_{17} k_{51} k_{45}\nonumber \\
& & + k_{23} k_{12} k_{51} k_{45} k_{71} k_{67}+k_{43} k_{54} k_{15} k_{21} k_{71} k_{67}\nonumber \\ 
& & +k_{63} k_{76} k_{17} k_{21} k_{51} k_{45}+k_{43} k_{63} k_{76} k_{17} k_{21} k_{51}\nonumber \\
& & + k_{43} k_{54} k_{63} k_{76} k_{17} k_{21}+k_{13} k_{21} k_{71} k_{67} k_{43} k_{54}\nonumber \\ 
& & +k_{23} k_{12} k_{71} k_{67} k_{43} k_{54}+k_{13} k_{71} k_{67} k_{23} k_{43} k_{54}\nonumber \\
& & + k_{23} k_{43} k_{54} k_{63} k_{76} k_{17}+k_{23} k_{12} k_{43} k_{54} k_{63} k_{76}\nonumber \\ 
& & +k_{23} k_{43} k_{54} k_{15} k_{63} k_{76}+k_{13} k_{23} k_{43} k_{54} k_{63} k_{76}\nonumber \\
& & + k_{13} k_{21} k_{43} k_{54} k_{63} k_{76}+k_{43} k_{54} k_{15} k_{21} k_{63} k_{76}\nonumber \\ 
& & +k_{23} k_{12} k_{51} k_{43} k_{63} k_{76}+k_{13} k_{21} k_{51} k_{43} k_{63} k_{76}\nonumber \\
& & + k_{13} k_{51} k_{23} k_{43} k_{63} k_{76}+k_{13} k_{51} k_{45} k_{23} k_{63} k_{76}\nonumber \\ 
& & +k_{13} k_{21} k_{51} k_{45} k_{63} k_{76}+k_{23} k_{12} k_{51} k_{45} k_{63} k_{76}\nonumber \\
& & + k_{23} k_{12} k_{51} k_{45} k_{71} k_{63}+k_{43} k_{54} k_{15} k_{21} k_{71} k_{63}\nonumber \\ 
& & +k_{13} k_{21} k_{51} k_{45} k_{71} k_{63}+k_{13} k_{51} k_{45} k_{71} k_{23} k_{63}\nonumber \\
& & + k_{23} k_{43} k_{54} k_{15} k_{71} k_{63}+k_{23} k_{12} k_{51} k_{71} k_{43} k_{63}\nonumber \\ 
& & +k_{13} k_{51} k_{71} k_{23} k_{43} k_{63}+k_{13} k_{21} k_{51} k_{71} k_{43} k_{63}\nonumber \\
& & + k_{13} k_{21} k_{71} k_{43} k_{54} k_{63}+k_{13} k_{71} k_{23} k_{43} k_{54} k_{63}\nonumber \\ 
& & +k_{23} k_{12} k_{71} k_{43} k_{54} k_{63},
\end{eqnarray}

\begin{eqnarray}
\tilde{P}_4 &=& k_{34} k_{13} k_{21} k_{51} k_{71} k_{67}+k_{54} k_{15} k_{21} k_{31} k_{71} k_{67}\nonumber \\ 
& & +k_{54} k_{15} k_{31} k_{23} k_{71} k_{67}+k_{34} k_{13} k_{51} k_{71} k_{67} k_{23}\nonumber \\
& & + k_{34} k_{23} k_{54} k_{15} k_{71} k_{67}+k_{34} k_{23} k_{12} k_{51} k_{71} k_{67}\nonumber \\ 
& & +k_{34} k_{23} k_{63} k_{76} k_{17} k_{51}+k_{54} k_{15} k_{71} k_{67} k_{36} k_{23}\nonumber \\
& & + k_{54} k_{15} k_{21} k_{32} k_{71} k_{67}+k_{34} k_{54} k_{15} k_{21} k_{71} k_{67}\nonumber \\ 
& & +k_{54} k_{15} k_{21} k_{71} k_{67} k_{36}+k_{34} k_{63} k_{76} k_{17} k_{21} k_{51}\nonumber \\
& & + k_{34} k_{63} k_{76} k_{17} k_{21} k_{54}+k_{34} k_{13} k_{21} k_{71} k_{67} k_{54}\nonumber \\ 
& & +k_{34} k_{23} k_{12} k_{71} k_{67} k_{54}+k_{34} k_{13} k_{71} k_{67} k_{23} k_{54}\nonumber \\
& & + k_{34} k_{23} k_{63} k_{76} k_{17} k_{54}+k_{34} k_{23} k_{12} k_{63} k_{76} k_{54}\nonumber \\ 
& & +k_{34} k_{23} k_{63} k_{76} k_{54} k_{15}+k_{34} k_{13} k_{23} k_{63} k_{76} k_{54}\nonumber \\
& & + k_{34} k_{13} k_{21} k_{63} k_{76} k_{54}+k_{34} k_{63} k_{76} k_{54} k_{15} k_{21}\nonumber \\ 
& & +k_{34} k_{23} k_{12} k_{51} k_{63} k_{76}+k_{34} k_{13} k_{21} k_{51} k_{63} k_{76}\nonumber \\
& & + k_{34} k_{13} k_{51} k_{23} k_{63} k_{76}+k_{54} k_{15} k_{31} k_{23} k_{63} k_{76}\nonumber \\ 
& & +k_{54} k_{15} k_{21} k_{31} k_{63} k_{76}+k_{54} k_{15} k_{21} k_{32} k_{63} k_{76}\nonumber \\
& & + k_{54} k_{15} k_{21} k_{32} k_{63} k_{71}+k_{34} k_{63} k_{54} k_{15} k_{21} k_{71}\nonumber \\ 
& & +k_{54} k_{15} k_{21} k_{31} k_{63} k_{71}+k_{54} k_{15} k_{31} k_{23} k_{63} k_{71}\nonumber \\
& & + k_{34} k_{23} k_{63} k_{54} k_{15} k_{71}+k_{34} k_{23} k_{12} k_{51} k_{71} k_{63}\nonumber \\ 
& & +k_{34} k_{13} k_{51} k_{71} k_{23} k_{63}+k_{34} k_{13} k_{21} k_{51} k_{71} k_{63}\nonumber \\
& & + k_{34} k_{13} k_{21} k_{71} k_{63} k_{54}+k_{34} k_{13} k_{71} k_{23} k_{63} k_{54}\nonumber \\ 
& & +k_{34} k_{23} k_{12} k_{71} k_{63} k_{54},
\end{eqnarray}

\begin{eqnarray}
\tilde{P}_5 &=& k_{15} k_{21} k_{31} k_{43} k_{71} k_{67}+k_{15} k_{21} k_{31} k_{71} k_{67} k_{45}\nonumber \\ 
& & +k_{15} k_{31} k_{23} k_{71} k_{67} k_{45}+k_{15} k_{31} k_{23} k_{43} k_{71} k_{67}\nonumber \\
& & + k_{15} k_{71} k_{67} k_{45} k_{34} k_{23}+k_{15} k_{21} k_{32} k_{43} k_{71} k_{67}\nonumber \\ 
& & +k_{15} k_{71} k_{67} k_{36} k_{23} k_{43}+k_{15} k_{71} k_{67} k_{36} k_{23} k_{45}\nonumber \\
& & + k_{15} k_{21} k_{32} k_{71} k_{67} k_{45}+k_{15} k_{21} k_{71} k_{67} k_{45} k_{34}\nonumber \\ 
& & +k_{15} k_{21} k_{71} k_{67} k_{36} k_{45}+k_{15} k_{21} k_{71} k_{67} k_{36} k_{43}\nonumber \\
& & + k_{45} k_{34} k_{63} k_{76} k_{17} k_{21}+k_{45} k_{34} k_{13} k_{21} k_{71} k_{67}\nonumber \\ 
& & +k_{45} k_{34} k_{23} k_{12} k_{71} k_{67}+k_{45} k_{34} k_{13} k_{71} k_{67} k_{23}\nonumber \\
& & + k_{45} k_{34} k_{23} k_{63} k_{76} k_{17}+k_{45} k_{34} k_{23} k_{12} k_{63} k_{76}\nonumber \\ 
& & +k_{15} k_{45} k_{34} k_{23} k_{63} k_{76}+k_{45} k_{34} k_{13} k_{23} k_{63} k_{76}\nonumber \\
& & + k_{45} k_{34} k_{13} k_{21} k_{63} k_{76}+k_{15} k_{21} k_{45} k_{34} k_{63} k_{76}\nonumber \\ 
& & +k_{15} k_{21} k_{32} k_{43} k_{63} k_{76}+k_{15} k_{21} k_{31} k_{43} k_{63} k_{76}\nonumber \\
& & + k_{15} k_{31} k_{23} k_{43} k_{63} k_{76}+k_{15} k_{31} k_{23} k_{63} k_{76} k_{45}\nonumber \\ 
& & +k_{15} k_{21} k_{31} k_{63} k_{76} k_{45}+k_{15} k_{21} k_{32} k_{63} k_{76} k_{45}\nonumber \\
& & + k_{15} k_{21} k_{32} k_{63} k_{71} k_{45}+k_{15} k_{21} k_{71} k_{45} k_{34} k_{63}\nonumber \\ 
& & +k_{15} k_{21} k_{31} k_{63} k_{71} k_{45}+k_{15} k_{31} k_{23} k_{63} k_{71} k_{45}\nonumber \\
& & + k_{15} k_{71} k_{45} k_{34} k_{23} k_{63}+k_{15} k_{21} k_{32} k_{43} k_{63} k_{71}\nonumber \\ 
& & +k_{15} k_{31} k_{23} k_{43} k_{63} k_{71}+k_{15} k_{21} k_{31} k_{43} k_{63} k_{71}\nonumber \\
& & + k_{45} k_{34} k_{13} k_{21} k_{71} k_{63}+k_{45} k_{34} k_{13} k_{71} k_{23} k_{63}\nonumber \\ 
& & +k_{45} k_{34} k_{23} k_{12} k_{71} k_{63},
\end{eqnarray}

\begin{eqnarray}
\tilde{P}_6 &=& k_{76} k_{17} k_{21} k_{31} k_{43} k_{51}+k_{76} k_{17} k_{21} k_{31} k_{51} k_{45}\nonumber \\ 
& & +k_{76} k_{17} k_{31} k_{23} k_{51} k_{45}+k_{76} k_{17} k_{31} k_{23} k_{43} k_{51}\nonumber \\
& & + k_{76} k_{17} k_{51} k_{45} k_{34} k_{23}+k_{76} k_{17} k_{21} k_{32} k_{43} k_{51}\nonumber \\ 
& & +k_{36} k_{23} k_{43} k_{76} k_{17} k_{51}+k_{36} k_{23} k_{76} k_{17} k_{51} k_{45}\nonumber \\
& & + k_{76} k_{17} k_{21} k_{32} k_{51} k_{45}+k_{76} k_{17} k_{21} k_{51} k_{45} k_{34}\nonumber \\ 
& & +k_{36} k_{76} k_{17} k_{21} k_{51} k_{45}+k_{36} k_{43} k_{76} k_{17} k_{21} k_{51}\nonumber \\
& & + k_{36} k_{43} k_{54} k_{76} k_{17} k_{21}+k_{76} k_{17} k_{21} k_{31} k_{43} k_{54}\nonumber \\ 
& & +k_{76} k_{17} k_{21} k_{32} k_{43} k_{54}+k_{76} k_{17} k_{31} k_{23} k_{43} k_{54}\nonumber \\
& & + k_{36} k_{23} k_{43} k_{54} k_{76} k_{17}+k_{36} k_{23} k_{12} k_{43} k_{54} k_{76}\nonumber \\ 
& & +k_{36} k_{23} k_{43} k_{54} k_{15} k_{76}+k_{36} k_{13} k_{23} k_{43} k_{54} k_{76}\nonumber \\
& & + k_{36} k_{13} k_{21} k_{43} k_{54} k_{76}+k_{36} k_{43} k_{54} k_{15} k_{21} k_{76}\nonumber \\ 
& & +k_{36} k_{23} k_{12} k_{51} k_{43} k_{76}+k_{36} k_{13} k_{21} k_{51} k_{43} k_{76}\nonumber \\
& & + k_{36} k_{13} k_{51} k_{23} k_{43} k_{76}+k_{36} k_{13} k_{51} k_{45} k_{23} k_{76}\nonumber \\ 
& & +k_{36} k_{13} k_{21} k_{51} k_{45} k_{76}+k_{36} k_{23} k_{12} k_{51} k_{45} k_{76}\nonumber \\
& & + k_{36} k_{23} k_{12} k_{51} k_{45} k_{71}+k_{36} k_{43} k_{54} k_{15} k_{21} k_{71}\nonumber \\ 
& & +k_{36} k_{13} k_{21} k_{51} k_{45} k_{71}+k_{36} k_{13} k_{51} k_{45} k_{71} k_{23}\nonumber \\
& & + k_{36} k_{23} k_{43} k_{54} k_{15} k_{71}+k_{36} k_{23} k_{12} k_{51} k_{71} k_{43}\nonumber \\ 
& & +k_{36} k_{13} k_{51} k_{71} k_{23} k_{43}+k_{36} k_{13} k_{21} k_{51} k_{71} k_{43}\nonumber \\
& & + k_{36} k_{13} k_{21} k_{71} k_{43} k_{54}+k_{36} k_{13} k_{71} k_{23} k_{43} k_{54}\nonumber \\ 
& & +k_{36} k_{23} k_{12} k_{71} k_{43} k_{54},
\end{eqnarray}

\begin{eqnarray}
\tilde{P}_7 &=& k_{17} k_{21} k_{31} k_{43} k_{51} k_{67}+k_{17} k_{21} k_{31} k_{51} k_{45} k_{67}\nonumber \\ 
& & +k_{17} k_{31} k_{23} k_{51} k_{45} k_{67}+k_{17} k_{31} k_{23} k_{43} k_{51} k_{67}\nonumber \\
& & + k_{17} k_{51} k_{45} k_{34} k_{23} k_{67}+k_{17} k_{21} k_{32} k_{43} k_{51} k_{67}\nonumber \\ 
& & +k_{17} k_{51} k_{67} k_{36} k_{23} k_{43}+k_{17} k_{51} k_{45} k_{67} k_{36} k_{23}\nonumber \\
& & + k_{17} k_{21} k_{32} k_{51} k_{45} k_{67}+k_{17} k_{21} k_{51} k_{45} k_{34} k_{67}\nonumber \\ 
& & +k_{17} k_{21} k_{51} k_{45} k_{67} k_{36}+k_{17} k_{21} k_{51} k_{67} k_{36} k_{43}\nonumber \\
& & + k_{17} k_{21} k_{67} k_{36} k_{43} k_{54}+k_{17} k_{21} k_{31} k_{43} k_{54} k_{67}\nonumber \\ 
& & +k_{17} k_{21} k_{32} k_{43} k_{54} k_{67}+k_{17} k_{31} k_{23} k_{43} k_{54} k_{67}\nonumber \\
& & + k_{17} k_{67} k_{36} k_{23} k_{43} k_{54}+k_{67} k_{36} k_{23} k_{12} k_{43} k_{54}\nonumber \\ 
& & +k_{67} k_{36} k_{23} k_{43} k_{54} k_{15}+k_{67} k_{36} k_{13} k_{23} k_{43} k_{54}\nonumber \\
& & + k_{67} k_{36} k_{13} k_{21} k_{43} k_{54}+k_{67} k_{36} k_{43} k_{54} k_{15} k_{21}\nonumber \\ 
& & +k_{67} k_{36} k_{23} k_{12} k_{51} k_{43}+k_{67} k_{36} k_{13} k_{21} k_{51} k_{43}\nonumber \\
& & + k_{67} k_{36} k_{13} k_{51} k_{23} k_{43}+k_{67} k_{36} k_{13} k_{51} k_{45} k_{23}\nonumber \\ 
& & +k_{67} k_{36} k_{13} k_{21} k_{51} k_{45}+k_{67} k_{36} k_{23} k_{12} k_{51} k_{45}\nonumber \\
& & + k_{17} k_{21} k_{32} k_{63} k_{51} k_{45}+k_{17} k_{21} k_{51} k_{45} k_{34} k_{63}\nonumber \\ 
& & +k_{17} k_{21} k_{31} k_{63} k_{51} k_{45}+k_{17} k_{31} k_{23} k_{63} k_{51} k_{45}\nonumber \\
& & + k_{17} k_{51} k_{45} k_{34} k_{23} k_{63}+k_{17} k_{21} k_{32} k_{43} k_{63} k_{51}\nonumber \\ 
& & +k_{17} k_{31} k_{23} k_{43} k_{63} k_{51}+k_{17} k_{21} k_{31} k_{43} k_{63} k_{51}\nonumber \\
& & + k_{17} k_{21} k_{31} k_{43} k_{54} k_{63}+k_{17} k_{31} k_{23} k_{43} k_{54} k_{63}\nonumber \\ 
& & +k_{17} k_{21} k_{32} k_{43} k_{54} k_{63},
\end{eqnarray}
From these exact expressions all stationary properties can be computed analytically or numerically exactly.

\section{Graphical representations of transition and cycle fluxes: Flux diagrams}
\label{app-flux_diag_Ribo}

Using the decomposition (\ref{eq:p_st}) of the stationary probabilities into contributions from the directed spanning trees one obtains the decomposition
\begin{eqnarray} 
\label{prob_current_st}
J_{ij} &=& k_{ij} P_{i} - k_{ji} P_{j} \nonumber \\
&=&\mathcal{Z}^{-1} \sum_{\mu=1}^{39} \left[k_{ij}A(T^{\mu}_{i})-k_{ji}A(T^{\mu}_{j})\right]
\end{eqnarray}
in terms of the unnormalized transition flux contributions 
$J^{\mu}_{ij}:= k_{ij}A(T^{\mu}_{i}) - k_{ji}A(T^{\mu}_{j})$ and the normalization factor (\ref{eq:part_func2}).

\begin{figure}[h]
\begin{center}
\includegraphics[width=0.7\columnwidth]{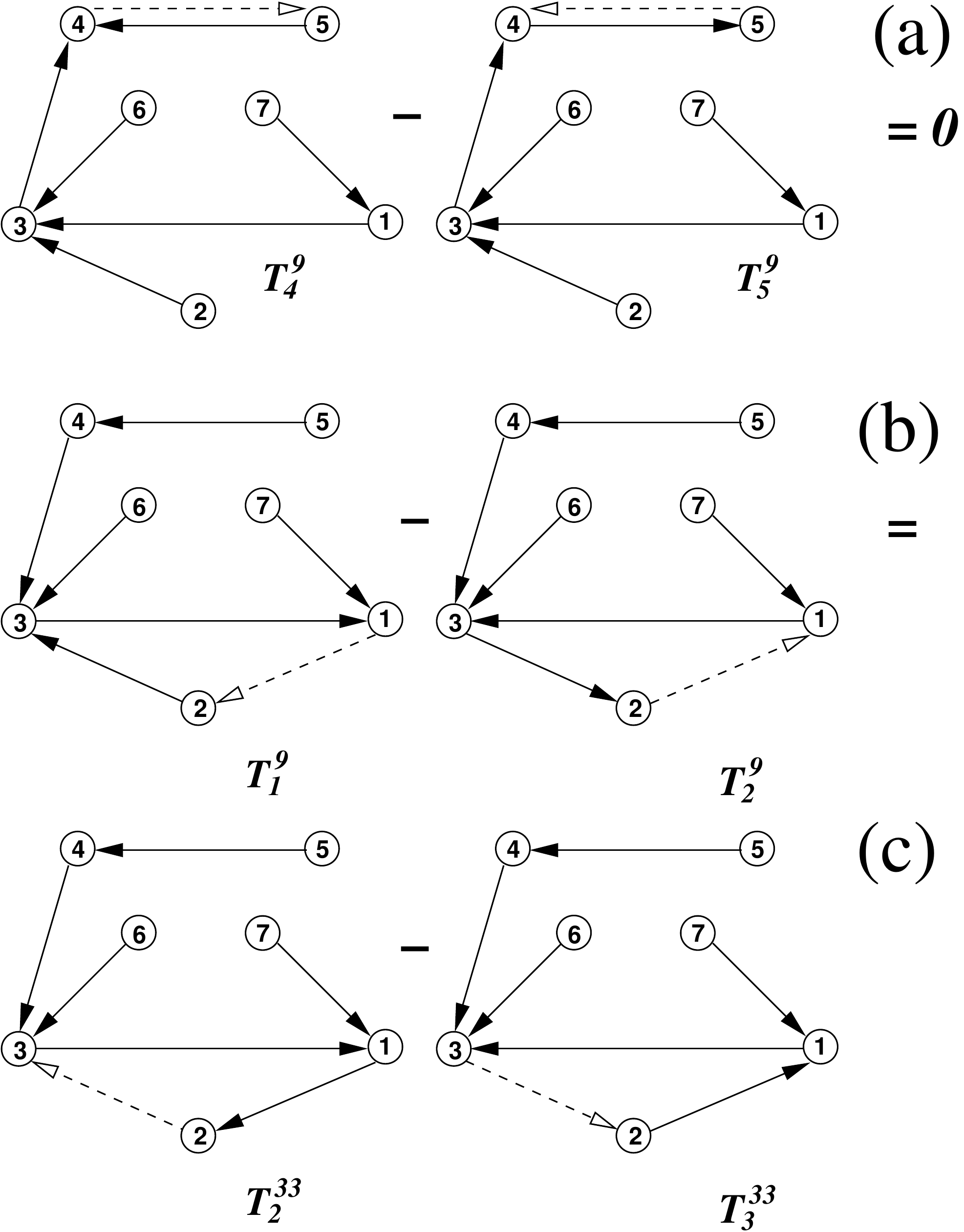}
\end{center}
\caption{Forming graphs with a single cycle: (a) Adding to the oriented tree $T^9_4$ an oriented edge from 
vertex 4 to the neighbouring vertex 5 yields the same tree (without any cycle) as adding to the oriented tree 
$T^9_5$ an oriented edge from vertex 5 to vertex 4, thus rendering $J^{9}_{45}=J^{9}_{54}=0$. 
(b) Adding to the directed tree $T^9_1$ an oriented edge from vertex 1 to vertex 2 generates a
cycle $(231)$, while adding to the oriented tree $T^9_2$ an oriented edge from vertex 2 to vertex 1
generates a cycle $(132)$ where generically $J^{9}_{12} = - J^{9}_{21} \neq 0$
(viz. unless $k_{12}k_{23}k_{31} = k_{21}k_{13}k_{32}$ for the cycle products of the transition rates).
(c) Adding oriented edges between vertices 2 and 3 of the oriented trees $T^{33}_{2}$ and 
$T^{33}_{3}$ resp., yields identical graphs so that $J^{9}_{12} = J^{33}_{23}$.}
\label{fig:form_loop}
\end{figure}

We observe that the directed trees $T^{\mu}_{i}$ and $T^{\mu}_{j}$ differ for the same $\mu$ 
only in the orientation of the edges connecting the vertices $i$ and $j$. Therefore, if 
vertices $i$ and $j$ are neighbours, i.e., if they are connected by an edge in the 
undirected spanning tree $\mu$, then the flux contribution
$J^{\mu}_{ij}$ from $T^{\mu}_{i}$ and $T^{\mu}_{j}$ vanishes  (see Fig. \ref{fig:form_loop}(a)). 
On the other hand, if $i$ and $j$ are not neighbours, then the multiplication of 
$A(T^{\mu}_{i})$ by the rate $k_{ij}$ yields a product of rates that can be obtained from a 
new directed graph by adding to $T^{\mu}_{i}$ an oriented edge from $i$ to $j$, thus converting 
the oriented tree $T^{\mu}_{i}$ into a graph $T^{\mu}_{ij}$ with a single oriented cycle 
(see Fig. \ref{fig:form_loop}(b)) whose original arrows are all directed towards vertex $i$ 
and with a new arrow from $i$ to $j$. Similarly, the product $k_{ji} A(T^{\mu}_{ji})$ 
corresponds to a graph $T^{\mu}_{ji}$ with the same cycle, but oriented in the opposite 
direction. The rest of both the one-cycle graphs (the side branches of the cycles) are identical, 
so that the product of the rates in the side branches $R(T^{\mu}_{ij})=R(T^{\mu}_{ji})$ is the 
same for both orientations. Notice that directed trees with different indices $\mu$ might lead to the same one-cycle graphs, see Fig.~\ref{fig:form_loop}(c) for two examples $T^9_{12} = T^{33}_{23}$ and 
$T^9_{21} = T^{33}_{32}$. 

\begin{figure}[h!]
\begin{center}
\includegraphics[width=0.9\columnwidth]{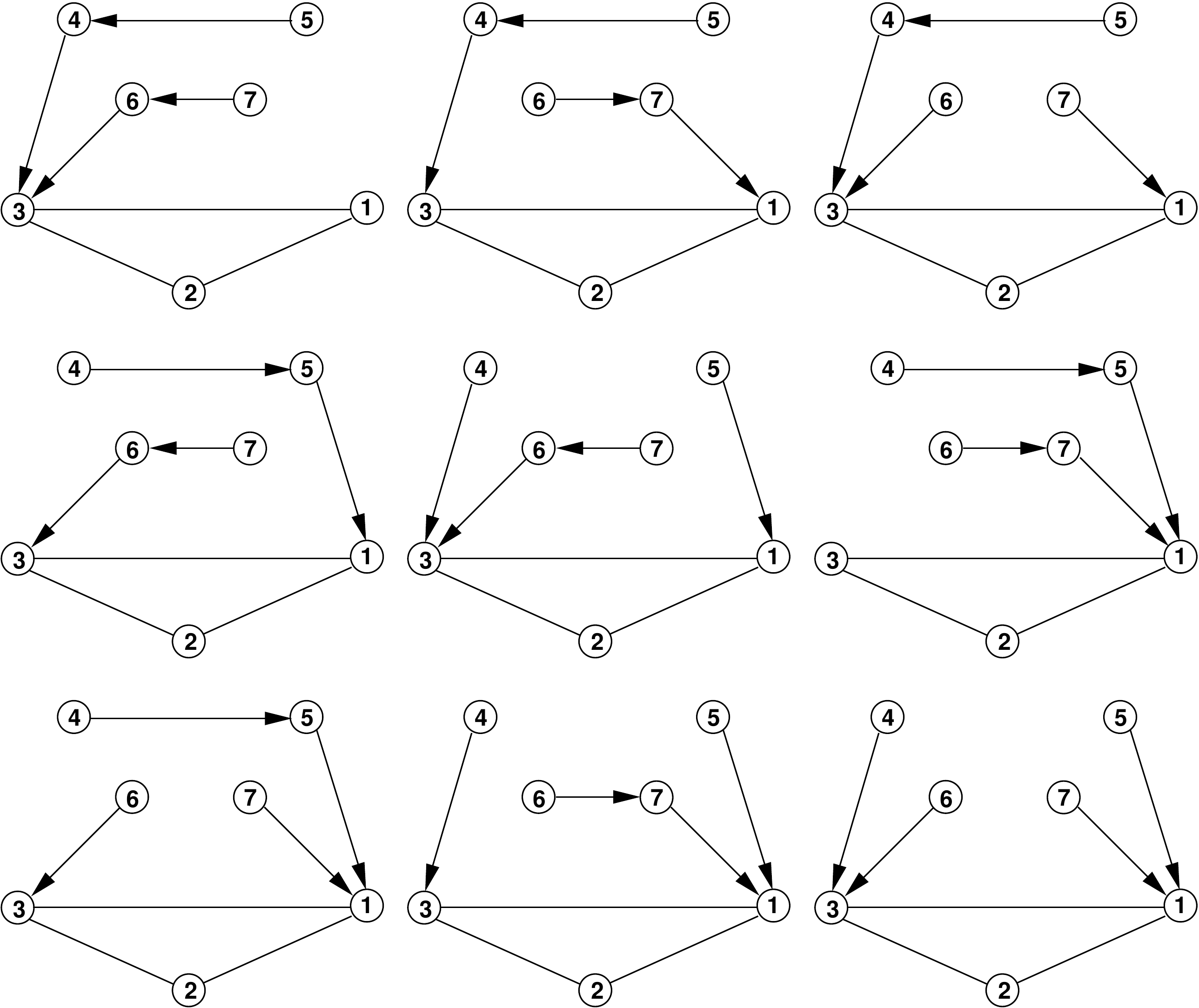}
\end{center}
\caption{Graphs with loop (a)}
\label{fig:cycle_a1}
\end{figure}

\begin{figure}[h!]
\begin{center}
\includegraphics[width=0.9\columnwidth]{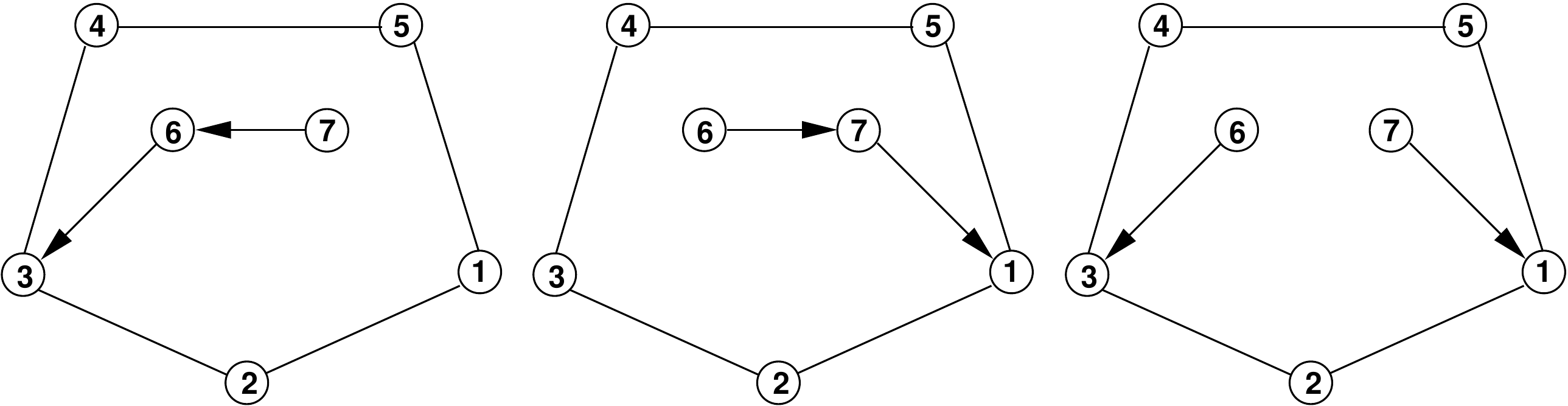}
\end{center}
\caption{Graphs with loop (b)}
\label{fig:cycle_b}
\end{figure}

\begin{figure}[h!]
\begin{center}
\includegraphics[width=0.9\columnwidth]{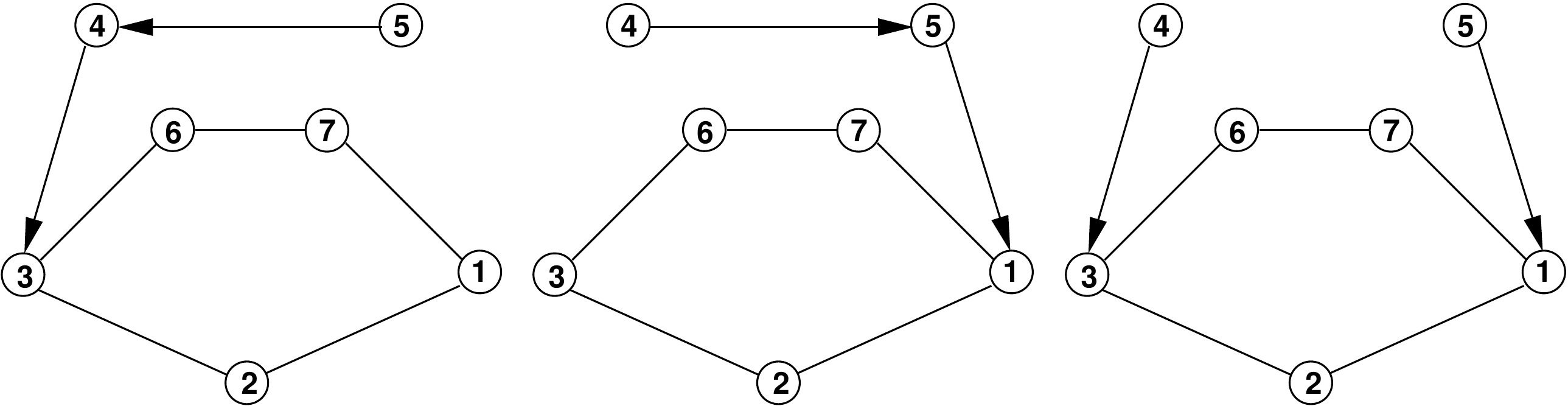}
\end{center}
\caption{Graphs with loop (c)}
\label{fig:cycle_c}
\end{figure}

\begin{figure}[h!]
\begin{center}
\includegraphics[width=0.9\columnwidth]{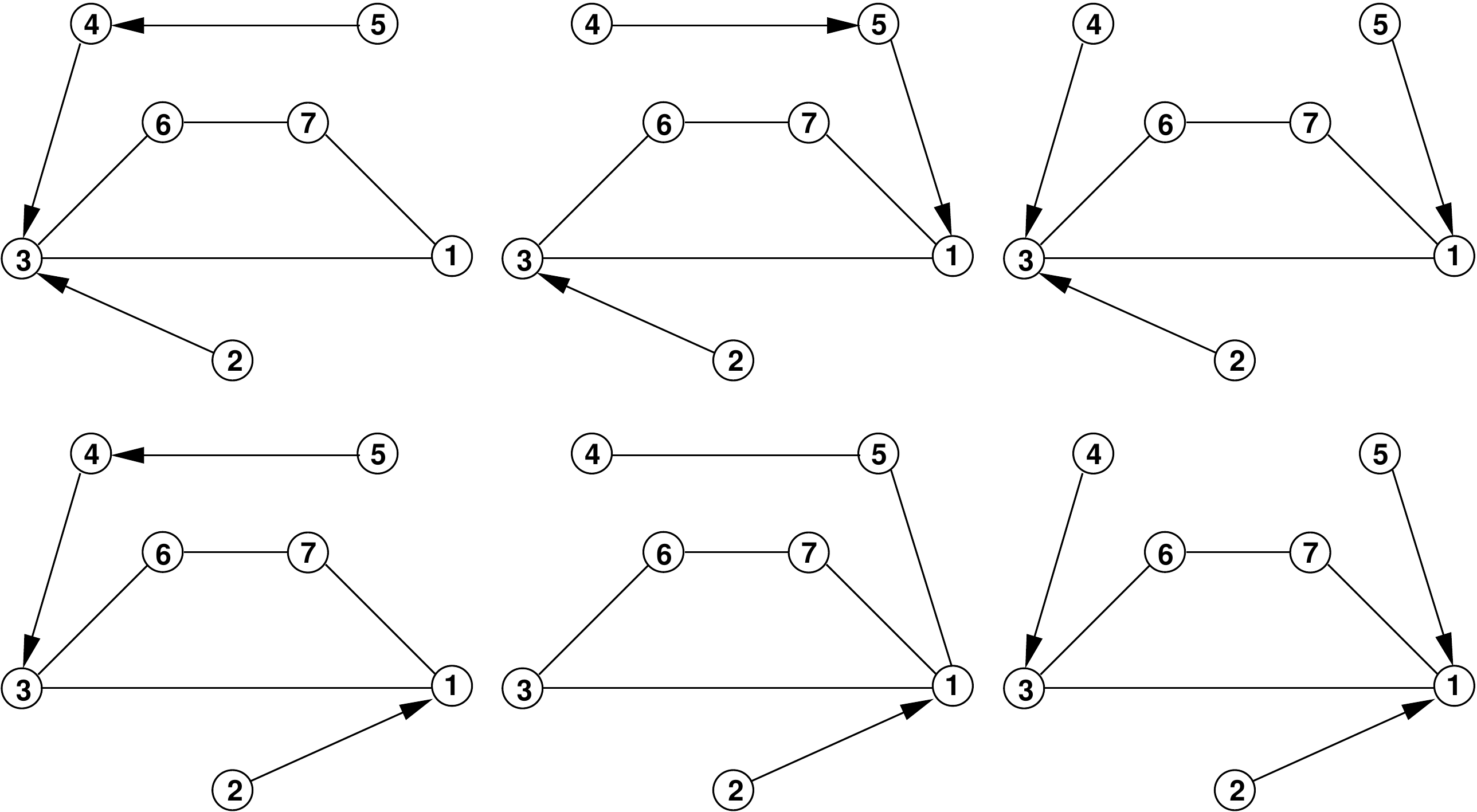}
\end{center}
\caption{Graphs with loop (d)}
\label{fig:cycle_d1}
\end{figure}

\begin{figure}[h!]
\begin{center}
\includegraphics[width=0.9\columnwidth]{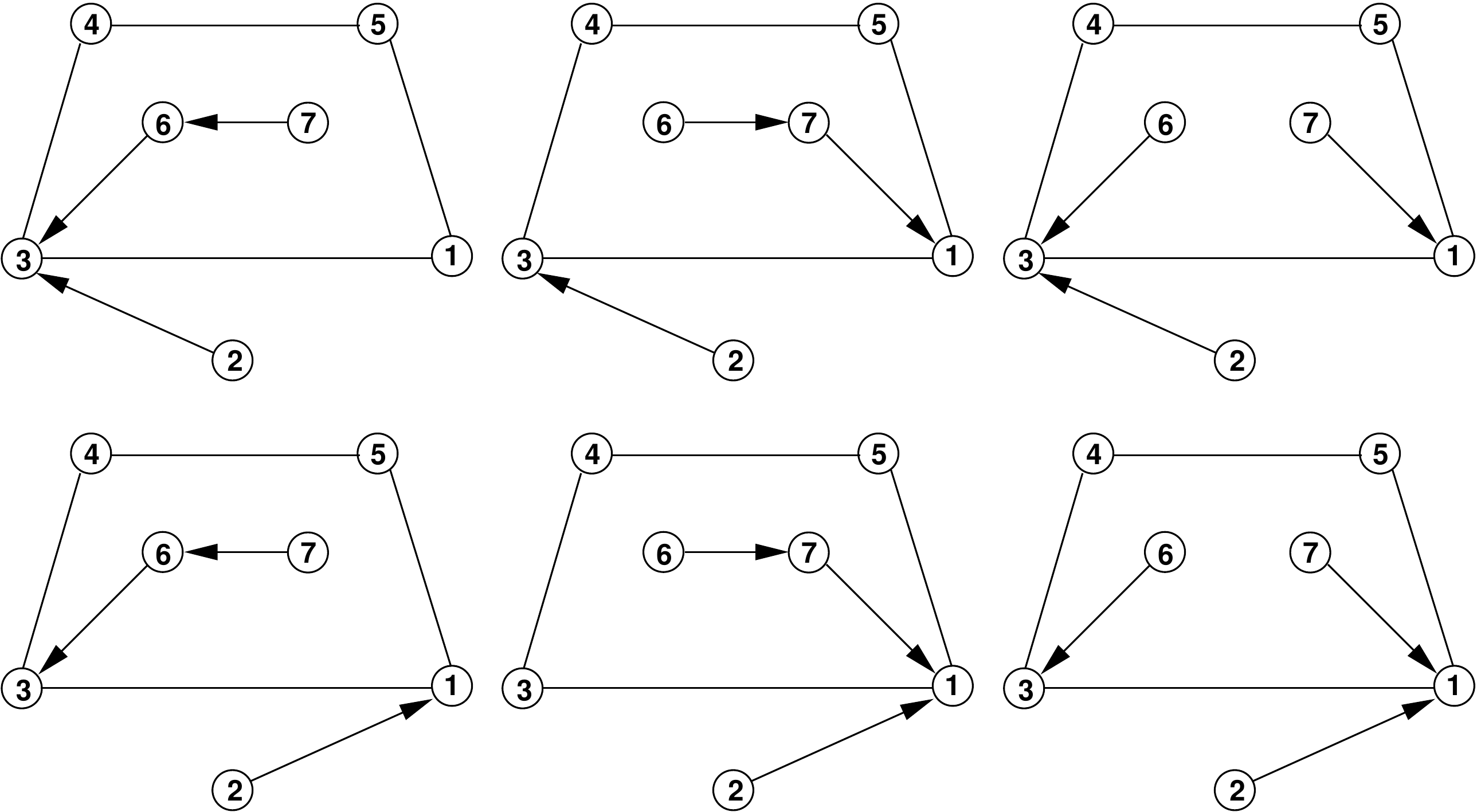}
\end{center}
\caption{Graphs with loop (e)}
\label{fig:cycle_e1}
\end{figure}

\begin{figure}[h!]
\begin{center}
\includegraphics[width=0.9\columnwidth]{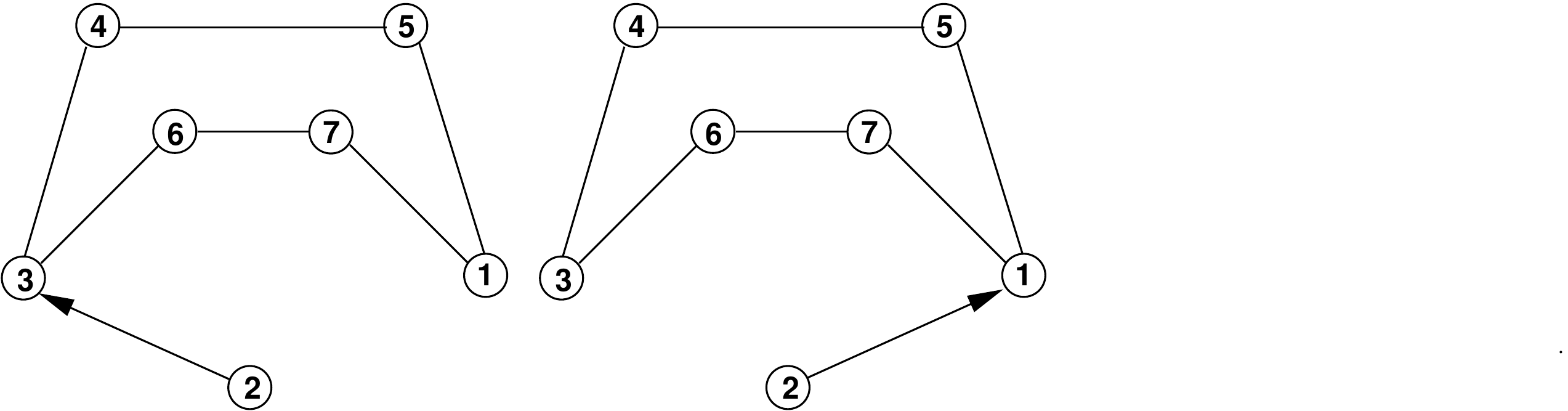}
\end{center}
\caption{Graphs with loop (f).}
\label{fig:cycle_f}
\end{figure}

From the undirected spanning trees one can construct the {\it flux diagrams} each of which has a single undirected cycle and directed branches that feed into this cycle. In total, there are 29 distinct flux diagrams for the 7-state model; all these 29 graphs are displayed in Figs.\ref{fig:cycle_a1}-\ref{fig:cycle_f}.

Using these flux diagrams, the exact expressions we obtain the contribution of the branches for every cycle 
\begin{eqnarray}
\label{Ra}
R_{a} &=&  k_{43}k_{67}k_{71}k_{51} + k_{43}k_{76}k_{63}k_{51} + k_{54}k_{43}k_{76}k_{63} \nonumber
\\ 
& & +k_{45}k_{51}k_{67}k_{71}+k_{54}k_{43}k_{67}k_{71} +k_{45}k_{51}k_{76}k_{63}\nonumber \\
& & +k_{45}k_{51}k_{63}k_{71} +k_{54}k_{43}k_{63}k_{71}+k_{43}k_{63}k_{51}k_{71} \\ 
\label{Rb}
R_{b} &=& k_{67}k_{71}+k_{76}k_{63}+k_{63}k_{71} \\ 
\label{Rc}
R_{c} &=& k_{43}k_{51}+k_{45}k_{51}+k_{54}k_{43} \\ 
\label{Rd}
R_{d} &=& k_{43}k_{51}k_{21}+k_{43}k_{51}k_{23}+k_{54}k_{43}k_{21}
\nonumber \\
& & +k_{54}k_{43}k_{23}+k_{45}k_{51}k_{21}+k_{45}k_{51}k_{23} \\ 
\label{Re}
R_{e} &=& k_{67}k_{71}k_{21}+k_{67}k_{71}k_{23}+k_{76}k_{63}k_{23} \nonumber \\ 
& & +k_{76}k_{63}k_{21}+k_{63}k_{71}k_{21}+k_{63}k_{71}k_{23} \\ 
\label{Rf}
R_{f} &=& k_{23}+k_{21}.
\end{eqnarray}

\begin{widetext}

\section{Entropy production rate in terms of cycle flux}
\label{app-EntropyProd}

The explicit derivation of the Eq.(\ref{eq:ss_entropy_c1}) is given below. 

\begin{eqnarray}
\sigma^{st}_{pr} &=& \dfrac{1}{2}k_{B} \sum_{ij} J_{ij} ln\bigg(\dfrac{k_{ij}}{k_{ji}}\bigg) \nonumber \\
&=& \dfrac{1}{2}k_{B}[J_{12} ln\bigg(\dfrac{k_{12}}{k_{21}}\bigg)+J_{23} ln\bigg(\dfrac{k_{23}}{k_{32}}\bigg)+J_{31} ln\bigg(\dfrac{k_{31}}{k_{31}}\bigg)+ \nonumber \\
&& J_{21} ln\bigg(\dfrac{k_{21}}{k_{12}}\bigg)+J_{32} ln\bigg(\dfrac{k_{32}}{k_{23}}\bigg)+J_{13} ln\bigg(\dfrac{k_{13}}{k_{31}}\bigg)+...] \nonumber \\
&=& \dfrac{1}{2}k_{B}[(J_{a}+J_{c}+J_{d}) ln\bigg(\dfrac{k_{12}}{k_{21}}\bigg)+(J_{a}+J_{c}+J_{d}) ln\bigg(\dfrac{k_{23}}{k_{32}}\bigg)+(J_{a}-J_{d}-J_{e}) ln\bigg(\dfrac{k_{31}}{k_{31}}\bigg)+ \nonumber \\
&& (-J_{a}-J_{c}-J_{d}) ln\bigg(\dfrac{k_{21}}{k_{12}}\bigg)+(-J_{a}-J_{c}-J_{d}) ln\bigg(\dfrac{k_{32}}{k_{23}}\bigg)+(-J_{a}+J_{d}+J_{e}) ln\bigg(\dfrac{k_{13}}{k_{31}}\bigg)+...]  \nonumber \\
&& \text{(using equation (28)-(31))} \nonumber \\
&=& \dfrac{1}{2}k_{B}[J_{a} ln \bigg(\dfrac{k_{12}k_{23}k_{31}}{k_{21}k_{32}k_{13}}\bigg)-J_{a} ln \bigg(\dfrac{k_{21}k_{32}k_{13}}{k_{12}k_{23}k_{31}}\bigg) +...] \nonumber \\
&=& \dfrac{1}{2}k_{B}[J_{a} ln \bigg(\dfrac{\Pi_{a,+}}{\Pi_{a,-}}\bigg)-J_{a} ln \bigg(\dfrac{\Pi_{a,-}}{\Pi_{a,+}}\bigg) +...] \nonumber \\
&& \text{(using equation(23))} \nonumber \\
&=& \dfrac{1}{2}[J_{a} k_{B} ln \bigg(\dfrac{\Pi_{a,+}}{\Pi_{a,-}}\bigg)-J_{a} k_{B} ln \bigg(\dfrac{\Pi_{a,-}}{\Pi_{a,+}}\bigg) +...] \nonumber \\
&=& \dfrac{1}{2}[J_{a} \Delta S_{a,+}-J_{a} \Delta S_{a,-} +...] \nonumber \\
&=& \dfrac{1}{2}[(J_{a,+}-J_{a,-}) \Delta S_{a,+}-(J_{a,+}-J_{a,-}) \Delta S_{a,-} +...] \text{(using equation(23))} \nonumber \\
&=& \dfrac{1}{2}[J_{a,+}\Delta S_{a,+}-J_{a,-}\Delta S_{a,+} -J_{a,+}\Delta S_{a,-}+J_{a,-}\Delta S_{a,-}+...] \nonumber \\
&=& \dfrac{1}{2}[J_{a,+}\Delta S_{a,+}+J_{a,-}\Delta S_{a,-} +J_{a,+}\Delta S_{a,+}+J_{a,-}\Delta S_{a,-}+...] \text{(using} \Delta S_{\kappa,-} = - \Delta S_{\kappa,+} )  \nonumber \\
&=&[J_{a,+}\Delta S_{a,+}+J_{a,-}\Delta S_{a,-}+...] \nonumber \\
&=& \sum_{\kappa} \left(J_{\kappa,+} \Delta S_{\kappa,+}+ J_{\kappa,-} \Delta S_{\kappa,-} \right) \nonumber \\
\end{eqnarray}
\end{widetext}


\end{document}